# Tutorial: Time-domain thermoreflectance (TDTR) for thermal property characterization of bulk and thin film materials

Puqing Jiang, Xin Qian, and Ronggui Yang[a]

Department of Mechanical Engineering, University of Colorado,

Boulder, Colorado 80309, USA

**Abstract:**

Measuring thermal properties of materials is not only of fundamental importance in understanding the transport processes of energy carriers (electrons and phonons in solids) but also of practical interest in developing novel materials with desired thermal properties for applications in energy conversion and storage, electronics, and photonic systems. Over the past two decades, ultrafast laser-based time-domain thermoreflectance (TDTR) has emerged and evolved as a reliable, powerful, and versatile technique to measure the thermal properties of a wide range of bulk and thin film materials and their interfaces. This tutorial discusses the basics as well as the recent advances of the TDTR technique and its applications in the thermal characterization of a variety of materials. The tutorial begins with the fundamentals of the TDTR technique, serving as a guideline for understanding the basic principles of this technique. Several variations of the TDTR technique that function similarly as the standard TDTR but with their own unique features are introduced, followed by introducing different advanced TDTR configurations that were developed to meet different measurement conditions. This tutorial closes with a summary that discusses the current limitations and proposes some directions for future development.

---

[a] Ronggui.Yang@Colorado.Edu

**Outline:**

Number of pages: 82; number of figures: 23; number of references: 244

# I. INTRODUCTION

Thermal conductivity $K$ is an important physical property that measures the capability of a material in conducting heat from high temperatures to low temperatures. Fourier's law of heat conduction states that the heat flux is proportional to the temperature gradient by the thermal conductivity. The thermal conductivity of a material is usually temperature-dependent and can be directionally dependent, *i.e.,* anisotropic. A temperature difference might also exist when heat flows across an interface of two materials. Interfacial thermal conductance (denoted as $G$) is thus defined as the ratio of the heat flux to the temperature drop across the interface to characterize the resistance to the heat flow.

The thin film or membrane form of many solid materials with a thickness ranging from one atomic layer to hundreds of microns has been extensively used in engineering systems to improve their mechanical, optical, electrical, and thermal functionalities, for example, in microelectronics,[1] photonics,[2] optical coatings,[3] solar cells, and thermoelectrics.[4] When the thickness of a thin film is smaller than the mean free paths or wavelengths of its heat carriers (electrons or phonons), the thermal conductivity of the film is significantly different from its bulk counterpart due to the geometric constraints. As a result, the thermal conductivities of most crystalline thin films are thickness-dependent and anisotropic, even if the thermal conductivities of their bulk counterparts are isotropic. Many other factors such as sample preparation/processing also strongly affect the thermal conductivity of thin films. Over the last two decades, there have been quite a few textbooks,[5, 6] monographs[7, 8] and review articles[9-15] published, indicating the vibrant research in this field driven by the technology needs.

Measuring thermal properties of materials is not only of fundamental importance in understanding the transport processes of energy carriers but also of practical interest in their wide applications relating to energy and information systems. Extensive efforts have been made since the 1950s for the characterization of thermal conductivity and thermal contact resistance in bulk materials.[16-21] However, most of the conventional thermal conductivity measurement techniques lack the spatial resolution to measure the temperature gradient/difference or the heat flux across a length scale below tens of microns.

Over the past two decades, ultrafast laser-based transient thermoreflectance (TTR) methods have emerged and evolved into a powerful and versatile technique[22-25] for characterizing thermal properties of a large variety of samples including both thin films[26-31] with a thickness down to ~20 nm,[26, 27] and bulk materials[32-35]. In particular, time-domain thermoreflectance (TDTR) technique has been applied for measuring materials with thermal conductivity values ranging from a high end at ~2000 W m$^{-1}$ K$^{-1}$ (like diamond[36] and graphite[24, 37]) to as low as 0.03 W m$^{-1}$ K$^{-1}$ (e.g., disordered WSe$_2$ films[26]). The heat capacity of new materials[38, 39] and thermal conductance of various solid/solid interfaces[40-47] have also been frequently measured using TDTR. As a pump-probe technique, TDTR has many advantages over other thermal conductivity measurement techniques. For example, it requires minimal sample preparation and no delicate design of electrical heaters or temperature sensors, and it works equally well in regular ambient conditions or through the window of a vacuum chamber.[48] Significant efforts have been devoted to advance the TDTR technique itself as well as its applications in thermal and phonon property characterization of various kinds of samples.

This tutorial reviews the fundamentals as well as the advanced configurations of the TDTR technique for measuring thermal properties of bulk and thin film materials, which should be of interest to the first-year graduate students as well as veteran researchers who are interested in developing new materials, understanding fundamental physics of thermal transport, or learning measurement techniques. This paper is organized as follows. Section II discusses the basics of the time-domain thermoreflectance technique, serving as a guideline in understanding the basic principles and implementations of this technique. Section III discusses several variations of the standard TDTR technique that have their own unique features while sharing the similarity with TDTR. Section IV discusses a variety of advanced TDTR configurations that have been developed to meet different measurement conditions of various kinds of samples. Finally, Section V gives a summary that discusses both the limitations and directions for future development.

## II. BASICS OF TDTR

### A. Basic principles and implementation

TDTR measures thermal properties through the reflectance change with temperature, known as the thermoreflectance. Samples are usually coated with a thin metal film acting as a transducer (see Figure 1(a) for an illustration), whose surface reflectance changes linearly with temperature when the temperature rise is small (typically <10% of the absolute temperature or <10 K, whichever is smaller). The thermoreflectance technique was first developed in the 1970s and 1980s, where continuous wave (CW) light sources were used for the heating and sensing.[49, 50] With the advancement of pico- and femtosecond pulsed laser in the 1980s, this technique was widely used for studying non-equilibrium electron-phonon interaction,[51-56] coherent phonon transport,[27, 57-61] picosecond acoustics,[62-72] optical properties,[73-75] thermal expansion coefficient[76] and thermal transport properties of thin films[77, 78] and interfaces.[40, 41, 43, 44, 46, 47, 79-85] Among these, Paddock and Eesley[77] were the first to measure thermal diffusivity of metal films using picosecond transient thermoreflectance, and Humphrey Maris' group have done a series of original research in developing picosecond ultrasonics using picosecond light pulses.[62-71] This technique has been further developed over the last two decades for measuring anisotropic thermal conductivity[24, 37, 86, 87] and probing spectral phonon transport.[88-92] The transient thermoreflectance technique can be implemented as both the TDTR method[22, 24, 25] and the frequency-domain thermoreflectance (FDTR) method[93, 94] (see Section III(A) for more details of FDTR). The TDTR method measures the thermoreflectance response as a function of delay time between the arrival of the pump and probe pulses on the sample surface, where the pump beam deposits a periodic heat flux on the sample surface and the probe beam detects the corresponding temperature change through the reflectance change. A schematic diagram of a typical TDTR setup is shown in Figure 1(b), while similar systems can also be found in some other References.[23, 25, 39, 95, 96]

In a typical TDTR setup, a mode-locked Ti:sapphire laser oscillator is used as the light source, which generates a train of 150 fs laser pulses at an 80-MHz repetition rate, with wavelengths centered around 800 nm. Other types of pulsed lasers may also be used as the light source, such as a dye laser with wavelengths centered at 632 nm[97] and a Yb:doped fiber laser with wavelengths centered at 1030 nm[98]. However, the Ti:sapphire laser is commonly used in the literature because of its ultrafast nature and excellent beam quality. A broadband Faraday optical isolator is installed at the outlet of the laser oscillator to prevent the laser beam from reflecting back into the oscillator. A half-wave plate installed before the isolator can be used to adjust the laser power for TDTR measurements. The laser beam is then split into a pump beam and a probe beam through a polarizing beam splitter (PBS), with the pump and the probe beams cross-polarized to each other. Another half-wave plate before the PBS can be used to adjust the power ratio between the pump and the probe beams. The pump beam is usually modulated at a frequency in the range 0.2-20 MHz using an electro-optic modulator (EOM) before being directed onto the sample through an objective lens. Alternatively, some TDTR setups use acousto-optic modulators (AOM) but with a generally limited modulation frequency of ≤1 MHz [78, 97, 99, 100] due to the much longer rise time of AOM. The EOM modulation frequency serves as the reference for lock-in detection. The probe beam is delayed with respect to the pump beam via a mechanical delay stage before being directed onto the sample through the same objective lens. The probe beam is usually expanded before the delay stage to minimize divergence over the long propagation distance. In some other configurations, the optical delay between the pump and the probe is achieved by advancing the pump instead.[22] This introduces a phase shift of $\exp(i2\pi f_{mod} t_d)$ in the signal, where $f_{mod}$ is the modulation frequency and $t_d$ is the delay time. Advancing the pump rather than delaying the probe has the advantage that the reflected probe beam as received by the detector would not be affected by the movement of the delay stage.[22]

The reflected probe beam is collected by a fast-response photodiode detector, which converts the optical signals into electrical signals. A radio-frequency (RF) lock-in amplifier is then used to pick up the signal from the strong background noise. In some early versions of TDTR systems, an inductor (not shown in the diagram) was inserted in the signal line between the photodiode detector and the lock-in amplifier with a 50-Ω input resistance. The reason is that the pump beam was usually modulated by a square-wave function (for example, using the Model 350-160 EOM and 25D amplifier from Conoptics) and all the undesired odd harmonics of the square wave were detected by the lock-in amplifier using square-wave multipliers (such as Model SR844 from Stanford Research Systems). The inductor thus serves as a resonant bandpass filter to remove the higher harmonics of the square-wave modulation function.[101] This kind of resonant filter becomes unnecessary if the pump beam is modulated by a sine-wave function instead, or if a digital lock-in amplifier with clean sine-wave multipliers (such as the Model HF2LI from Zurich Instruments) is used for the lock-in detection, both of which are inherently free of unwanted harmonics.

Alternatively, the thermal responses from both the fundamental frequency and the higher harmonics of the square wave can be both detected and simulated, which works equally well to derive thermal properties of the sample, see Ref. 102 for an example. The RF lock-in amplifier has outputs of an in-phase ($V_{in}$) signal and an out-of-phase ($V_{out}$) signal at the modulation frequency. In some TDTR systems, a double-modulation scheme is implemented,[23, 39] *i.e.*, a mechanical chopper is added in the probe path and two computer-based audio-frequency lock-in detections are implemented on the outputs of the RF lock-in amplifier. This double lock-in scheme removes coherent pick-up contributions to the signal at the radio frequency.[39]

To make the lock-in detection effective, the reflected pump beam must be blocked from the photodiode detector. Since the pump and probe beams are cross-polarized, the PBS between the objective lens and the detector can suppress > 99% of the reflected pump beam. However, due to the small value of thermoreflectance coefficient $dR/dT$ (~ $10^{-4}$ K$^{-1}$), a tiny amount of reflected pump beam even with an intensity less than 0.01% of the reflected probe beam is significant enough to distort the TDTR measurements.[95] Optical techniques are used to further suppress the reflected pump beam. One commonly used approach is the spatial separation of the pump and the probe beams, as depicted in the schematic in Figure 1(b). In this approach, the pump and the probe beams are in parallel and vertically separated by ~4 mm when entering the objective lens so that the reflected pump beam can be blocked by an aperture while only the specularly reflected probe beam is allowed to pass through the aperture and be received by the photodiode detector. Care should be taken to make sure that the reflected probe beam is not clipped by the aperture; otherwise, it would induce an amplitude modulation of the probe beam due to thermal expansion of the sample surface and affect the measurements. This approach has the merit of simple configuration and works well for optically smooth samples with surface roughness <15 nm.[12] For optically rough samples, the pump and the probe beams need to be spectrally separated so that the reflected pump beam can be eliminated using highly efficient optical filters. Spectral separation of the pump and the probe can also be accomplished through the "two-tint" approach[23] using sharp-edged filters that separate the spectra of the pump and probe by a small amount of ~7 nm in the wavelength, or through the "two-color" approach[25, 103] using second harmonic generation to double the frequency of either the pump or the probe beam. Both the two-tint and the two-color approach can effectively filter the reflected pump beam but at a significant sacrifice of the laser power. Such a laser power loss is a critical issue for measurements of highly conductive materials such as diamond, where a sufficiently high laser power intensity is desired to generate a high enough signal-to-noise ratio (SNR). In addition, the two-tint method has a high demand on the laser stability in both the intensity and the wavelength (like the one reported in Ref. [23]). Otherwise, the fluctuations in the laser wavelength would translate into fluctuations in the power intensities of the pump and probe beams and appear as noise in TDTR measurements (like the one reported in Ref. [95]). Alternatively, Sun and Koh[95] developed a simple empirical method to correct artifact signals induced by the reflected pump beam. Sun and Koh[95] observed that the leaked pump beam mostly affects the $V_{out}$ signal linearly as $V_{out} = V_{out,0} + aV_{leak}$, where $V_{out,0}$ is the out-of-phase thermal signal unaffected by the leaked pump beam, $a$ is a proportional constant, and $V_{leak}$ is the detected signal solely caused by the leaked pump beam. The constant $a$, which depends on factors including the photocurrent of the detector and the Q-factor of the resonant circuit, can be determined from a calibration experiment by measuring $V_{out}$ as a function of $V_{leak}$ and taking the gradient as $a = \Delta V_{out}/\Delta V_{leak}$. The TDTR signals can thus be conveniently corrected from the artifacts by monitoring the $V_{leak}$ signal during TDTR experiments. This approach was claimed to work well for all rough samples, as long as the Gaussian laser profiles on the sample surface are not distorted by the surface roughness.[95]

In most TDTR experiments, the signals are taken as the phase of the detected temperature response as carried by the reflected probe beam, computed as $\varphi = \tan^{-1}(V_{out}/V_{in})$, or equivalently the ratio $R = -V_{in}/V_{out}$, to derive the thermal properties of the sample. Using the phase signal for data reduction has the advantage over the amplitude signal as it does not require normalization.[22] In addition, the normalized amplitude signal loses sensitivity to the thermal properties of the sample especially when the $V_{out}$ signal is small. However, cautions need to be taken to correct the additional phase shift introduced by cables, electronic instruments, and optical components into the measured signals. Usually, this phase shift due to the instrumentation can be conveniently canceled based on the fact that the $V_{out}$ signal in TDTR experiments should be constant across the zero delay time.

For a Ti:sapphire oscillator centered around 800 nm, an Al thin layer with a thickness of ~100 nm is a common choice as the transducer layer due to its strong absorption (the 99% absorption depth is <60 nm)[104] and an exceptionally large thermoreflectance coefficient $dR/dT \sim 10^{-4}$ K$^{-1}$ at the 800-nm wavelength.[98, 105, 106] The thin metallic transducer layer not only makes it possible to assume a surface heat flux boundary condition and thus simplifying the analysis but also improves the overall SNR due to its large thermoreflectance coefficient.

In TDTR experiments, besides the unknown thermal properties of the sample layer to be derived, there are also many other input parameters (such as the laser spot size, thermal conductivity of the transducer layer, thickness and heat capacity of each layer in the sample stack) that affect the signals and need to be pre-determined as accurately as possible. The commonly adopted approaches to determine these input parameters are described briefly below.

For the characterization of the laser spot size, while many techniques such as the knife-edge method[25, 107] and patterned sample image method[94] can be used, an accurate and easy-to-use approach for TDTR is to measure the spatial correlation between the pump and probe focal spots.[39] In this method, the focused pump spot is swept across the probe spot at a positive delay time of ~100 ps and a high modulation frequency of 10 MHz. The root-mean-square (RMS) average of the pump and probe spots $w_0$ is derived by fitting the $V_{in}$ profile to a Gaussian function $V_{in} \sim \exp(-x_c^2/w_0^2)$, where $x_c$ is the offset distance. This simple approach works well for TDTR experiments because the TDTR signals are affected by the RMS average of the pump and probe spot sizes rather than their individual sizes, see more details in the thermal modeling in Section II(B). The thermal conductivity of the metal transducer layer can be characterized using four-point probe measurement of the electrical resistance and the Wiedemann-Franz law as $K_m = L_m T / RC'h_m$, where $L_m$ is the Lorenz number for the metal film, $T$ is the temperature, $R$ is the electrical resistance measured using four-point probe, $C'$ is a constant related to the sample lateral size and the spacing between the probe needles. The metal transducer thickness $h_m$ can be determined by picosecond acoustics[72]. To measure the electrical resistance $R$, a large and clean glass slide is usually placed next to the sample when metal film is deposited on the sample and the four-point probe measurement is performed on the metal film deposited on the glass slide instead. In most cases, the four-point probe measurements is performed only at room temperature. To obtain thermal conductivity of the metal film at other temperatures, the additional electrical resistivity of the metal film as compared to its pure bulk counterpart is assumed to be dominantly caused by impurity scattering, which is temperature-independent. The additional electrical resistivity of the metal film can thus be determined by the four-point probe at room temperature, which, combined with the temperature-dependent thermal conductivity of its bulk counterpart from the literature, gives the information of the temperature-dependent thermal conductivity of the metal film. The heat capacities of the metal transducer film and the sub-layers are usually assumed to be the same as their bulk counterparts[5] and can be obtained from the literature. For those samples whose heat capacities are not readily available from the literature or not measurable using standard techniques such as the differential scanning calorimetry, frequency-dependent TDTR experiments can simultaneously determine the thermal conductivity and heat capacity of the sample, see details in Section IV(A), the advanced TDTR configurations.

An estimation of the steady-state temperature rise is also important for TDTR experiments, as the temperature rise needs to be small (usually <10 K) to ensure the reflectance change is linearly proportional to the temperature change. Cahill[22] has derived the expression of steady-state temperature rise for a semi-infinite solid as $\Delta T = \frac{A_0}{2\sqrt{\pi} w_0} \frac{1}{K}$, where $A_0$ is the laser power absorbed by the sample. For thin layers and interfaces, the temperature rise can be estimated as $\Delta T = \frac{A_0}{\pi w_0^2} \frac{h}{K}$ and $\Delta T = \frac{A_0}{\pi w_0^2} \frac{1}{G}$, respectively. Note that these formulas are only for a rather crude estimation. A more in-depth discussion on the steady-state temperature rise in thermoreflectance experiments can be found in Ref. 108.

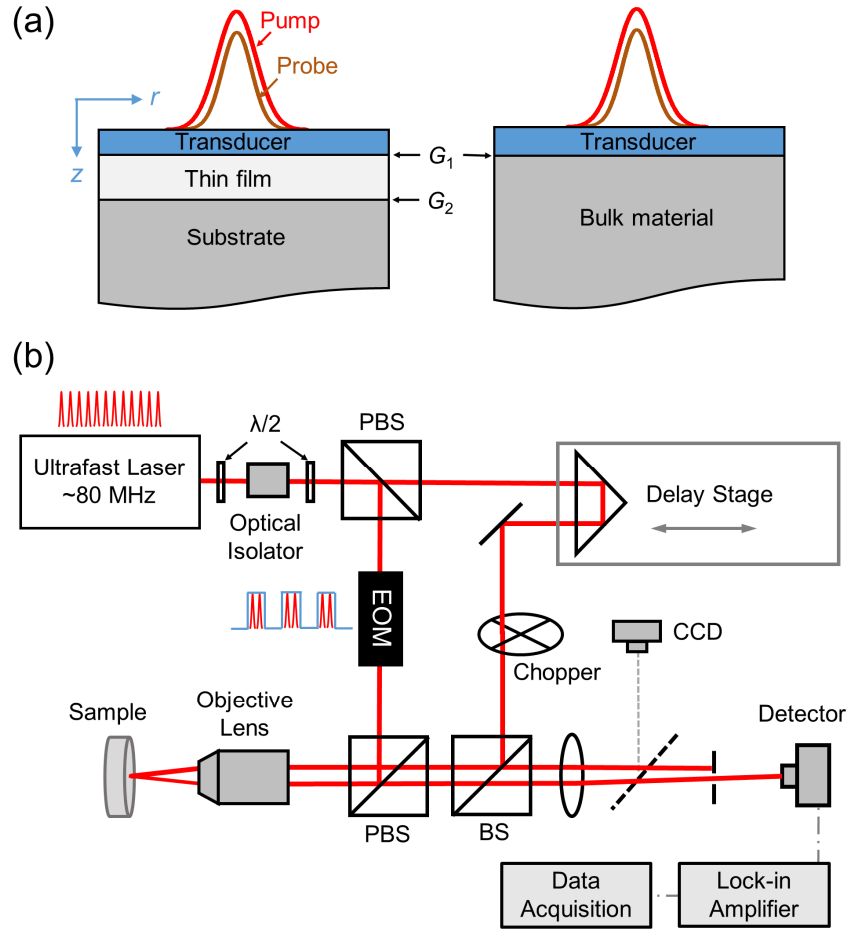

Figure 1. (a) Typical sample configurations of thin film and bulk materials measured using the TDTR technique with concentric pump and probe beams. Samples are usually coated with a thin metal transducer layer. (b) Schematic of a typical transient thermoreflectance setup. The acronyms PBS, BS, EOM, and λ/2 stand for polarizing beam splitter, beam splitter, electro-optic modulator, and half-wave plate, respectively.

## B. Thermal transport modeling and signal processing

The determination of thermal transport properties by TDTR is typically accomplished by adjusting free parameters (the unknown thermal properties) in a thermal transport model to obtain the best fit between the model prediction and the experimental data. Although significant advances have been made on observing the interesting or even counter-intuitive phenomena in nanoscale thermal transport, most studies still use the effective thermal conductivity of nanostructured materials as a common language to interpret the findings and to communicate across different communities in physics, materials, and engineering. Similarly, while TDTR can be used to study non-equilibrium phenomena such as quasi-ballistic phonon transport[90, 91, 109-111] and electron-phonon coupling[55] due to its ability to detect temperature evolutions at micrometer-scale and picosecond-scale resolutions, the majority of the experimental work in the literature deduced thermal properties based on Fourier's law of heat conduction with an effective thermal conductivity. However, care should be taken when applying a diffusive thermal transport model to systems where a significant fraction of heat is carried by phonons with long mean-free paths (MFPs).[90-92, 112]

The early versions of TDTR data reduction schemes can only be applied to some simplified cases with many assumptions.[77, 78] The key advance in the data analysis of TDTR experiments was made by Cahill in 2004.[22] The three-dimensional heat diffusion equation through multi-layered structures was solved with the consideration of the so-called "pulse accumulation" effect, *i.e.*, the response of a new pulse should account for the previous pulse that has not fallen to a negligible value. Schmidt[24] further extended the model for anisotropic heat conduction where thin films are very likely anisotropic due to the size effect of energy carriers. While more modeling efforts are needed for TDTR measurements for various sophisticated situations (see Section IV for details), the thermal model for the most commonly encountered case in TDTR experiments is outlined here. A schematic of the sample structure in TDTR experiments is shown in Figure 2. In this basic model, both the laser profile and the thermal properties of the sample are assumed to be axisymmetric rather than in-plane anisotropic (see Section IV(C)); both the heating and sensing occur on the top surface of the sample; and the different phonon modes are assumed to be in thermal equilibrium so that effective thermal properties $K_r$, $K_z$, $C$, and $G$ can be assumed for each layer. Thermal modeling of TDTR experiments involves two steps: one is to solve heat diffusion in a multilayer stack, and the other is to model the data acquired from TDTR experiments.

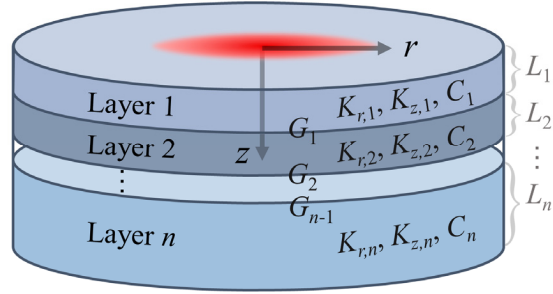

Figure 2. Schematic of a multi-layered sample configuration for thermal modeling of TDTR experiments.

## B1. Solution of heat diffusion equation in a multilayer stack

There have been several publications on the solution of one-dimensional heat diffusion equation through layered structures.[113-115] Cahill[22] extended Feldman's algorithm[114, 115] to three-dimensional heat conduction and applied it to calculate thermal responses in TDTR. Schmidt,[101] on the other hand, adopted the approach described in *Conduction of Heat in Solids* by Carslaw and Jaegar[113] and extended it to account for three-dimensional and anisotropic heat conduction. Here, we summarize the solution of the anisotropic heat diffusion equations in cylindrical coordinates for a multilayered system using a quadrupole approach.[116] The governing equation, which is based on Fourier's law of heat conduction with apparent thermal property values assumed, is written as:

$$C\frac{\partial T}{\partial t} = \frac{\eta K_z}{r}\frac{\partial}{\partial r}\left(r\frac{\partial T}{\partial r}\right) + K_z\frac{\partial^2 T}{\partial z^2} \tag{2.1}$$

where $K_r$ and $K_z$ are the thermal conductivities of the sample in the radial and through-plane directions, respectively, $\eta = K_r/K_z$ is the anisotropic parameter for the thermal conductivity, and $C$ is the volumetric heat capacity. Applying the Fourier transform to the time variable $t$ and the Hankel transform to the radial coordinate $r$,[117] this parabolic partial differential equation can be simplified into an ordinary differential equation as:

$$\frac{\partial^2 \Theta}{\partial z^2} = \lambda^2 \Theta \qquad (2.2)$$

where $\Theta$ is the temperature in frequency domain, $\lambda^2 = 4\pi^2 k^2 \eta + i\omega C/K_z$, $k$ is the Hankel transform variable, and $\omega$ is the angular frequency.[117] The general solution of Eq. (2.2) can be written as:

$$\Theta = e^{\lambda z} B^+ + e^{-\lambda z} B^- \qquad (2.3)$$

where $B^+$ and $B^-$ are complex constants to be determined based on the boundary conditions.

The heat flux can be obtained from the temperature (Eq. (2.3)) and Fourier's law of heat conduction $Q = -K_z(d\Theta/dz)$ as:

$$Q = \gamma(-e^{\lambda z} B^+ + e^{-\lambda z} B^-) \qquad (2.4)$$

where $\gamma = K_z \lambda$.

It would be convenient to re-write Eqs. (2.3) and (2.4) in the matrices (quadrupoles) as:

$$\begin{bmatrix} \Theta \\ Q \end{bmatrix}_{i,z=L} = \begin{bmatrix} 1 & 1 \\ -\gamma_i & \gamma_i \end{bmatrix} \begin{bmatrix} e^{\lambda L} & 0 \\ 0 & e^{-\lambda L} \end{bmatrix}_i \begin{bmatrix} B^+ \\ B^- \end{bmatrix}_i = [N]_i \begin{bmatrix} B^+ \\ B^- \end{bmatrix}_i \qquad (2.5)$$

The constants $B^+$ and $B^-$, which can be viewed as the properties of the $i$-th layer, can also be obtained from the surface temperature and heat flux of that layer by setting $L = 0$ in Eq. (2.5) and performing its matrix inversion:

$$\begin{bmatrix} B^+ \\ B^- \end{bmatrix}_i = \frac{1}{2\gamma_i} \begin{bmatrix} \gamma_i & -1 \\ \gamma_i & 1 \end{bmatrix} \begin{bmatrix} \Theta \\ Q \end{bmatrix}_{i,z=0} = [M]_i \begin{bmatrix} \Theta \\ Q \end{bmatrix}_{i,z=0} \qquad (2.6)$$

For heat flow across the interface, there is another matrix to relate the temperature and heat flux at the bottom of the upper layer to those at the top of the underlayer as

$$\begin{bmatrix} \Theta \\ Q \end{bmatrix}_{i+1,z=0} = \begin{bmatrix} 1 & -1/G \\ 0 & 1 \end{bmatrix}_i \begin{bmatrix} \Theta \\ Q \end{bmatrix}_{i,z=L} = [R]_i \begin{bmatrix} \Theta \\ Q \end{bmatrix}_{i,z=L} \qquad (2.7)$$

where $G$ is the interface conductance between the two layers.

The temperature and heat flux on the top surface of the multilayer stack can thus be related to those at the bottom of the substrate as:

$$\begin{bmatrix} \Theta \\ Q \end{bmatrix}_{i=n,z=L_n} = [N]_n [M]_n \cdots [R]_1 [N]_1 [M]_1 \begin{bmatrix} \Theta \\ Q \end{bmatrix}_{i=1,z=0} = \begin{bmatrix} A & B \\ C & D \end{bmatrix} \begin{bmatrix} \Theta \\ Q \end{bmatrix}_{i=1,z=0} \qquad (2.8)$$

Applying the boundary condition that at the bottom of the substrate $Q_{z \to \infty} = 0$ yields $0 = C\Theta_{i=1,z=0} + DQ_{i=1,z=0}$. The Green's function $\hat{G}$, which is essentially the detected temperature response due to the applied heat flux of unit strength,[22] can thus be solved as

$$\hat{G}(k,\omega) = \frac{\Theta_{i=1,z=0}}{Q_{i=1,z=0}} = -\frac{D}{C} \qquad (2.9)$$

With the Green's function $\hat{G}$ determined, the detected temperature response is simply the product of $\hat{G}$ and the heat source function in the frequency domain. See details the following section.

### *B2. Modeling of signals acquired in TDTR experiments*

The next step is to understand and simulate the signals acquired in TDTR experiments. Assuming a Dirac delta function for the laser pulses, which is a valid assumption considering the ultrashort pulse duration (<500 fs) compared to the interval between pulses (~12.5 ns), the intensity of the pump beam in real space and time domain is expressed as:

$$p_1(r,t) = \frac{2A_1}{\pi w_1^2} \exp\left(-\frac{2r^2}{w_1^2}\right) e^{i\omega_0 t} \sum_{n=-\infty}^{\infty} \delta(t - nT_s - t_0) \qquad (2.10)$$

It is a train of delta functions modulated by a sinusoidal function at frequency $\omega_0$. The laser repetition frequency is $f_{rep}$ with a period $T_s = 1/f_{rep} = 2\pi/\omega_s$; $A_1$ is the average power of the pump beam, which has a Gaussian distribution in space with a $1/e^2$ radius of $w_1$; $t_0$ is the arbitrary time shift of laser pulses. In most TDTR implementations, the pump beam is modulated by a square wave but with the higher harmonics of the square wave being removed by a bandpass filter, while in some other TDTR configurations, the pump beam is modulated by a sine wave. Thus, only the fundamental frequency of the modulation is of interest, as illustrated in Figure 3(a).

The frequency domain expression of the pump beam intensity (illustrated in Figure 3(b)) can thus be obtained by the Hankel transform on space and the Fourier transform on time[117] of Eq. (2.10):

$$P_1(k,\omega) = A_1 \exp(-\pi^2 k^2 w_1^2 / 2) \omega_s \sum_{n=-\infty}^{\infty} \delta(\omega - \omega_0 - n\omega_s) e^{-in\omega_s t_0} \qquad (2.11)$$

The surface temperature response in the frequency domain is the product of the heat input $P_1$ (Eq.(2.11)) and the thermal response function of the system $\hat{G}$ (Eq.(2.9)):

$$\Theta(k,\omega) = P_1(k,\omega)\hat{G}(k,\omega) \qquad (2.12)$$

Inverse Hankel transform on the equation above gives the temperature distribution on the surface as a result of modulated pump heating

$$\Theta(r,\omega) = \int_0^\infty P_1(k,\omega)\hat{G}(k,\omega) J_0(2\pi kr) 2\pi k dk \qquad (2.13)$$

Inverse Fourier transform of $\Theta(r,\omega)$ gives the surface temperature response $\theta(r,t)$. An example of $\theta(r,t)$ for a simplified case of one-dimensional pulse heating of a semi-infinite solid is shown by the blue solid curve in Figure 3(c). Note that some pulses appear to have a negative contribution to the temperature response in Figure 3(c). The reason is that the DC component of the pump heating is neglected in the mathematical model here for simplicity (similar to the practice in Ref.

24). This simplification is justified because the DC component is rejected by the lock-in amplifier in TDTR experiments. An example of temperature responses with the DC component being considered can also be found in Ref. 102.

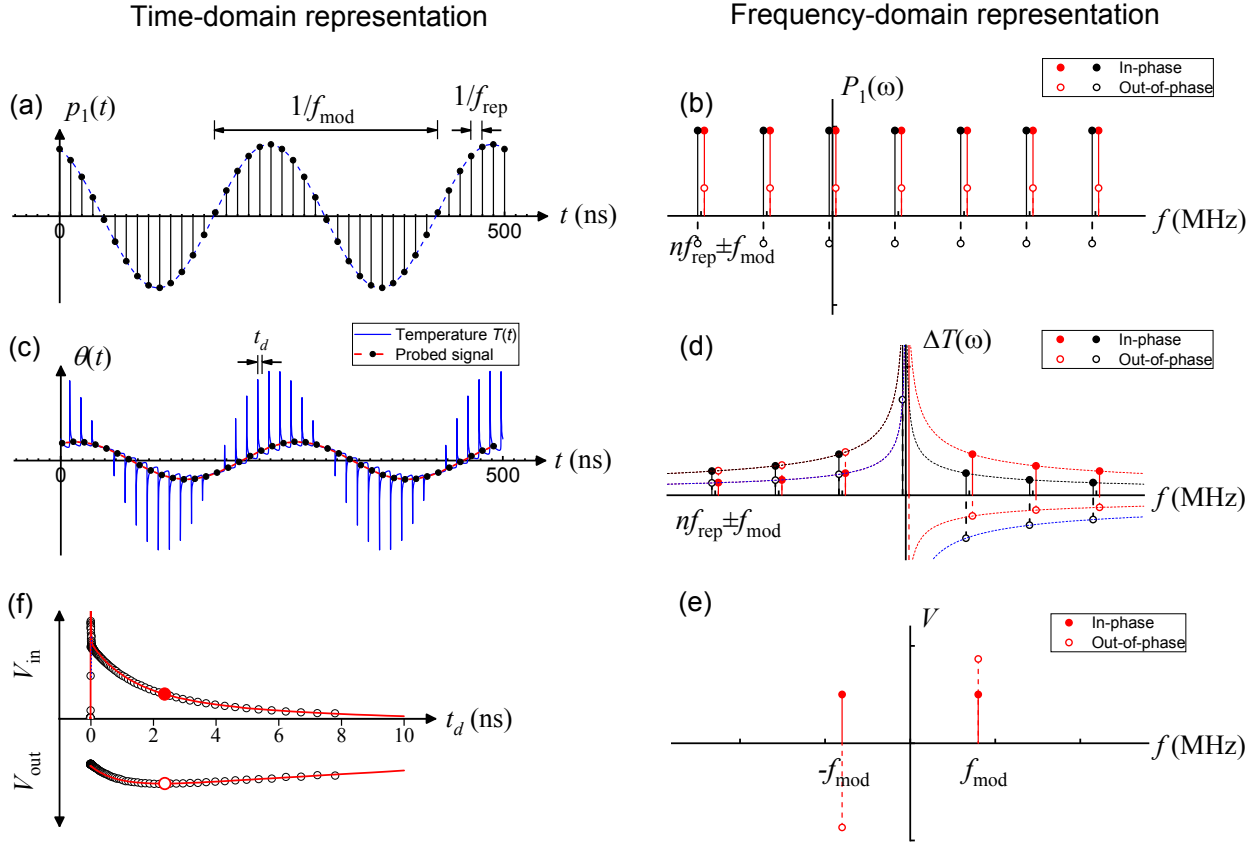

Figure 3. The signal detection mechanism in TDTR experiments in both the time and frequency domains. (a) Modulated pump pulses by a sine wave in the time domain. (b) Modulated pump pulses in the frequency domain. (c) Surface temperature response (solid curves) and probed signals at a fixed delay time (dashed curve) in the time domain. (d) Temperature response at a fixed delay time that contains in-phase components (solid symbols) and out-of-phase components (open symbols) in the frequency domain. (e) Frequency-domain representation of the in-phase and out-of-phase signals detected by the lock-in amplifier at a fixed delay time. (f) The measurements are repeated at different delay positions, yielding the $V_{in}$ and $V_{out}$ signals as a function of the delay time. Figures similar to plots (a) and (c) can also be found in Ref. 24.

Another train of time-delayed laser pulses is used to detect the temperature change due to periodic heating. The probe beam also has a Gaussian distribution of intensity in space and is delayed by time $t_d$ with respect to the pump beam:

$$p_2(r,t) = \frac{2A_2}{\pi w_2^2} \exp\left(-\frac{2r^2}{w_2^2}\right) \sum_{m=-\infty}^{\infty} \delta(t - mT_s - t_0 - t_d) \quad (2.14)$$

Fourier transform of the probe beam is

$$P_2(r,\omega) = \frac{2A_2}{\pi w_2^2} \exp\left(-\frac{2r^2}{w_2^2}\right) \sum_{m=-\infty}^{\infty} \delta(\omega - m\omega_s) e^{-im\omega_s(t_0 + t_d)} \quad (2.15)$$

The probe beam samples a weighted average of the temperature distribution in real space as a convolution between the temperature profile and the probe beam profile:

$$\Delta\Theta(\omega) = \int_0^{\infty} \left( \frac{1}{2\pi} \int_{-\infty}^{\infty} \Theta(r,\zeta) P_2(r, \omega - \zeta) d\zeta \right) 2\pi r \, dr \quad (2.16)$$

which, after some derivations (see Ref. [101] for more details), can be simplified as:

$$\Delta\Theta(\omega) = A_1 \int_0^{\infty} \sum_{n=-\infty}^{\infty} \delta(\omega - \omega_0) \hat{G}(k, \omega_0 + n\omega_s) \exp(in\omega_s t_d) \exp\left(-\pi^2 k^2 (w_1^2 + w_2^2)/2\right) 2\pi k \, dk \quad (2.17)$$

If we define the temperature response due to the harmonic heating at frequency $\omega$ as

$$\Delta T(\omega) = A_1 \int_0^{\infty} \hat{G}(k,\omega) \exp\left(-\pi^2 k^2 w_0^2\right) 2\pi k \, dk \quad (2.18)$$

where $w_0 = \sqrt{(w_1^2 + w_2^2)/2}$ is the root mean square (RMS) average of the pump and probe $1/e^2$ radii, Eq. (2.17) can then be re-written as

$$\Delta\Theta(\omega) = \delta(\omega - \omega_0) \sum_{n=-\infty}^{\infty} \Delta T(\omega_0 + n\omega_s) \exp(in\omega_s t_d) \quad (2.19)$$

Inverse Fourier transform of Eq. (2.19) gives the probed signal in the time domain at delay time $t_d$ as:

$$\Delta T(t) = e^{i\omega_0 t} \sum_{n=-\infty}^{\infty} \Delta T(\omega_0 + n\omega_s) \exp(in\omega_s t_d) \qquad (2.20)$$

This is a sinusoidal function as depicted by the red curve in Figure 3(c). Its amplitude is the summation of temperature responses $\Delta T(\omega)$ at the multiple frequencies $\omega_0 + n\omega_s$. The temperature responses $\Delta T(\omega)$ in the frequency domain are illustrated in Figure 3(d). By multiplying the input signal $\Delta T(t)$ by two lock-in reference signals with a phase offset of $\pi/2$ and the same frequency $\omega_0$, and using low-pass filters to remove the AC components of the outputs of the multipliers,[118] the lock-in amplifier picks up the in-phase and out-of-phase components of the signal $\Delta T(t)$ at the modulation frequency $\omega_0$ at a fixed delay time $t_d$ as:

$$V_{in} = \frac{1}{2} \sum_{n=-\infty}^{\infty} \left[ \Delta T(\omega_0 + n\omega_s) + \Delta T(-\omega_0 + n\omega_s) \right] \exp(in\omega_s t_d) \qquad (2.21)$$

$$V_{out} = -\frac{i}{2} \sum_{n=-\infty}^{\infty} \left[ \Delta T(\omega_0 + n\omega_s) - \Delta T(-\omega_0 + n\omega_s) \right] \exp(in\omega_s t_d) \qquad (2.22),$$

as illustrated in Figure 3(e)). By changing the delay time and repeating the lock-in detection, the TDTR data acquisition is completed, as shown in Figure 3(f), which contains an in-phase $V_{in}$ and an out-of-phase $V_{out}$ component. The in-phase signal $V_{in}$ represents the surface temperature change because of the pulse heating. The decay rate of $V_{in}$, which represents the cooling process of sample surface due to heat dissipation, directly relates to the thermal diffusivity of the sample. The length scale affected by the pulsed heating is characterized by a thermal diffusion length $d_f = \sqrt{Kt_d/C}$, where $K$ and $C$ are the thermal conductivity and volumetric heat capacity of the sample, respectively.[119] The out-of-phase $V_{out}$, on the other hand, is mainly due to the modulated continuous heating of the sample at the modulation frequency $\omega_0$. The depth over which the modulated continuous heating penetrates through can be estimated by a thermal penetration depth $d_p = \sqrt{K/\pi f_{mod} C}$, where $f_{mod} = \omega_0/2\pi$ is the modulation frequency.[90] When processing the experimental data, the ratio between the in-phase and out-of-phase signals $-V_{in}/V_{out}$ is usually fitted[22] by the thermal transport model to derive the thermal properties of the sample.

## C. Measurements of through-plane thermal conductivity and interface thermal conductance

The complexity of the thermal model described above potentially also enables TDTR for determining many thermal properties involved in the model, including the heat capacity $C$, in-plane thermal conductivity $K_r$, cross-plane thermal conductivity $K_z$, and interface conductance $G$. However, it imposes great challenges to extract the target property with multiple unknown parameters. Therefore, the number of unknowns needs to be reduced when performing TDTR measurements. For example, the heat capacity is usually pre-determined from separate measurements using differential scanning calorimetry or known from literature values, with the unknown parameters often reduced to $K_r$, $K_z$, and $G$. The number of unknown parameters can be further reduced if the samples are isotropic, so that there is no need to differentiate the in-plane and through-plane thermal properties. Furthermore, most practitioners use a large laser spot size ($w_0$=5-20 μm) and a high modulation frequency ($f$=1-10 MHz), resulting in quasi-one-dimensional thermal transport. In this case, the TDTR signals are most sensitive to the thermal properties in the through-plane direction ($K_z$ and $G$). Therefore, TDTR is often used to measure the through-plane thermal conductivity $K_z$ of the sample and the transducer/sample interface conductance $G$ without much elaboration.[12, 40, 120-125]

Figure 4(a, b) shows an example of TDTR in-phase and out-of-phase data measured on a GaN sample with a 100 nm Al transducer as a function of the delay time. The modulation frequency was $f$ = 10 MHz and the laser spot size was $w_0$ = 12 μm. As mentioned before, the ratio between in-phase and out-of-phase signals, $-V_{in}/V_{out}$, is usually compared with the thermal model simulation

over the delay time range 0.1-8 ns to extract the thermal properties, as shown in Figure 4(c, d). The through-plane thermal conductivity $K_z$ of the sample and the transducer/sample interface conductance $G$ affect the ratio signal $R=-V_{in}/V_{out}$ in different manners over the 0.1-8 ns delay time range. As indicated by the ±30% bounds of the best-fitted value in Figure 4(c, d), $K_z$ of the sample mainly affects the amplitude of the ratio signal, whereas the interface conductance $G$ mainly affects the gradient of the ratio signal. Both $K_z$ and $G$ can thus be simultaneously determined from one single set of TDTR measurements.

Whether an unknown thermal property can be measured with good accuracy from the experiments can be quantitatively analyzed through the definition of a sensitivity coefficient:[126]

$$S_\xi = \frac{\partial \ln R}{\partial \ln \xi} = \frac{\xi}{R}\frac{\partial R}{\partial \xi} \tag{2.23}$$

Here the sensitivity coefficient $S_\xi$ represents how sensitive the measured signal $R=-V_{in}/V_{out}$ is to the parameter $\xi$. Both the magnitude and the sign of $S_\xi$ have specific meanings, i.e., 1% increase in the parameter $\xi$ will result in $S_\xi \times 1\%$ increase in the ratio signal $R$. Therefore, $S_\xi = 0$ means that the TDTR signals are not affected by the parameter $\xi$, while a larger amplitude of $S_\xi$ indicates that the TDTR measurements are more strongly dependent on $\xi$.[37] Sensitivity analysis is a powerful tool to guide us on how the experiments should be best designed to achieve the most accurate results. Figure 4(e) shows the sensitivity coefficients of the ratio signal $R$ to $K_z$ and $G$ of the 100 nm Al/GaN sample as a function of the delay time at different modulation frequencies of 1 and 10 MHz. Comparison of the sensitivity coefficients at 1 and 10 MHz suggests that the modulation frequency has a very little effect on the sensitivity to $K_z$ but affects the sensitivity to $G$ dramatically. At 10 MHz, the sensitivity to $G$ changes from positive to negative over the delay time range of 0.1-10 ns, indicating that $G$ mainly affects the gradient of the ratio signal. On the other hand, the sensitivity to $K_z$ remains positive over the whole delay time range, indicating that $K_z$ mainly affects the amplitude of the ratio signal.

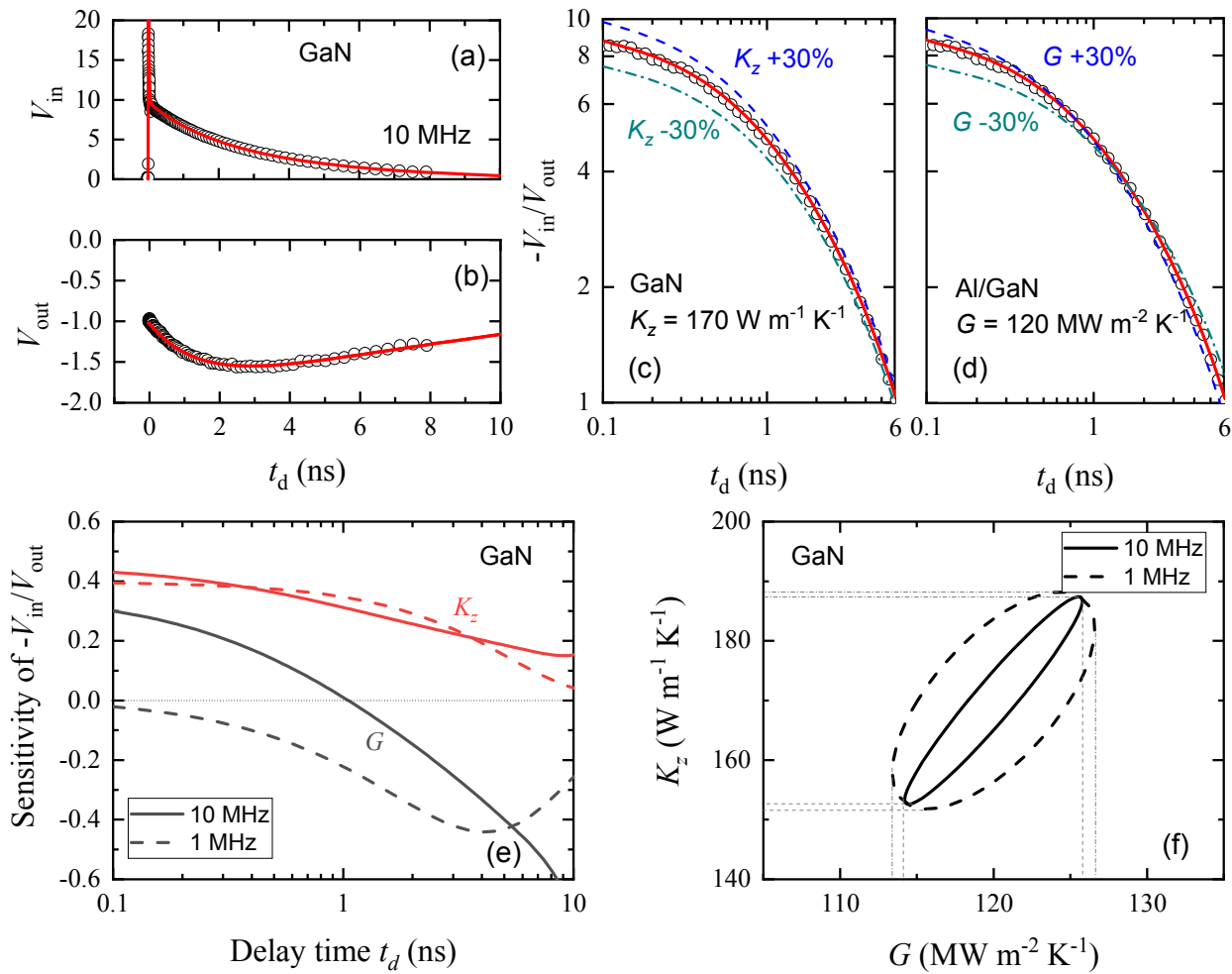

Figure 4. (a, b) An example of TDTR in-phase and out-of-phase data measured on a GaN substrate with a 100 nm Al transducer as a function of the delay time. (c, d) The ratio between in-phase and out-of-phase signals, $-V_{in}/V_{out}$, as a function of delay time is compared with the thermal modeling to extract the thermal properties including thermal conductivity of the substrate and thermal conductance of the metal/substrate interface. (e) Sensitivity coefficients of the TDTR signal $R = -V_{in}/V_{out}$ to the thermal conductivity of the substrate and interface thermal conductance of the GaN sample with 100 nm Al transducer as a function of delay time at a modulation frequency of 1 and 10 MHz, respectively. (f) Confidence ranges of $K_z$ and $G$ of the GaN substrate with a 100 nm Al transducer determined using the least-squares regression method when these two parameters are simultaneously determined from TDTR experiments at modulation frequencies of 10 MHz (solid curve) and 1 MHz (dashed curve), respectively.

The uncertainties of multiple fitting parameters can be estimated using a multivariate error-propagation formula based on Jacobian matrices derived by Yang et al.[127] This multivariate error-propagation formula has been validated by the Monte Carlo method[127] but is much more computationally efficient: it takes only a few seconds to complete the analysis on a desktop computer using MATLAB scripts rather than a few days as needed by the Monte Carlo method. Figure 4(f) shows the confidence range of $K_z$ and $G$ of the GaN sample coated with 100 nm Al estimated by the least-squares regression method when these two parameters are fitted simultaneously from TDTR measurements. The results show that $K_z$ and $G$ can be determined with an uncertainty of ~10% and ~5%, respectively, at both frequencies of 1 and 10 MHz. The major sources of measurement uncertainty come from the thermal capacitance (volumetric heat capacity times the film thickness, $C*h$) of the transducer for high-frequency measurements and from the laser spot size $w_0$ for low-frequency measurements.

The reliability of the TDTR technique for thermal conductivity characterization has been verified on a series of standard bulk material samples over a wide range of thermal conductivity values. Figure 5 shows the benchmark studies of TDTR conducted by Zheng et al.[122] and Wilson and Cahill[128] over a wide range of thermal conductivity from 0.2 to 2000 W m$^{-1}$ K$^{-1}$. The through-plane thermal conductivities of these materials determined by TDTR are in good agreement with the literature values with an overall experimental uncertainty of ±8%.

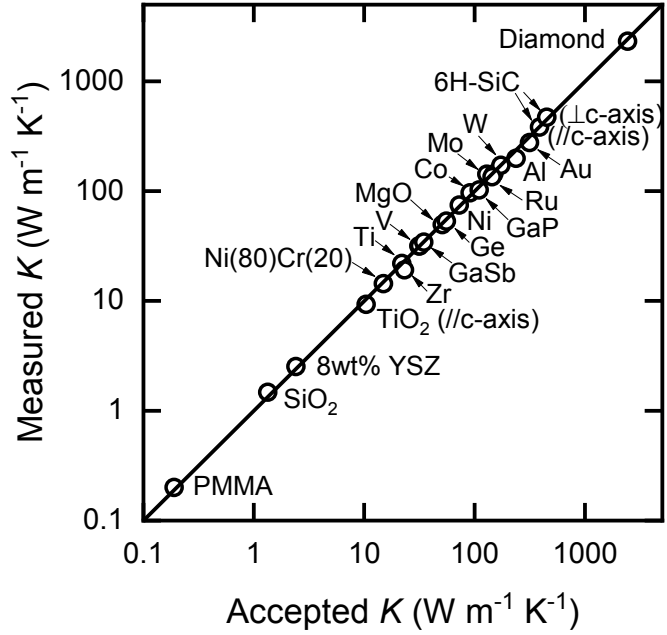

Figure 5. Benchmark TDTR measurements of the through-plane thermal conductivity of a wide range of metals and semiconductors. The thermal conductivity measured using TDTR are in good agreement with accepted literature values, generally within ±8%. The data for PMMA measured using TDTR is from Ref. [129], with the accepted literature value from Ref. [130]. The data for TiO$_2$ measured using TDTR is from Ref. [37], with the accepted literature value from Ref. [131]. The other data are reproduced with permission from Zheng *et al.*[122] Copyright 2007 by Acta Materialia Inc. and from Wilson and Cahill[128] Copyright 2015 by AIP Publishing.

Besides the bulk materials, TDTR has also been widely employed to measure the through-plane thermal conductivity of many thin film materials. Note that TDTR cannot measure any arbitrarily thin films but usually requires the films to be "thermally thick", *i.e.*, the film thickness should be larger than the through-plane thermal penetration depth, $h_{\text{film}} > d_{p,z}$, defined as $d_{p,z} = \sqrt{K_z/\pi f_{mod} C}$. For thermally thin films with $h_{\text{film}} < d_{p,z}$, the measurements of $K_z$ of the film would be challenging due to the low sensitivity to $K_z$. An advanced dual-frequency TDTR configuration can be employed to improve the accuracy in determining $K_z$ of thermally thin films, see Section IV(B) for more details.

The thermal conductivity of superlattices and nanolaminates[27, 59, 78, 79, 123, 132-138] (see Figure 6(b)) for an example) has been extensively characterized using TDTR over the last 15 years. To conduct the TDTR measurements, superlattice samples are usually configured in a scheme as shown in Figure 6(a). The superlattice sample is treated as a homogeneous thin film with thickness $h_{film}$ and an effective thermal conductivity. The through-plane thermal conductivity of superlattices has been found to depend on several factors including the interface density (=$2/d_{SL}$, where $d_{SL}$ is the period thickness), the total film thickness, and the thickness ratio between the two materials. Figure 6(c-e) summarizes a few typical results of $K_z$ of superlattices measured using TDTR reported in the literature. A higher interface density generally reduces the $K_z$ of superlattices (see Figure 6(c)) due to the interface thermal resistance introduced by the periodic boundaries.[79, 137] However, the $K_z$ of superlattices was also found to depend on the total film thickness, as shown in Figure 6(d), indicating that a significant fraction of phonons can coherently transport across the interfaces and are only scattered by the film boundary.[27, 137] Besides the two boundary scattering length scales, the thickness ratio of the two materials was also found to affect the thermal conductivity significantly, as illustrated in Figure 6(e) for the example of InAlAs/InGaAs superlattices.[136]

Over the past two decades, TDTR has been applied to characterize the thermal properties of a wide range of bulk and thin film materials and their interfaces. Figure 7 summarizes the thermal conductivity and interface conductance of some typical samples measured as a function of temperature using TDTR. It shows that TDTR has been applied across a broad range of thermal conductivity and interface thermal conductance. Ultra-low thermal conductivities of full-dense solid materials were reported for fullerene derivatives at room temperature (~0.05 W m$^{-1}$ K$^{-1}$)[125] and disordered layered WSe$_2$ thin films at cryogenic temperatures (~0.03 W m$^{-1}$ K$^{-1}$)[26], both measured using TDTR. The lowest thermal conductance was reported for the interface between Bi and hydrogen-terminated diamond, with a value of ~8 MW m$^{-2}$ K$^{-1}$ at room temperature,[41] while the highest thermal conductance observed to-date is for metal/metal interfaces, such as Al/Cu, with a $G$ of ~3700 MW m$^{-2}$ K$^{-1}$ at room temperature.[40]

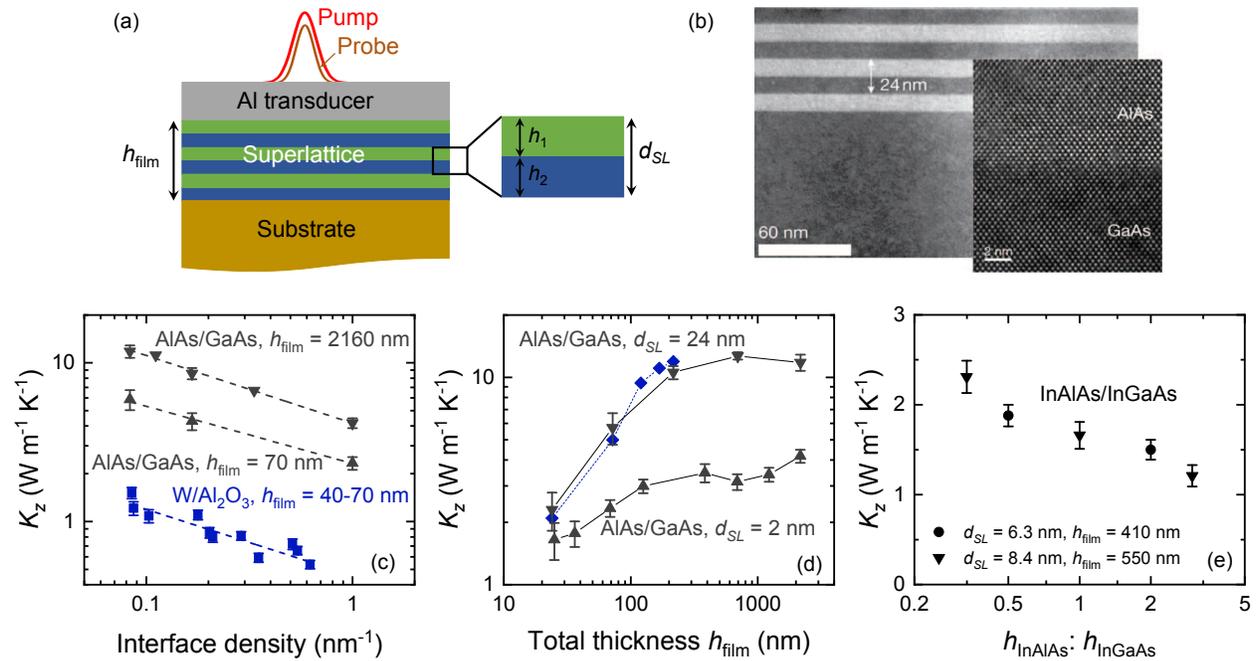

Figure 6. (a) Schematic of a sample configuration for TDTR measurements of $K_z$ in superlattices. (b) An example of TEM images of AlAs/GaAs superlattice from Luckyanova et al.[27]. (c-e) $K_z$ of superlattices measured using TDTR were found to depend on interface density, total film thickness, and the thickness ratio of the two materials. Among these, the data for AlAs/GaAs superlattices are from Cheaito et al.[137] (up and down triangles in (c) and (d)) and Luckyanova et al.[27] (diamonds in (d)), the data for W/Al$_2$O$_3$ nanolaminates are from Costescu et al.[79], while the data for InAlAs/InGaAs superlattices are from Sood et al.[136].

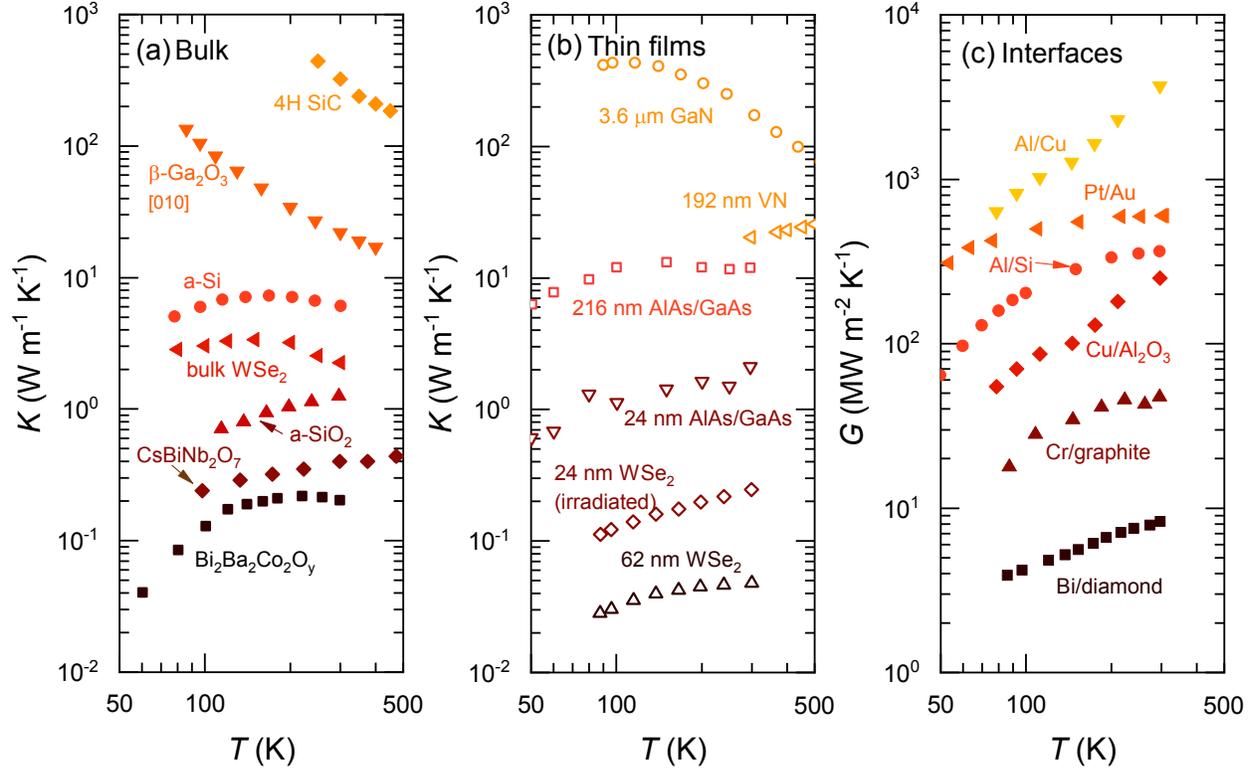

Figure 7. A compilation of the thermal conductivity values of (a) bulk and (b) thin film materials and (c) interface thermal conductance, measured using TDTR as a function of temperature, showing that TDTR can measure a broad range of thermal conductivity and interface thermal conductance. Among the bulk thermal conductivity values, the 4H SiC data are from Qian *et al.*[33], the β-Ga$_2$O$_3$ data are from Jiang *et al.*, the amorphous Si and SiO$_2$ (a-Si and a-SiO$_2$) data are from Yang *et al.*[139], the bulk WSe$_2$ data are from Jiang *et al.*[35], the CsBiNb$_2$O$_7$ data are from Cahill *et al.*[140], and the Bi$_2$Ba$_2$Co$_2$O$_y$ data are from Li *et al.*[141]. Among the thin films, the 3.6-μm-thick GaN data are from Koh *et al.*[134], the data for the VN film are from Zheng *et al.*[142], the data for the AlAs/GaAs superlattices are from Luckyanova *et al.*[27], and the data for the WSe$_2$ thin films are from Chiritescu *et al.*[26]. Among the interface thermal conductance values, the Al/Cu and Cu/Al$_2$O$_3$ data are from Gundrum *et al.*[40], the Pt/Au data are from Wang and Cahill[143], the Al/Si data are from Minnich *et al.*[91], the Cr/graphite data are from Schmidt *et al.*[144], while the Bi/diamond data are from Lyeo and Cahill[41].

## III. Variations of TDTR

In addition to the "standard" TDTR configuration discussed in Section II, there are a large variety of related TDTR techniques. For example, instead of using femtosecond laser pulses for both the pump and probe beams, some other transient thermoreflectance techniques use picosecond or nanosecond laser pulses as the pump beam, and either the synchronized and time-delayed laser pulses or a continuous wave (CW) laser as the probe beam, which is monitored with either a lock-in amplifier or an oscilloscope.[100, 145, 146] In what follows, the unique features and challenges of three important variations of TDTR, *i.e.*, frequency-domain thermoreflectance (FDTR), time-resolved magneto-optic Kerr effect (TR-MOKE), and asynchronous optical sampling (ASOPS), are briefly discussed.

### A. Frequency-domain thermoreflectance (FDTR)

Frequency-domain thermoreflectance, shorten as FDTR, is a variation of TDTR where the thermoreflectance signal as a function of the modulation frequency of the pump beam is collected instead of monitoring the thermoreflectance signal as a function of the delay between the arrival time of the pump and probe pulses. FDTR is thus much easier to implement because it avoids the complexity of a long mechanical delay for the time delay and it can use inexpensive CW laser sources. By holding the delay stage at a fixed location, an ultrafast-laser-based TDTR can also be implemented as FDTR.[93] Both the pulsed and CW FDTR are discussed and compared below.

The pulsed FDTR uses a similar setup as the conventional TDTR (see Figure 1(b) for a schematic of the TDTR setup). The only difference is that the resonant circuit used to eliminate the higher harmonic signals in TDTR cannot be used in FDTR experiments since the data are acquired as a continuous function of modulation frequency while the resonant circuits are usually at fixed cutoff frequencies. The use of the resonant filter, however, is not always necessary for TDTR, if the pump beam is modulated by a pure sine wave or if a digital lock-in amplifier using clean sine-wave multipliers is used for the lock-in detection, or if the higher harmonics are also considered in the thermal modeling.

The CW FDTR can be configured in a much easier way as shown in Figure 8. One major challenge of CW-laser-based FDTR is the accurate determination of the phase signal in FDTR experiments. Besides the desired thermal phase signal $\varphi_{therm}$, an additional frequency-dependent phase shift, collectively written as $\varphi_{instrum}$, would be introduced by components such as the photodetector, the cables, instruments, and the different optical path lengths of the beams. In ultrafast-laser-based TDTR and FDTR, this instrumentation phase can be conveniently corrected by the fact that the $V_{out}$ signal should be constant across the zero delay time. For CW-laser-based FDTR, one commonly adopted approach is to split a portion of the pump beam after the EOM and send it to a reference photodetector that is identical to the primary photodetector, as illustrated in Figure 8.[93] Note that the "identical" here means not only the same detector model but also the same operational parameters such as the applied reverse bias, the incident beam intensity, and the laser wavelength, all of which affect the phase shift introduced by the detector.[101] Besides, the optical path length between the EOM and the reference detector should also be identical to the sum of the path lengths from the EOM to the sample and from the sample to the probe detector. In such a condition, the signal of the primary detector would be $\varphi_1 = \varphi_{therm} + \varphi_{instrum}$, while the signal of the reference detector would be $\varphi_2 = \varphi_{instrum}$. The lock-in amplifier measures the phase difference between the two signal channels as $\varphi_{therm} = \varphi_1 - \varphi_2$. An alternative approach is to use the same photodetector but to conduct two separate experiments: one is the phase signal of the probe beam with the reflected pump beam being filtered and the other is the phase signal of the pump beam with the reflected probe beam being filtered. In this case, the measured phase of the pump is $\varphi_{pump} = \varphi_{instrum} - \varphi_{ref}$ and the measured phase of the probe is $\varphi_{probe} = \varphi_{therm} + \varphi_{instrum} - \varphi_{ref}$. The thermal signal can thus be deducted by subtracting the phase responses of the two experiments as $\varphi_{therm} = \varphi_{pump} - \varphi_{probe}$. However, extreme care should be taken to make sure that factors such as the laser beam intensity and wavelength at the detector are identical for the two experiments to avoid unintentional systematic errors.

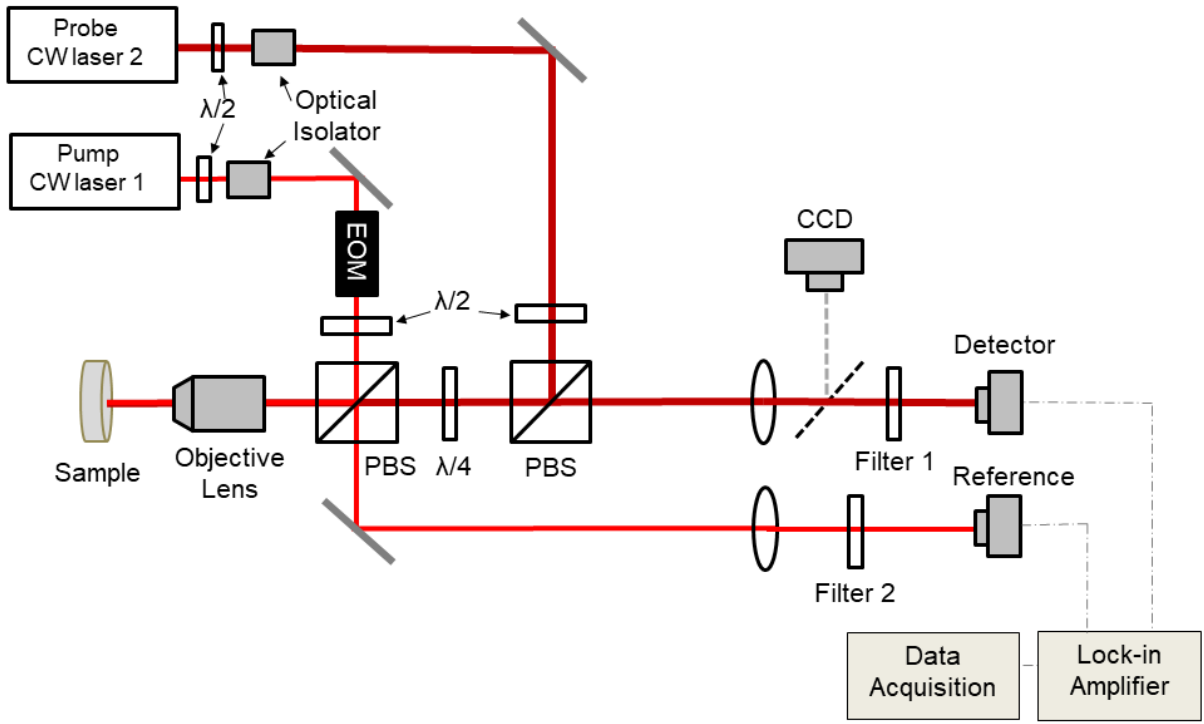

Figure 8. Schematic of an FDTR system based on two CW lasers with different wavelengths.

The thermal transport model for TDTR (shown in Section II(B)) still applies to FDTR, with only one difference that the signals are computed as a function of the modulation frequency instead of the delay time. When CW lasers are used instead of a pulsed laser, the thermal analysis is similar, with the $V_{in}, V_{out}$ signals still given by Eq. (2.21) (2.22), only that the "$n$" in Eq. (2.21) (2.22) should be zero:

$$V_{\text{in}} = \text{Re}(\Delta T(\omega_0)); \quad V_{\text{out}} = \text{Im}(\Delta T(\omega_0)) \quad (3.1)$$

with $\Delta T(\omega)$ given by Eq. (2.18).

Figure 9 shows an example of the calculated signals for CW and pulsed FDTR measurements of a sapphire sample with a 100 nm Al transducer over the modulation frequency range of 0.05-20 MHz, using an RMS spot size of 5 μm. The delay time is fixed at 100 ps for the pulsed FDTR. The results show that both CW and pulsed FDTR configurations have almost the identical $V_{\text{out}}$ signals but the $V_{\text{in}}$ signals differ significantly, suggesting that the $V_{\text{out}}$ signals in TDTR experiments mainly come from the continuous heating at the modulation frequency. The phase signal $\varphi$ of the pulsed FDTR changes relatively little over the modulation frequency range of 0.05-20 MHz, while the phase signal $\varphi$ of the CW FDTR changes dramatically as a function of frequency.

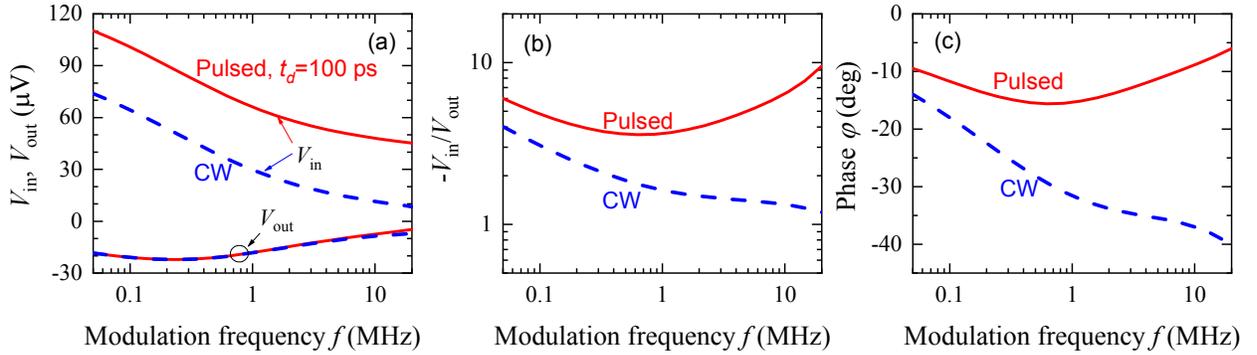

Figure 9. The calculated signals for CW and pulsed FDTR measurements of a sapphire substrate covered with a 100 nm Al transducer over the range 0.05-20 MHz, using a laser spot size $w_0 = 5$ μm. For the pulsed solution, the delay time was fixed at 100 ps.

The sensitivity analysis for TDTR experiments can be applied similarly to FDTR. While there are cases in the literature using the phase signal $\varphi$ to derive thermal properties,[93, 94] for consistency, the ratio signal $R = -V_{in}/V_{out}$, which is equivalent to the phase signal $\varphi = \arctan(V_{out}/V_{in})$, is used here. Figure 10 compares the sensitivities of the ratio signals from CW and pulsed FDTR to different parameters of the sapphire sample over the modulation frequency range of 0.05-20 MHz, using a spot size of $w_0 = 5$ μm (see Eq. (2.23) for the definition of the sensitivity coefficient). The results show that both the CW and pulsed FDTR have similar sensitivities to the thermal conductivity and heat capacity of the substrate; however, the signals of pulsed FDTR are much more sensitive to the spot size $w_0$ and the heat capacitance ($C_{Al}h_{Al}$) of the transducer film.

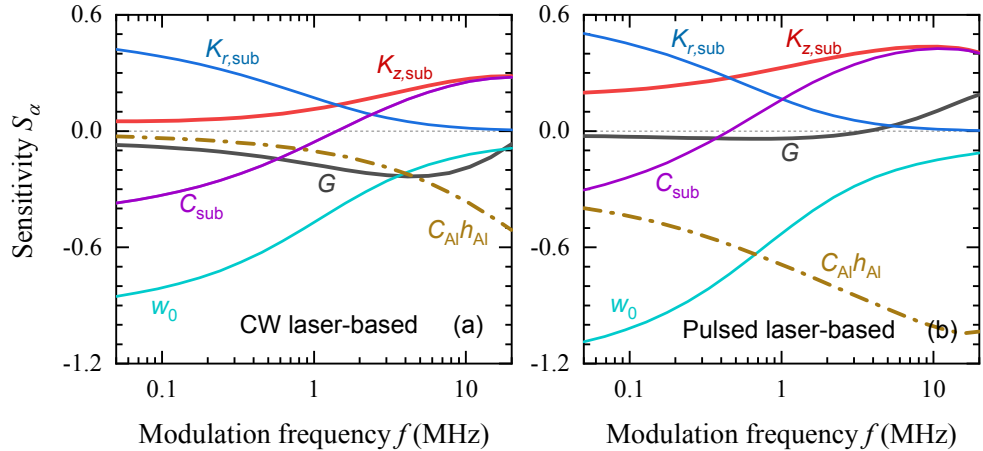

Figure 10. Sensitivity coefficients of the ratio signals $R=-V_{in}/V_{out}$ from CW and pulsed FDTR on different parameters of a sapphire substrate covered by a 100 nm Al transducer, using a spot size of $w_0 = 5$ μm. The delay time for the pulsed FDTR is fixed at 100 ps.

Theoretically, the continuous-wave laser beam in CW FDTR can be modulated at any frequency. However, in practice, the modulation frequency is limited to be < 20 MHz due to the poor SNR at higher modulation frequencies. A heterodyne technique called broadband frequency-domain thermoreflectance (BB-FDTR)[147] has thus been implemented to extend the modulation frequency up to 200 MHz . However, measurements at such a high modulation frequency can be challenging due to the small thermal penetration depth and severe heating effect. A schematic of the BB-FDTR setup is shown in Figure 11. The major difference between FDTR and BB-FDTR is that an additional modulation at frequency $f_2$ is added on the reflected probe beam. Regardless of how high the modulation frequency of the pump beam $f_1$ is, the lock-in amplifier only measures the signals at a much lower frequency, $f_1 - f_2$, which can be chosen to be in a proper range to achieve very high SNR and retain great fidelity of the thermal signal. Another benefit is that the frequency difference $f_1 - f_2$ can also be chosen to be close to the upper limit of the frequency range of the lock-in amplifier so that the higher harmonic components are naturally excluded from the lock-in detection.[147] Extending modulation frequency range greatly expands the capability of FDTR. For example, it was introduced to study phonon MFP spectra in semiconductors,[92, 147, 148] similar to the frequency-dependent TDTR that is based on the physical picture of non-diffusive phonon transport, see more details in Section IV(D).

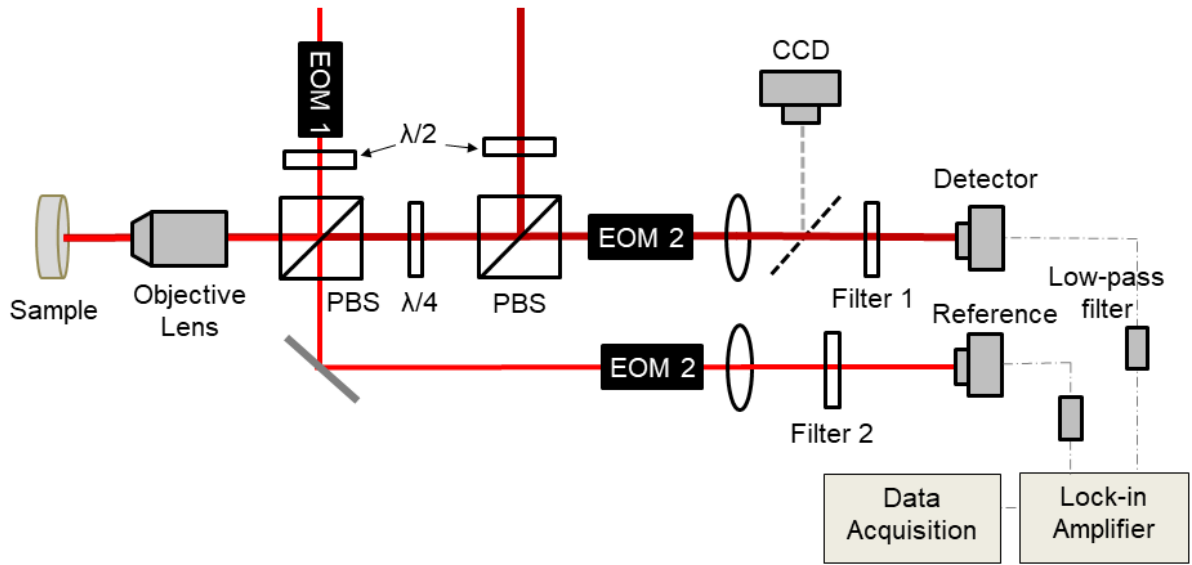

Figure 11. Schematic of the BB-FDTR setup. The remainder of the experimental setup is the same as the FDTR system. Compared to the FDTR setup, a second EOM is added in the reflected probe path. Similar schematics of BB-FDTR can be found in Ref. [147].

## B. Time-resolved magneto-optic Kerr effect (TR-MOKE)

TDTR can also incorporate a novel temperature sensing technique utilizing time-resolved magneto-optic Kerr effect (TR-MOKE).[34, 149, 150] Instead of thermoreflectance, TR-MOKE relies on the temperature-dependent transient polar Kerr rotation to detect the temperature response of a magnetic transducer under pump laser heating, from which the thermal properties of the material underneath the transducer film can be derived, following the same data reduction scheme of TDTR. Thermometry based on the thermo-magneto-optic Kerr effect allows the use of much thinner magnetic transducer films that are not necessarily optically opaque as required in conventional TDTR method. A thinner transducer with a lower thermal conductivity cam minimize lateral heat flow in the transducer layer and thus enhancing the measurement sensitivity to the in-plane thermal conductivity.[34, 87] Meanwhile, the small thermal mass of the transducer layer also enables an enhanced sensitivity to the interface thermal conductance[150] when the thermal conductivity of the substrate is small.

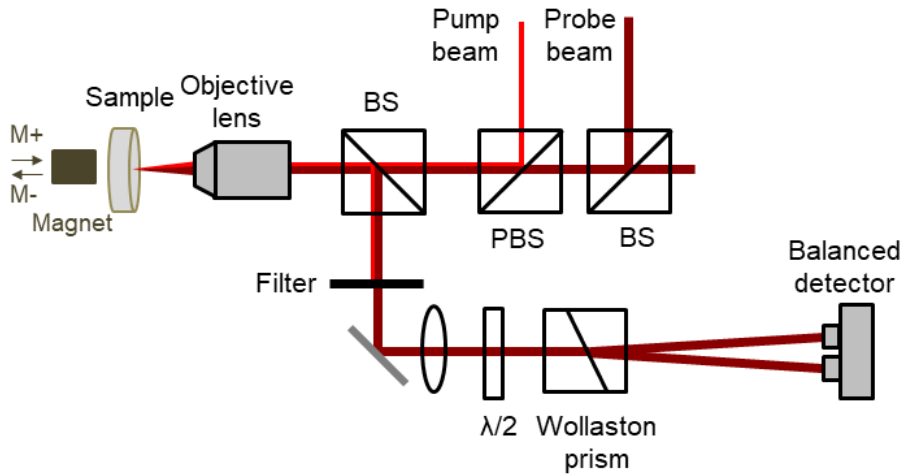

Figure 12. Schematic illustrating the TR-MOKE detection scheme. The remainder of the experimental setup is the same as the TDTR system described in Section II(A). The polarization states of the reflected probe beam are split by a Wollaston prism and detected by a balanced detector. A half-wave plate placed before the Wollaston prism is used to balance the average intensities.

Figure 12 shows the TR-MOKE signal detection scheme. To conduct TR-MOKE measurements, the sample needs to be coated with a thin perpendicularly magnetized transducer, which is magnetized with an external magnet prior to the measurements. A non-polarizing beam splitter is inserted between the steering PBS and the microscope objective lens to divert the reflected pump and probe beams toward the detection path. In the detection path, the pump beam is removed by a filter, while the probe beam passes through a half-wave plate and then is split into two orthogonally polarized components by a Wollaston prism. The half-wave plate is adjusted such that the two components have approximately the same intensity. Transient changes in the polarization of the probe beam are monitored by detecting the changes in the relative intensities on the balanced detector.

In case of non-perfect balancing with the half-wave plate, thermoreflectance signals overlap the transient Kerr rotation. Since the TR-MOKE signals would change signs for oppositely aligned magnetization states of the magnetic transducers, the TR-MOKE signals can be isolated out by subtracting the in-phase and out-of-phase signals recorded for oppositely aligned magnetization states of the transducers as $V_{in} = (V_{in}^{M+} - V_{in}^{M-})/2$ and $V_{out} = (V_{out}^{M+} - V_{out}^{M-})/2$. Figure 13 shows an example of the TR-MOKE signals measured as a function of delay time for a bulk black phosphorus sample coated with a 26.9 nm TbFe transducer.[151] Note that both the M+ and M- signals in Figure 13(a) show observable oscillations in the short delay time range up to 500 ps (see the inset of Figure 13(a)), which was attributed by the authors to the Brillouin scattering.[151] The Brillouin scattering arises from the interaction between the reflected probe beam and the acoustic waves inside the black phosphorus sample caused by the birefringence originated from the in-plane anisotropy in black phosphorus.[152, 153] These oscillations are also cancelled out by the corrective subtraction (note that the corrected signals in Figure 13(a) are smooth without oscillations). This approach makes TR-MOKE thermometry less prone to errors because any spurious signals that are independent of the magnetization states of the transducer can be cancelled out.

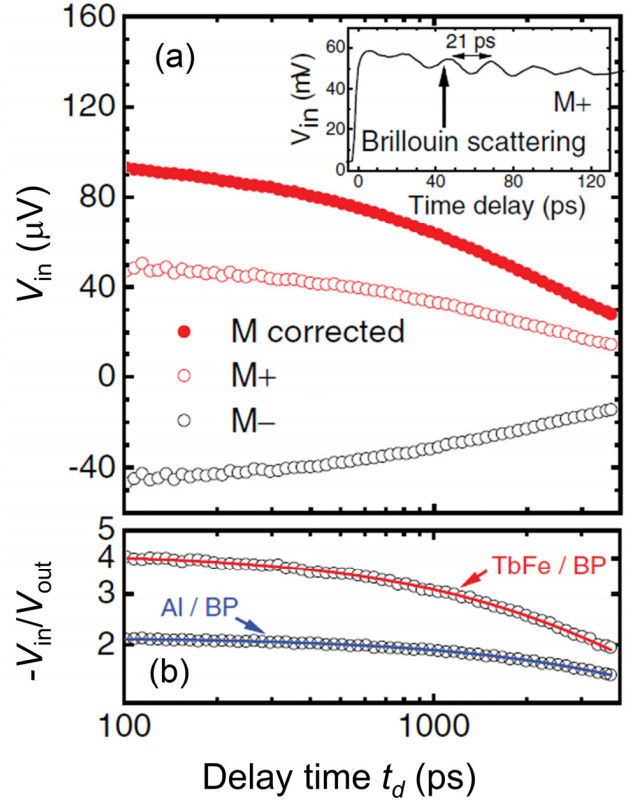

Figure 13. Example of TR-MOKE signals measured using 9 MHz modulation frequency and a laser spot size of $w_0 = 12$ μm on a black phosphorus sample coated with a 26.9-nm-thick TbFe layer. (a) The positive (M+), negative (M-) and corrected $V_{in}$ signals as a function of delay time. The inset plot shows the Brillouin scattering oscillations with a period of 21 ps occurring at the first few hundreds of ps. These oscillations are canceled out in the corrected $V_{in}$ signals. (b) The ratio signals $-V_{in}/V_{out}$ from experiments (symbols) and thermal model simulation (lines) for both TDTR measurements of a black phosphorus sample coated with 81 nm Al and TR-MOKE measurements of a black phosphorus sample coated with 26.9 nm TbFe. (Reproduced with permission from Zhu et al.[151] Copyright 2016 by Wiley.)

Some commonly used magnetic transducer films for TR-MOKE are Co/Pt, Co/Pd, CoFe/Pt, TbFe, GdFeCo, etc. For example, Feser et al.[87] used a Co/Pt multilayer with a total thickness of 19 nm as the TR-MOKE transducer with the nominal structure of the multilayer from top to bottom as Pt (1 nm)/[Co (0.5 nm) Pt (1 nm)] × 6/Pt (10 nm). Chen et al.[149] carefully evaluated the performance of four types of magnetic transducers, including TbFe, GdFeCo, Co/Pd, and CoFe/Pt, and found that TbFe with a thickness of ~18.5 nm gives the optimal performance. Magnetic transducers as thin as 4.2 nm were also used for TR-MOKE experiments.[150] Given the small thermal masses of the magnetic transducers, adsorbents such as hydrocarbons and water molecules on the sample surface can possibly change the effective heat capacity of the transducer and need to be carefully considered in the thermal model. Kimling *et al.*[150] empirically made the correction by adding a 1-nm-thick transparent layer with a thermal conductivity of 2 W m$^{-1}$ K$^{-1}$ and a volumetric heat capacity of 2.8 J cm$^{-3}$ K$^{-1}$ on top of the transducer layer in their thermal model for data reduction.

Besides thermal characterization, TR-MOKE has also been widely employed to study spin dynamics and ultrafast magnetization processes,[154-161] which, however, is beyond the scope of this tutorial.

## C. Asynchronous optical sampling (ASOPS)

In conventional TDTR, the pump and probe beams are two synchronous pulse trains and the measured data are recorded in the time domain as a function of the delay time. This approach is known as synchronous sampling, with the speed of data acquisition often limited by the traveling speed of the delay stage as well as the time required for the system to reach an equilibrium state for each new delay time (typically ~1 s is needed to acquire one data point). Much faster data acquisition is possible with asynchronous optical sampling (ASOPS), using two different mode-locked lasers with slightly different pulse repetition rates. This automatically provides a temporally varying delay between the two pulses.

Figure 14(a) illustrates the principle of signal detection in ASOPS. The curves shown on the top of Figure 14(a) represent the surface temperature of the sample in response to the pump input, which has a repetition rate of $f_{pump}$ and a period of $1/f_{pump}$. Each successive probe pulse, represented by the dots, is delayed with respect to the pump pulse by the time $\Delta t = \Delta f/(f_{pump} f_{probe})$, where $\Delta f = f_{pump} - f_{probe}$ is also known as the beat frequency. The total number of points sampled during one full period of the signal is $N = f_{probe}/\Delta f$. More importantly, this sampling process automatically repeats at a period of $t = 1/\Delta f$. The ASOPS technique is thus an optical analog of the electronic oscilloscope. For example, given $f_{pump} = 80$ MHz and frequency offset $\Delta f = 1$ kHz, ASOPS would achieve a delay time increment of $\Delta t = 0.16$ ps, and the measurement time to finish one full period scan is only 1 ms, which means that the ASOPS instrument can obtain 1000 acquisitions within 1 second. This fast data acquisition allows ASOPS to average many repeated measurements, significantly reducing the detection noise to a very low level. Note that the temporal resolution eventually achieved by ASOPS depends not only on the delay time increment but also on the pulse duration and the detection bandwidth. Nevertheless, a temporal resolution of 1 ps is already sufficient for TDTR experiments, while ultrafine temporal resolution as small as 50 fs has been reported for the ASOPS technique.[162]

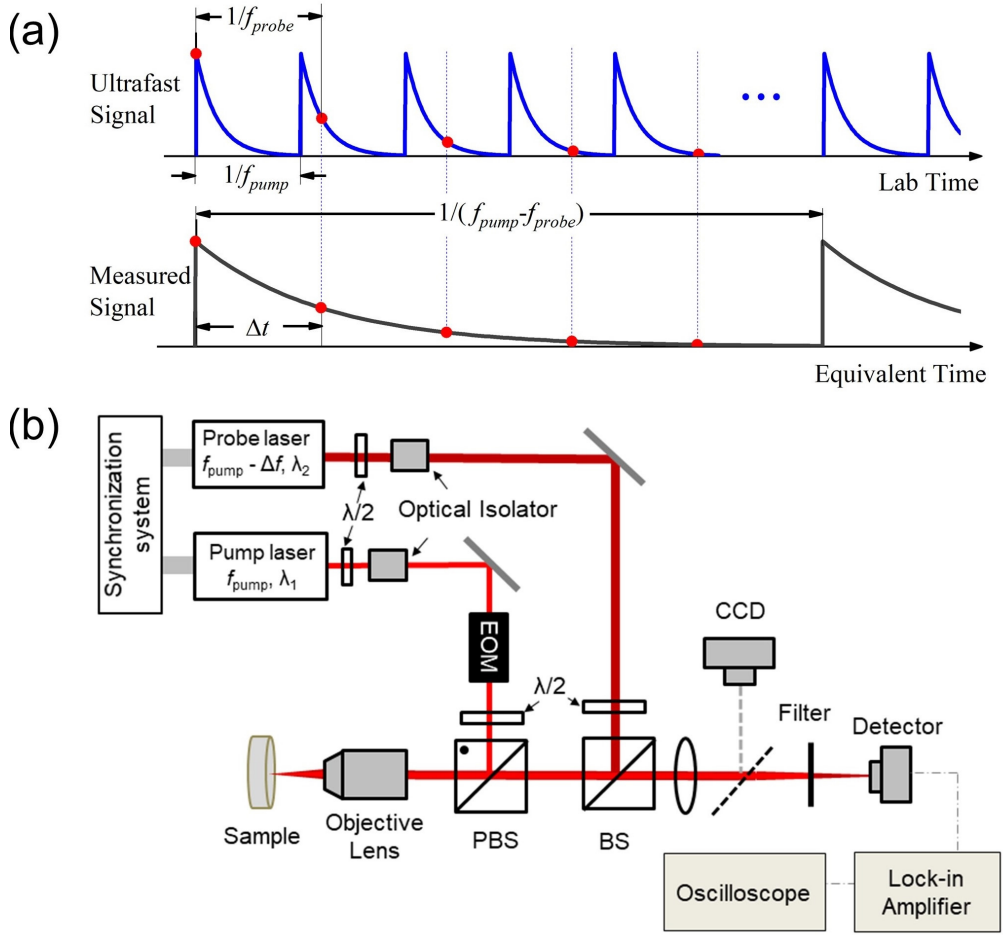

Figure 14. (a) The signal detection mechanism in ASOPS. (b) A schematic of ASOPS-based TDTR system.

ASOPS was first reported by Elzinga *et al*. in 1987,[163] who used a Nd:YAG laser with an 81.597 MHz repetition rate as the pump beam and a 10 kHz offset rate for the probe and measured the fluorescence lifetime of rhodamine B to be 2.3 ns. Fiechtner *et al*.[164, 165] later employed ASOPS for rapid diagnosis of turbulent flames in combustion. Kafka *et al*.[166] built a two-color ASOPS system using two regeneratively mode-locked Ti:sapphire lasers and achieved a temporal resolution of 150 fs. In recent years, researchers started using the ASOPS technique for THz spectroscopy.[162, 167-170] For example, Cuffe *et al*.[171] studied the lifetimes of confined acoustic phonon modes in free-standing silicon membranes, using an ASOPS system with two Ti:sapphire oscillators at 1 GHz repetition rate and a frequency offset of 10 kHz.

ASOPS clearly has many advantages over the traditional mechanical delay-based pump-probe systems: 1). ASOPS enables much faster data acquisition. The noise level can thus be significantly reduced by averaging over a large amount of data acquired within a short period of time. 2). ASOPS eliminates the mechanical delay stage and the associated systematic errors due to beam pointing instability and spot size variation. 3). ASOPS has access to the full delay range, whereas in the traditional pump-probe metrology the delay time is limited by the length of the delay stage. Admittedly, all the advantages of ASOPS come at the cost of requiring two ultrafast lasers instead of one.

ASOPS normally does not require amplitude modulation of the pump beam for data acquisition, while the traditional pump-probe systems usually require modulation of the pump beam for lock-in detection. However, without modulating the pump beam, the signals acquired in a typical ASOPS experiment are composed of frequency responses only at multiples of the laser repetition rate (e.g., $f_{rep}$, $2f_{rep}$, …) but lack the frequency components at the modulation frequency (e.g., $\pm f_{mod}$, $\pm f_{mod}+f_{rep}$, $\pm f_{mod}+2f_{rep}$, …). Since the laser repetition rate is usually a fixed value and is much larger than the modulation frequency, $f_{rep} \gg f_{mod}$, the functionality of ASOPS in thermal measurements would be greatly compromised without modulation (note that many advanced TDTR configurations discussed in Section III are achieved through the variation of modulation frequency in the range 0.1-20 MHz). To overcome this problem, Dilhaire and co-workers[172, 173] proposed a high-throughput time-domain thermoreflectance (HT-TDTR) technique that combines ASOPS with high-frequency modulation of the pump beam, enabling fast and accurate measurements of thermal properties. Figure 14(b) shows a schematic of the ASOPS-based TDTR system. In this system, the pump beam is modulated at a frequency $\omega_0 = 2\pi f_{mod}$ by an EOM. The probed signal by the photodiode detector can be expressed as:

$$\Delta T(t'_d) = e^{i\omega_0 t'_d} \sum_{n=-\infty}^{\infty} \Delta T(\omega_0 + n\omega_s)\exp(in\Delta\omega t'_d) \tag{3.2}$$

and the expressions of the in-phase and out-of-phase outputs of the lock-in amplifier are:

$$V_{in} = \frac{1}{2}\sum_{n=-\infty}^{\infty}\left[\Delta T(\omega_0 + n\omega_s) + \Delta T(-\omega_0 + n\omega_s)\right]\exp(in\Delta\omega t'_d) \tag{3.3}$$

$$V_{out} = -\frac{i}{2}\sum_{n=-\infty}^{\infty}\left[\Delta T(\omega_0 + n\omega_s) - \Delta T(-\omega_0 + n\omega_s)\right]\exp(in\Delta\omega t'_d) \tag{3.4}$$

Here $\omega_s = 2\pi f_{pump}$ and $\Delta\omega = 2\pi\Delta f$, and $t'_d$ is the equivalent delay time. Compared to the signals in conventional TDTR (Eqs. (2.21) and (2.22)), the only difference is in the exponential term, with $\omega_s t_d$ in conventional TDTR replaced by $\Delta\omega t'_d$ in ASOPS-TDTR. The output signals of the lock-in amplifier in ASOPS-TDTR are thus time-varying with a period of $1/\Delta f$. The detected signals in ASOPS-TDTR are not pure sinusoidal as those in conventional TDTR but are centered at the modulation frequency with sidebands of $\Delta f$ in the frequency domain.[172] A small beat frequency of $\Delta f$=1 Hz is usually needed so that a time constant of 30 μs can be chosen for the lock-in amplifier to avoid signal aliasing.[173] An oscilloscope is then used to record the signal as a function of the scan time or lab time $t$. Since the lab time recorded by the oscilloscope has been stretched by a factor of $f_{pump}/\Delta f$ compared to the equivalent delay time $t'_d$ (see Figure 14(a)), the recorded lab time needs to be corrected in order to compare the recorded signals with the model prediction. ASOPS-TDTR and conventional TDTR should produce exactly the same signals as a function of the delay time, only in different manners of data acquisition. This ASOPS-TDTR technique, with its ~10 times faster data acquisition speed and free of artefacts associated with the moving delay stage, has been successfully employed in several studies, including the measurements of Si/Ge superlattices with various thicknesses (15-50 nm)[174] and the high-speed thermal conductivity mapping for a Fe-Si-Ge thin film alloy.[173]

Another important feature of ASOPS is that the probing rate can be many times slower than the pumping rate, with $f_{pump} = nf_{probe} + \Delta f$, where $n$ is an integer. This enables sampling of ultrafast phenomena at a much slower acquisition rate. For example, Pradere et al.[175] used this technique to sample 30 kHz thermal waves using an infrared camera at an acquisition rate of only 25 Hz.

## IV. ADVANCED TDTR CONFIGURATIONS

In addition to the frequent practice of measuring the through-plane thermal conductivity $K_z$ and the interface thermal conductance $G$, advanced TDTR configurations have also been developed over the past decade to measure specific thermal properties (e.g., heat capacity, anisotropic thermal conductivity, etc.) or to meet the requirements of different circumstances (e.g., measurements of liquids and rough coatings). Here we discuss advanced TDTR configurations that demonstrate the flexibility and versatility of the TDTR technique for thermal property characterization of novel materials. These advanced TDTR configurations are essentially based on varying the modulation frequency and/or the spot size, which alters the measurement sensitivity of different target properties, making TDTR more powerful and versatile.

### A. Simultaneous measurements of thermal conductivity and heat capacity

In TDTR measurements of thermal conductivity, the heat capacity of the sample is usually a prerequisite that affects the determined thermal conductivity values significantly and thus needs to be pre-determined accurately. Measurements of the heat capacity of nanostructured materials such as thin films, however, are very challenging.[176, 177] A common practice is thus to adopt the bulk values for the thin films, where the bulk values can be easily measured using conventional techniques such as the differential scanning calorimeter. This is usually a reasonable assumption for most of the well-studied condensed materials (e.g., Al, Si, SiO$_2$, etc.) since the nanostructures do not significantly change the heat capacity.[177] However, this approach is problematic for many novel materials either because it is questionable to assume bulk heat capacity for the nanostructures[178-180] or because it is difficult to obtain the bulk form of the new materials for a conventional heat capacity measurement. TDTR has been demonstrated to determine thermal conductivity $K$ and heat capacity $C$ simultaneously for both bulk and thin film materials by performing multiple measurements using different modulation frequencies,[38, 39] with the basic principles explained below.

Figure 15(a) shows an example of the sensitivity coefficients (defined in Eq. (2.23)) of the TDTR ratio signal to $K$ and $C$ of the substrate as a function of modulation frequency when the delay time is fixed at 100 ps. At a high modulation frequency of 10 MHz, the TDTR signal has the same sensitivity to $K$ and $C$, suggesting that the TDTR signal $-V_{in}/V_{out}$ is mostly determined by the thermal effusivity $e = \sqrt{KC}$ of the substrate at high modulation frequencies. As the modulation frequency decreases to sufficiently low of 0.1 MHz, the sensitivities to $K$ and $C$ have the similar amplitudes but opposite signs, suggesting that the TDTR signal is mainly affected by the thermal diffusivity $\alpha = K/C$ of the substrate at low modulation frequencies. Since the TDTR signals depend on $K$ and $C$ differently at high and low modulation frequencies, a $K$-$C$ diagram shown in Figure 15(b) suggests that the $K$-$C$ curves at the different modulation frequencies should cross at a certain point. These two properties can thus be determined simultaneously by conducting at least two sets of the measurements at a high and a low modulation frequency. Figure 15(c) shows an example of the frequency-dependent TDTR approach on simultaneous measurements of $K$ and $C$ of bulk Si.[38] A good agreement of ±5% in the determined $C$ of bulk Si was achieved with the literature value. Wei et al.[39] validated this approach for the heat capacity of several standard bulk samples, as shown in Figure 15(d), with an overall accuracy of ±8%.

Besides the bulk samples, this frequency-dependent TDTR approach also applies to thin film materials. Employing this frequency-dependent TDTR approach, Liu et al.[181] simultaneously determined the through-plane thermal conductivity and volumetric heat capacity of three types of hybrid organic-inorganic zincone thin films prepared by the atomic layer deposition (ALD) and molecular layer deposition (MLD) techniques, with the film thickness in the range 40-400 nm, see Figure 16 for a summary of the results. The heat capacities of the hybrid films were found to be independent of the film thickness, while the thermal conductivities changes slightly with the film thickness. Very low thermal conductivities of 0.15-0.35 W m$^{-1}$ K$^{-1}$ were measured for these thin films. The ultralow thermal conductivities are due to the alternating layered structures with very different atomic configurations between the ALD atomic layers and the MLD molecular layers that strongly scatter the phonons. Wang et al.[125] also used the frequency-dependent TDTR approach to simultaneously determine the heat capacity and thermal conductivity of fullerene derivative thin films with thicknesses in the range 50-120 nm. The same method has also been employed by Xie et al.[129, 182] to study both the thermal conductivity and heat capacity of 27 different polymers. Another application of this frequency-dependent TDTR approach was also conducted by Liu et al.[183] to study the film thickness effect on the effective thermal conductivity and volumetric heat capacity of polystyrene thin films over the thickness range 5-300 nm.

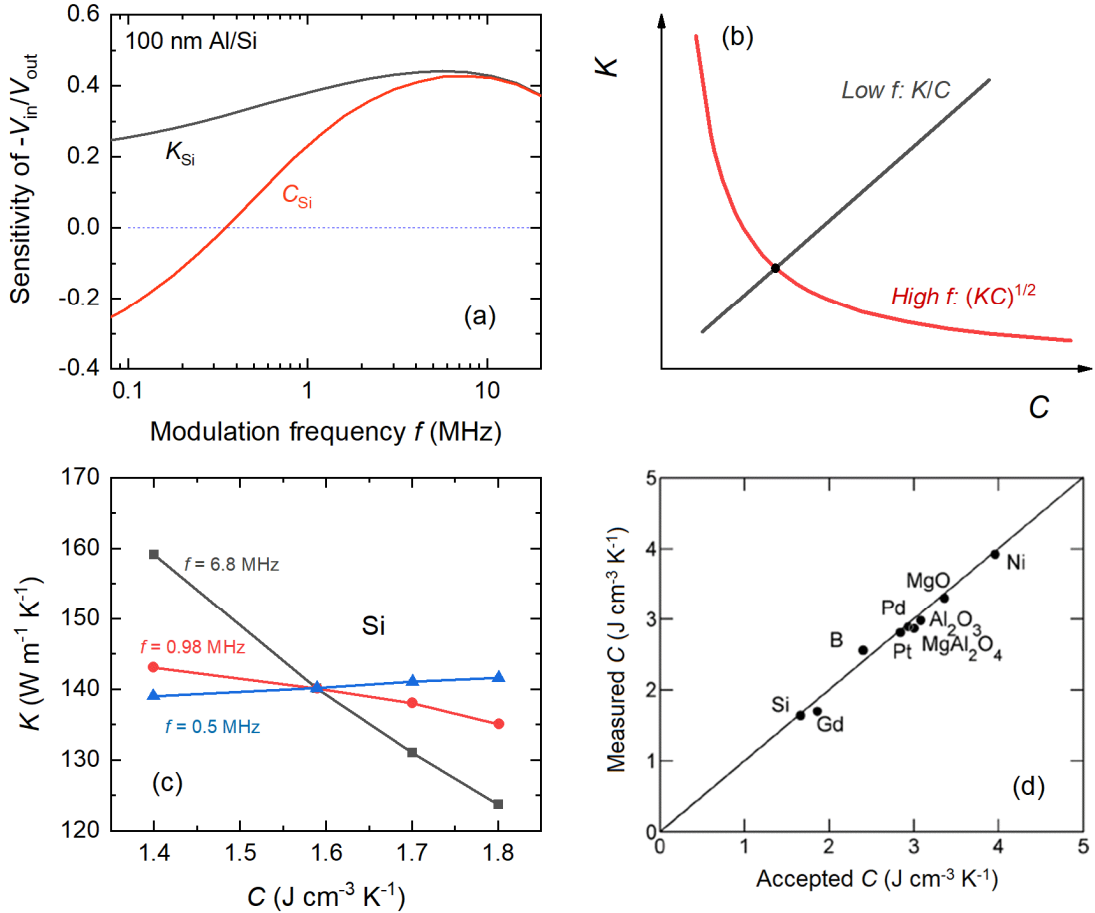

Figure 15. (a) The sensitivity coefficients of TDTR signal $R = -V_{in}/V_{out}$ to the substrate thermal conductivity $K_{Si}$ and heat capacity $C_{Si}$ of a Si substrate covered with 100 nm Al transducer, calculated as a function of modulation frequency with the delay time fixed at 100 ps. (b) A schematic illustration of the frequency-dependent TDTR approach to simultaneously determine thermal conductivity $K$ and heat capacity $C$ of the substrate. (c) The $K$-$C$ diagram of bulk Si, with the crossing point of $K$ & $C$ as the measured value of the Si sample, which agrees well (within 5%) with literature values of bulk Si. (Reproduced with permission from Liu et al.[38]. Copyright 2013 by AIP Publishing.) (d) Benchmark studies conducted by Wei et al.[39] using the frequency-dependent TDTR approach to determine heat capacity of various materials. (Reproduced with permission from Wei et al.[39] Copyright 2013 by AIP Publishing.)

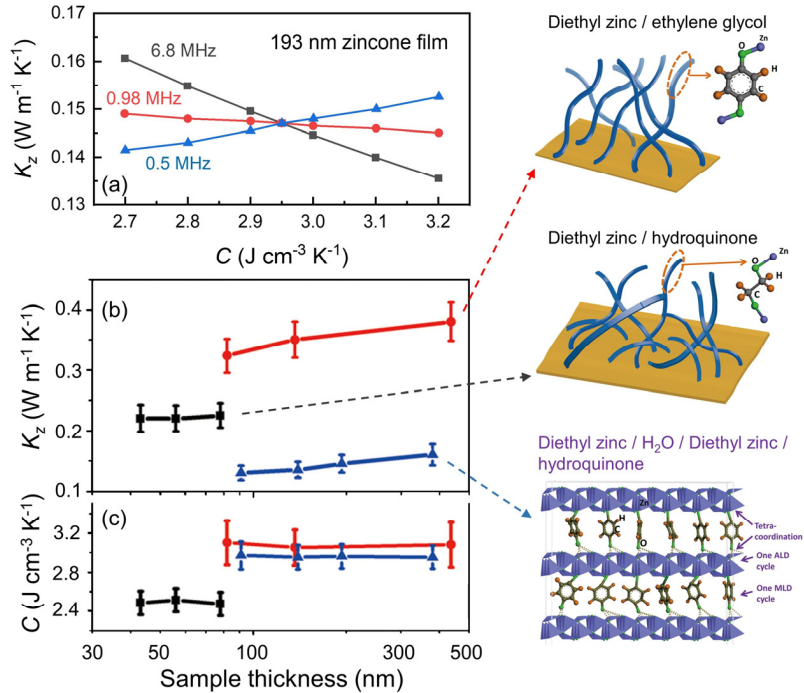

Figure 16. (a) The *K-C* diagram of a 193-nm-thick zincone film measured using frequency-dependent TDTR, with the crossing point of *K-C* as the measured value of the zincone film. (Reproduced with permission from Liu *et al.*[38]. Copyright 2013 by AIP Publishing.) (b, c) The through-plane thermal conductivity and volumetric heat capacity of three types of hybrid organic-inorganic zincone thin films determined using the frequency-dependent TDTR as a function of film thickness. (Reproduced with permission from Liu *et al.*[181] Copyright 2013 by American Chemical Society.)

## B. Measurements of thermal conductance in thermally thin films

TDTR can measure the through-plane thermal conductivity $K_z$ of thin films using a high modulation frequency and a large laser spot size, in which configuration the heat flow is mainly one-dimensional in the through-plane direction. However, using TDTR to measure $K_z$ of thermally thin films on low-thermal-conductivity substrates is very challenging. The reason is that the TDTR signals at high modulation frequencies are also highly sensitive to other input parameters such as the thickness and heat capacity of the Al transducer, $h_{Al}C_{Al}$, the uncertainty of which would significantly affect the accuracy of the determined $K_z$ of the thermally thin films. Here, the "thermally thin" films are referred to as the films with thickness $h_{film}$ less than the thermal penetration depth in the through-plane direction, defined as $d_{p,z} = \sqrt{K_z/\pi f C}$.[12] To overcome this challenge, a dual-frequency TDTR approach has been developed[184] that takes the advantage of the different frequency dependence in the sensitivities of the different parameters.

Figure 17(a) shows an example of the sensitivity coefficients for a sample composed of a 2.6-μm-thick Si film on a 100-nm-thick $SiO_2$ layer on Si substrate as a function of the modulation frequency. The sample is coated with a 100 nm Al transducer, and the delay time is fixed at 100 ps. The sensitivity plot shows that the TDTR signal has the highest sensitivity to the through-plane thermal conductivity of the Si film $K_{z,Si}$ at a high modulation frequency of 10 MHz. However, at such a high modulation frequency, the TDTR signals are much more sensitive to the thermal mass of the transducer layer, *i.e.*, the product of thickness and heat capacity of the transducer layer $h_{Al}C_{Al}$, the uncertainty of which can propagate and affect the accuracy of the determined $K_{z,Si}$. However, at a slightly lower modulation frequency of 2 MHz, the sensitivity to $K_{z,Si}$ decreases drastically to near zero while the sensitivity to $h_{Al}C_{Al}$ remains high. Noticing that the sensitivity of a ratio signal is the difference of the sensitivities of the two components as

$$S_\xi^{R_1/R_2} = \frac{\partial \ln(R_1/R_2)}{\partial \ln \xi} = \frac{\partial \ln R_1}{\partial \ln \xi} - \frac{\partial \ln R_2}{\partial \ln \xi} = S_\xi^{R_1} - S_\xi^{R_2} \qquad (4.1),$$

one can thus conduct two sets of measurements at the two different modulation frequencies and take the ratio of the two measurements as the signal instead. In this case, the sensitivity to $K_{z,Si}$ is maintained whereas the sensitivity to $h_{Al}C_{Al}$ is greatly reduced. The accuracy in determining $K_z$ of the thin film can thus be greatly improved. Figure 17(b) shows the TDTR signals measured on this Si film sample at different modulation frequencies of 9.8 and 1.8 MHz. The TDTR data measured at 9.8 MHz, although being highly sensitive to $K_z$ of the Si film, cannot be fitted well due to the uncertainties of input parameters. However, fitting the ratio of TDTR signals at the two frequencies of 9.8 and 1.8 MHz, as shown in Figure 17(c), determines the $K_z$ of the Si film accurately.

This dual-frequency TDTR approach was verified by measuring the through-plane thermal conductivity of a 3-μm-thick Cu film on a SiO$_2$ substrate.[184] The thermal conductivity of the Cu film is expected to be isotropic with negligible boundary scattering effect due to the much shorter mean free paths of the major heat carriers, *i.e.*, electrons, compared to the film thickness. The $K_z$ of the Cu film measured using dual-frequency TDTR can thus be verified against the $K_r$ of the film determined from a four-point probe measurement using the Wiedemann-Franz law. Excellent agreement was achieved between the dual-frequency TDTR measurements and the four-point measurements of this Cu film, whereas the conventional TDTR measurements showed ~20% discrepancy from the four-point measurements.[184] The dual-frequency TDTR approach has also been successfully applied to measure $K_z$ of single crystalline Si films with thickness in the range 1-10 μm,[185] which has shown strong anisotropy in the in-plane and through-plane directions[186] due to the long MFPs of phonons in crystalline silicon.[187]

Generally, in the dual-frequency approach, the high frequency should be chosen to have the highest sensitivity to $K_z$ of the thin film, and the low frequency should be chosen to have a near-zero sensitivity to $K_z$ of the thin film but sufficiently high sensitivity to $h_{Al}C_{Al}$ of the transducer layer. Through detailed modeling analysis, Jiang *et al.*[184] proposed a guideline that the high modulation frequency can be chosen so that the thermal penetration depth in the film $d_{p,z}$ is nearly half of the film thickness $h_f$, $d_{p,z} \approx 0.5 h_f$, and the low modulation frequency can be chosen so that $d_{p,z} \approx 1.5 h_f$. This dual-frequency approach has been shown to be able to improve the accuracy in determining $K_z$ of thin films by ≈3 times, and is very useful when the thermal conductance of the transducer/film interface is high ($G > K_z/h_f$) and the thermal conductivity of the underlying substrate $K_{sub}$ is low ($K_z/K_{sub} > 10$).[184]

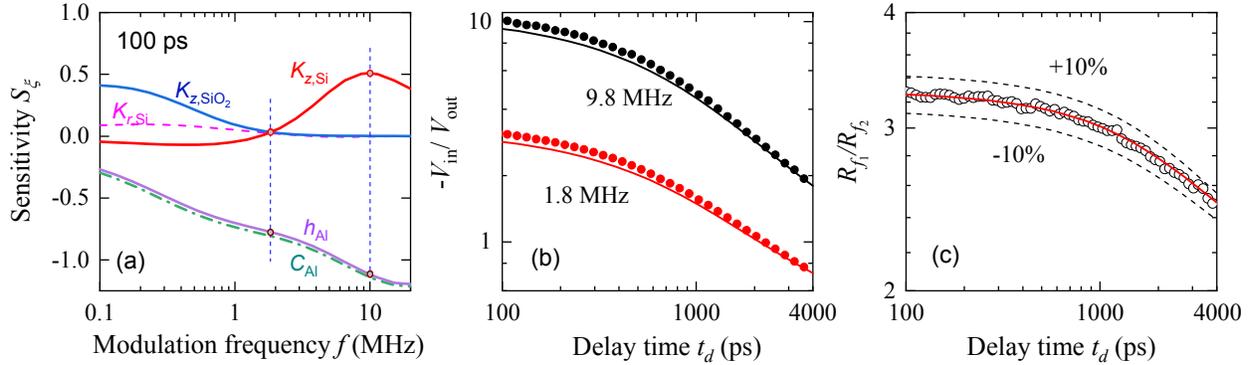

Figure 17. (a) The calculated sensitivity coefficients $S_\xi$ of a sample composed of a 2.6-μm-thick Si film on a 100-nm-thick SiO$_2$ layer on Si substrate, coated with 100 nm Al film transducer as a function of modulation frequency at a delay time of 100 ps. The parameters include the through-plane and in-plane thermal conductivity of Si film, $K_{z,Si}$ and $K_{r,Si}$, the thermal conductivity of the SiO$_2$ film $K_{SiO2}$, and the thickness and heat capacity of the Al film, $h_{Al}$ and $C_{Al}$. (b) Measured TDTR signals of this Si film sample at 9.8 MHz and 1.8 MHz cannot be fitted well independently. (c) Fitting the ratio of TDTR signals $R_{f_1}/R_{f_2}$ at 9.8 MHz and 1.8 MHz over the delay time range from 100 to 4000 ps.

## C. Probing anisotropic thermal transport

While TDTR signals are most sensitive to thermal conductivity in the through-plane direction, TDTR can also be configured to measure anisotropic thermal transport properties. Several advanced approaches based on TDTR have been developed, including a variable spot size configuration,[37] a beam-offset configuration,[86, 87] and an elliptical-beam configuration,[119] all with their own advantages. A summary and comparison of the three different configurations are presented in Table 1, with the details discussed below. Besides TDTR, some related techniques such as the 3ω method[188, 189] and the transient thermal grating[190, 191] can also measure anisotropic thermal transport properties.

Table 1. A summary/comparison of three different configurations for TDTR to measure anisotropic thermal properties of bulk and thin film materials.

| Configurations | Principles | Sensitive properties | Measurable range | Guidelines |
| --- | --- | --- | --- | --- |
| Variable spot size | Two sets of TDTR measurements are conducted: the one using a large $w_0$ is sensitive to $K_z$ only, while the other using a small $w_0$ is sensitive to both $K_r$ and $K_z$. | $K_r$, $K_z$ ($K_r$ needs to be isotropic) | $K_r > 5$ W m$^{-1}$ K$^{-1}$ | $w_0 > 5d_{p,r}$ to make sure the TDTR signals are only sensitive to $K_z$; $w_0 < 2d_{p,r}$ to make sure the TDTR signals are sufficiently sensitive to $K_r$. |
| Beam-offset | With the pump spot swept across the probe spot at a fixed delay time, the detected signal is exclusively sensitive to the in-plane thermal conductivity in the offset direction. Alternatively, the offset distance between the pump and probe spots can also be fixed while the signals are acquired as a function of delay time or modulation frequency. | $K_x$ ($x$ here stands for the offset direction in in-plane) | $K_x > 5$ W m$^{-1}$ K$^{-1}$ if using a low-thermal-conductivity transducer such as NbV | The laser spot size and modulation frequency should be chosen so that $w_0 \approx d_{p,x}$ to have the highest sensitivity to $K_x$. |
| Elliptical-beam | Using a highly elliptical pump beam for TDTR experiments, the signal is selectively sensitive to the in-plane thermal conductivity along the short axis direction. | $K_y$, $K_z$ ($y$ here stands for the short axis direction in in-plane) | $K_y > 5$ W m$^{-1}$ K$^{-1}$ | The long radius of the laser spot needs to be $w_x > 5d_{p,x}$ to suppress the sensitivity to $K_x$; The short radius of the laser spot needs to be $w_y < 2d_{p,y}$ to have a sufficient sensitivity to $K_y$. |

## *C.1. Variable spot size TDTR configuration*

By changing the relative size of laser spot as compared to the thermal penetration depth, as illustrated in Figure 18(a), TDTR can measure both the in-plane and the through-plane thermal conductivities. When TDTR experiments are conducted using a laser spot size $w_0$ much larger than the in-plane thermal penetration depth $d_{p,r}$, the heat mainly flows in the through-plane direction. In this case, the TDTR signals are sensitive to the through-plane thermal conductivity $K_z$ only. When TDTR experiments are conducted using $w_0$ similar or even smaller than $d_{p,r}$, the heat flux becomes three-dimensional. In this case, the TDTR signals are sensitive to both the in-plane and the through-plane thermal properties, $K_r$ and $K_z$. Therefore, both $K_r$ and $K_z$ of the substrate can be determined by conducting two sets of measurements under the different heat transfer regimes. Note that since the TDTR signals are sensitive to $K_z$ in both regimes, the validity of this approach relies on the assumption that $K_z$ is the same for the two sets of the measurements. Considering that $K_z$ of some materials measured by TDTR depends on the modulation frequency (see Section IV(D) for details), the same modulation frequency should be chosen for the two sets of the measurements. Caution should also be taken to make sure the laser spot sizes are large enough not to induce spot size dependence in the measurements.

Given the constraint that the same modulation frequency should be chosen for the variable spot size TDTR measurements, a question raised naturally is how to choose the laser spot sizes and the modulation frequency for the measurements. Jiang et al.[37] found that in the variable spot size TDTR measurements, the modulation frequency should be chosen based on the in-plane thermal diffusivity of the sample, with an empirical correlation as:

$$f = a\left(K_r/C\right)^{0.7} \tag{4.2}$$

In this formula, $f$ has a unit of MHz and $K_r/C$ has a unit of cm$^2$ s$^{-1}$, and the constant $a$ is in the range 2.2~4.4. Figure 18(b) shows an example of experimental data fitting for ZnO [0001], which has its $c$-axis along the through-plane direction. Given that the in-plane thermal diffusivity of ZnO [0001] to be $K_r/C = 0.157$ cm$^2$ s$^{-1}$ at room temperature,[37] the modulation frequency for the variable spot size TDTR measurements of ZnO should be in the range 0.6-1.2 MHz based on Eq.(3.2). The data in Figure 18(b) were taken at room temperature using two different laser spot sizes (1/$e^2$ radius $w_0 = 4$ μm and $w_0 = 16$ μm) under the same modulation frequency of 1 MHz. The measurement using the large spot size ($w_0 = 16$ μm) is only sensitive to $K_z$, while the data measured using the small spot size ($w_0 = 4$ μm) is sensitive to both $K_z$ and $K_r$. This variable spot size TDTR approach has been validated on some standard samples over a wide range of in-plane and through-plane thermal conductivities[37] (see Figure 18(c, d)), but requires $K_r > 5$ W m$^{-1}$ K$^{-1}$ for anisotropic thermal conductivity. This method has then been successfully applied in several studies of anisotropic materials including both layered two-dimensional materials (such as transition metal dichalcogenides[35] and their alloys[192] and boron nitride nanosheets[193]) and bulk crystals (such as silicon carbide[33]).

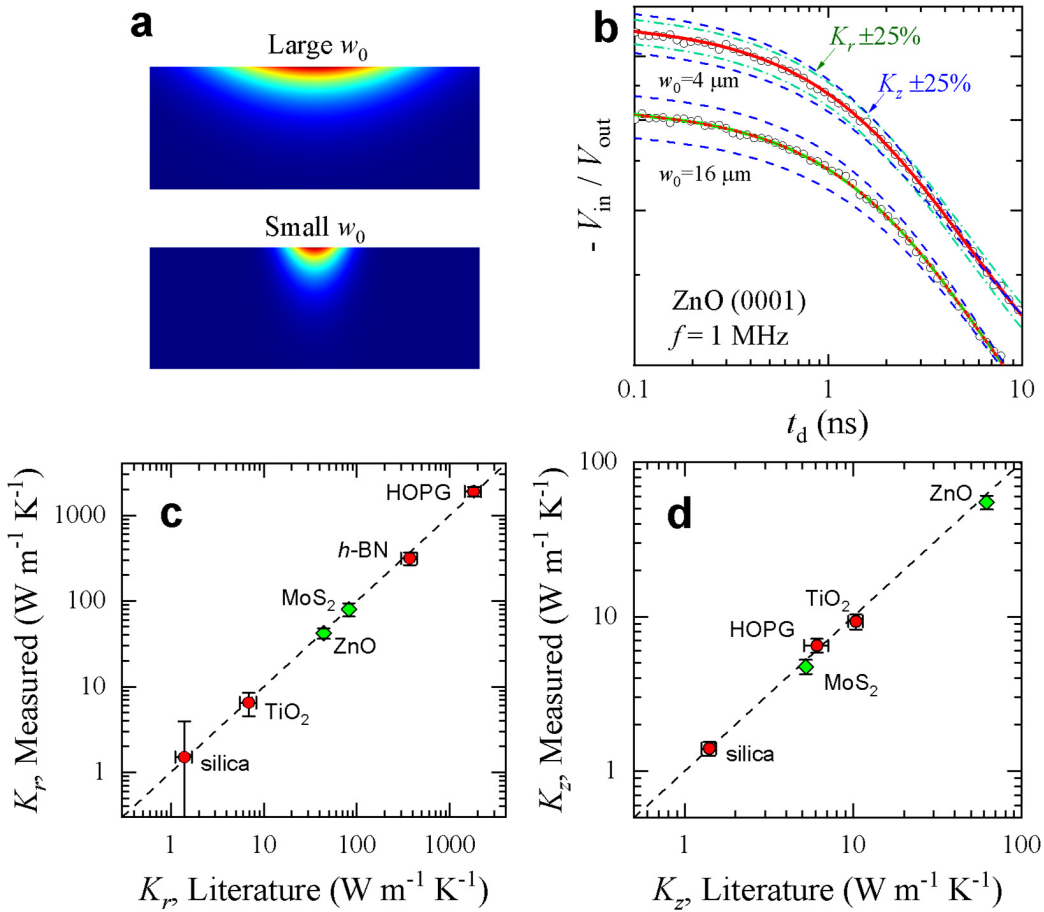

Figure 18. (a) Heat transfer in TDTR measurement is in different regimes when using a large laser spot size and a tightly focused spot size. (b) Representative data fitting using the variable spot size TDTR approach to measure the anisotropic thermal conductivity of ZnO [0001]. (c, d) Anisotropic thermal conductivity of several standard samples measured using the variable spot size TDTR approach compared with the literature values. (Reproduced with permission from Jiang *et al*.[37] Copyright 2017 by AIP Publishing.)

## C2. Beam-offset TDTR configuration

The variable spot size TDTR approach described above is limited to materials that are transversely isotropic because the thermal model assumes a cylindrical symmetry. Feser et al.[86, 87] developed a beam-offset TDTR approach to measure in-plane anisotropic materials. In this approach, the pump beam is swept across the probe beam and the full-width half-maximum (FWHM) of the out-of-phase signal at a negative delay time (e.g., -100 ps) is used to derive the in-plane thermal conductivity along the scanning direction.[194] Feser et al.[87] also recommended using a 70-nm-thick NbV film (typically with a thermal conductivity of ~20 W m$^{-1}$ K$^{-1}$) instead of the conventional 100-nm-thick Al film (with a thermal conductivity of ~200 W m$^{-1}$ K$^{-1}$) as the metal transducer for the beam-offset TDTR experiments. With a much-reduced thermal conductance of the metal film ($K_m h_m$), the FWHM signal is more sensitive to the thermal conductivity of the substrate and less sensitive to the properties of the transducer layer, thus yielding a smaller measurement uncertainty.[87] For this reason, TR-MOKE is more suitable to perform beam-offset experiments because much thinner transducer films (as thin as ~5 nm) can be used for the measurements.

For the thermal modeling of beam-offset TDTR experiments, since the in-plane symmetry is lost with the offset between the pump and the probe spots, writing the heat diffusion equation in the Cartesian coordinates rather than in the cylindrical coordinates would be more convenient. (Note that the cylindrical coordinates can still be used, but it requires some manipulation to transfer the offset Gaussian beam into an equivalent ring-shaped profile.[86]). The fully anisotropic heat diffusion equation in the Cartesian coordinates is written as:

$$C\frac{\partial T}{\partial t} = K_x \frac{\partial^2 T}{\partial x^2} + K_y \frac{\partial^2 T}{\partial y^2} + K_z \frac{\partial^2 T}{\partial z^2} + 2K_{xy}\frac{\partial^2 T}{\partial x \partial y} + 2K_{xz}\frac{\partial^2 T}{\partial x \partial z} + 2K_{yz}\frac{\partial^2 T}{\partial y \partial z} \quad (4.3)$$

By applying Fourier transform to the time variable $t$ and the in-plane coordinates $x$ and $y$, this parabolic partial differential equation can be simplified as:

$$\frac{\partial^2 \Theta}{\partial z^2} + \lambda_2 \frac{\partial \Theta}{\partial z} - \lambda_1 \Theta = 0 \tag{4.4}$$

where $\lambda_1 \equiv [iC\omega + 4\pi^2(K_x u^2 + 2K_{xy} uv + K_y v^2)]/K_z$, $\lambda_2 \equiv i4\pi(K_{xz} u + K_{yz} v)/K_z$, with $u$, $v$, and $\omega$ the Fourier transform variables of $x$-, $y$-coordinates and the time, respectively.[117]

The general solution of Eq. (4.4) is

$$\Theta = e^{u^+ z} B^+ + e^{u^- z} B^- \tag{4.5}$$

where $u^+$ and $u^-$ are the roots of the equation $x^2 + \lambda_2 x - \lambda_1 = 0$, and $B^+$ and $B^-$ are the complex constants to be determined based on the boundary conditions.

The same quadrupole approach outlined in Section II(B) can be employed for the solution of heat diffusion in a multilayered structure in the Cartesian coordinates, with the only changes that the matrix $[N]_i$ and $[M]_i$ in Eq. (2.5) and (2.6) should be updated as:

$$[N]_i = \begin{bmatrix} 1 & 1 \\ -K_z u^+ & -K_z u^- \end{bmatrix} \begin{bmatrix} e^{u^+ z} & 0 \\ 0 & e^{u^- z} \end{bmatrix}_i \tag{4.6}$$

$$[M]_i = \frac{1}{K_z(u^+ - u^-)} \begin{bmatrix} -K_z u^- & -1 \\ K_z u^+ & 1 \end{bmatrix} \tag{4.7}$$

For the heating and signal detection in beam-offset TDTR experiments, the pump and probe intensity distributions are expressed in the Cartesian coordinates as

$$p_1(x,y) = \frac{2A_1}{\pi w_{x_1} w_{y_1}} \exp\left(-\frac{2x^2}{w_{x_1}^2}\right) \exp\left(-\frac{2y^2}{w_{y_1}^2}\right) \tag{4.8}$$

$$p_2(x,y) = \frac{2A_2}{\pi w_{x_2} w_{y_2}} \exp\left(-\frac{2(x-x_c)^2}{w_{x_2}^2}\right) \exp\left(-\frac{2(y-y_c)^2}{w_{y_2}^2}\right) \tag{4.9}$$

where $w_{x_1}, w_{y_1}$ and $w_{x_2}, w_{y_2}$ are the $1/e^2$ radii of the pump and probe spots in the $x$ and $y$ directions, respectively, and $x_c, y_c$ are the offset distances between the pump and the probe in the $x$ and $y$ directions, respectively.

To adapt the thermal transport model outlined in Section II(B) for beam-offset TDTR experiments, the only other change that needs to be made is to update the expression of $\Delta T(\omega)$ in Eq. (2.18) as

$$\Delta T(\omega) = \int_{-\infty}^{\infty}\int_{-\infty}^{\infty} \hat{G}(u,v,\omega)\exp(-\pi^2(u^2 w_x^2 + v^2 w_y^2))\exp(i2\pi(ux_c + vy_c))dudv \quad (4.10)$$

where $w_x = \sqrt{(w_{x_1}^2 + w_{x_2}^2)/2}$ and $w_y = \sqrt{(w_{y_1}^2 + w_{y_2}^2)/2}$ are the RMS average of pump and probe spot sizes in the $x$ and $y$ directions, respectively.

When the beam-offset method was first developed, it was believed that the FWHM of $V_{\text{out}}$ was highly sensitive to $K_r$ of the substrate in the low-frequency limit with $d_{p,r} \gg w_0$. Note that the sensitivity of the FWHM signal is defined in the same way as that in the conventional TDTR (see Eq. (2.23)), only to have the ratio signal replaced with the FWHM signal:

$$S_\xi^{\text{FWHM}} = \frac{\partial \ln(\text{FWHM})}{\partial \ln \xi} \quad (4.11)$$

Therefore, the use of the lowest possible modulation frequency and the smallest possible laser spot size was recommended for beam-offset experiments to achieve the highest sensitivity to $K_r$ of the sample.[87] However, through the detailed sensitivity analysis, Jiang et al.[119] found that the FWHM of $V_{\text{out}}$ had the highest sensitivity to $K_r$ of the sample when $d_{p,r} \approx w_0$. The reason, as argued by Jiang et al.[119], is that although the $V_{\text{out}}$ signal is highly dependent on $K_r$ of the substrate in the limit $d_{p,r} \gg w_0$, the $V_{\text{out}}$ signal as a function of the offset distance is proportionally dependent on $K_r$, making the FWHM of $V_{\text{out}}$ not sensitive to the change in $K_r$. On the other hand, with $d_{p,r} \approx w_0$, only the $V_{\text{out}}$ signals in the short offset range $x_c \leq w_0$ are sensitive to $K_r$, making the FWHM of $V_{\text{out}}$ highly sensitivity to the change in $K_r$. Jiang et al.[119] also pointed out that the NbV transducer as a replacement of the conventional Al transducer is effective in reducing the measurement uncertainty in beam-offset experiments only when the substrate has a $K_r$ in the range 6-30 W m$^{-1}$ K$^{-1}$. For samples with $K_r > 30$ W m$^{-1}$ K$^{-1}$, both NbV and Al transducers work well, giving an uncertainty of <15% for the $K_r$ measurements. For samples with $K_r < 6$ W m$^{-1}$ K$^{-1}$, neither NbV nor Al transducer is likely to work, as the measurement uncertainty becomes very high for both transducers.

The procedure of extracting in-plane thermal conductivity from beam-offset TDTR experiments is demonstrated in Figure 19, where a ZnO [11-20] sample coated with 100 nm Al is taken as the example. Figure 19(a) shows the $V_{out}$ signal of the ZnO sample measured as a function of offset distance $x_c$ using a laser spot size $w_0 = 4.7$ μm at a modulation frequency of 0.35 MHz. ZnO is a hexagonal Wurtzite crystal with a higher thermal conductivity parallel to its c-axis (55-62 W m$^{-1}$ K$^{-1}$ from the literature[37, 195]) than the other directions (~44 W m$^{-1}$ K$^{-1}$ from the literature[37, 195]). The in-plane thermal penetration depth in ZnO, when measured at a modulation frequency of 0.35 MHz, is ~4.2 μm, which is close to the laser spot size $w_0$. Such a configuration of the laser spot size and modulation frequency is expected to have a high sensitivity to $K_r$ of the substrate. The FWHM is determined as 11.3 μm by fitting the $V_{out}$ signals in Figure 19(a) using a Gaussian function. Meanwhile, the FWHM signals are also simulated as a function of $K_r$, as shown by the red solid line in Figure 19(b). $K_r$ of the substrate along the offset direction can thus be determined by matching the measured FWHM to the simulated FWHM. The in-plane thermal conductivity of ZnO along the direction perpendicular to the c-axis is thus extracted as $K_r = 38.5$ W m$^{-1}$ K$^{-1}$ from Figure 19(b).

There are two major sources of uncertainty for the determined $K_r$ in beam-offset experiments, as shown in Figure 19(b). One is the measured FWHM signal that has an uncertainty of repeatability due to the experimental noise and the error in determining the reference phase of the lock-in detection; the other is the simulated FWHM that has an error propagated from the uncertainties of the input parameters in the thermal transport model. The uncertainty of the measured FWHM signal can be determined from the standard deviation of several individually measured FWHM signals and is usually 1%~2% for most cases (but it can reach up to ~4% at low-frequency measurements due to the high noise level). The uncertainty of the simulated FWHM is determined using the formula

$$\eta_{\text{FWHM}} = \sqrt{\sum_{\xi}\left(S_{\xi}\eta_{\xi}\right)^2} \qquad (4.12)$$

where $\xi$ is any input parameter except $K_r$. For the current case, the simulated FWHM has an estimated uncertainty of ±1.7%. Figure 19(b) shows that the ±3.5% uncertainty from the measured FWHM causes ±18.5% uncertainty in $K_r$, while the ±1.7% uncertainty in the simulated FWHM propagated from the input parameters especially the laser spot size causes ±9% uncertainty in $K_r$. Since these two sources of uncertainty are independent of each other, the total uncertainty of $K_r$ is determined as $\eta_{K_r} = \pm\sqrt{18.5^2 + 9^2}\% = \pm 21\%$.

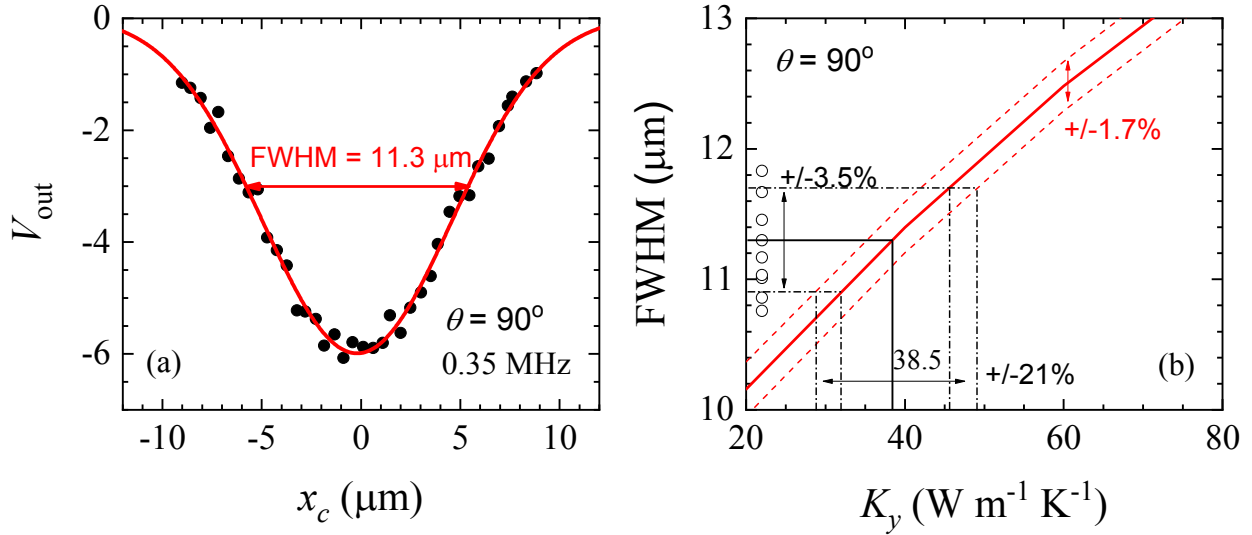

Figure 19. (a) Beam-offset experimental data (symbols) for a ZnO [11-20] sample covered with a 100nm Al transducer measured using 0.35 MHz modulation frequency and laser spot size 4.7 μm at -100 ps delay time, with the FWHM determined from the fitted Gaussian function (curves). The offset direction is perpendicular to the *c*-axis of ZnO [11-20]. (b) Determination of $K_r$ and its uncertainty by comparing the measured FWHM with the simulated FWHM. (Reproduced with permission from Jiang *et al.*[119] Copyright 2018 by AIP Publishing.)

### *C3. Elliptical-beam TDTR configuration*

The beam-offset TDTR approach is subject to measurement uncertainty from the FWHM signals and from the input parameters, especially the laser spot size. Alternatively, based on the physical picture that the sensitivity of TDTR signals to the in-plane thermal conductivity depends on how the spot size $w_0$ is compared to $d_{p,r}$, an elliptical-beam TDTR approach was recently developed to measure the in-plane thermal conductivity of transversely anisotropic samples.[119]

In the elliptical-beam TDTR approach, the experiments are conducted following the same procedure as in the conventional TDTR, *i.e.*, the pump and the probe spots are concentrically aligned on the sample surface, the ratio signals $R = -V_{in}/V_{out}$ acquired as a function of delay time are used to derive the thermal properties, except that the pump beam is of a highly elliptical shape. A schematic of the elliptical-beam method is shown in Figure 20. By using a highly elliptical pump beam for TDTR experiments, a quasi-one-dimensional temperature profile that has a fast decay along the short axis of the pump beam is induced on the sample surface. The detected TDTR signal is thus exclusively sensitive to the in-plane thermal conductivity along the short axis of the elliptical beam.

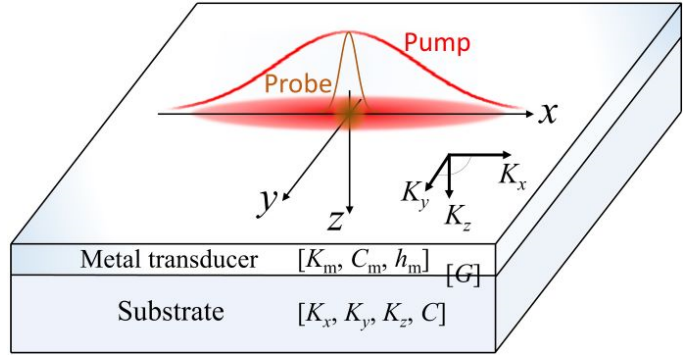

Figure 20. Schematic of the elliptical-beam method for measuring in-plane anisotropic thermal conductivity. (Reproduced with permission from Jiang *et al.*[119] Copyright 2018 by AIP Publishing.)

In the elliptical-beam configuration, the shape and the orientation of the elliptical pump beam can be controlled using a pair of cylindrical lenses. The RMS average of the pump and probe spot sizes can be characterized using the dependence of $V_{in}$ on the spatial offset of the pump and probe beams at a short positive delay time (e.g., 100 ps) at a high modulation frequency of 10 MHz.[39] Generally, the size of the elliptical laser spot should be chosen so that the sensitivity of TDTR signals to the in-plane thermal conductivity is sufficiently high along the short axis direction but is effectively suppressed along the long axis direction. A general guideline is that the major radius of the elliptical spot $w_x$ should be at least five times the in-plane thermal penetration depth in the same direction, $w_x > 5d_{p,x}$, and the minor radius of the elliptical spot $w_y$ needs to be less than two times the in-plane thermal penetration depth in the same direction, $w_y < 2d_{p,y}$.[119] For low-thermal-conductivity materials such as quartz, a more stringent criterion of $w_y < d_{p,y}$ would be needed to have a sufficiently high sensitivity to $K_y$. Other than that, the data acquisition, data reduction, and uncertainty analysis in the elliptical-beam configuration are all the same as in the conventional TDTR approach. The thermal model for the beam-offset TDTR configuration can also be applied to the elliptical-beam TDTR configuration but with the offset distance set as zero, since both employ the analysis in the Cartesian coordinates.

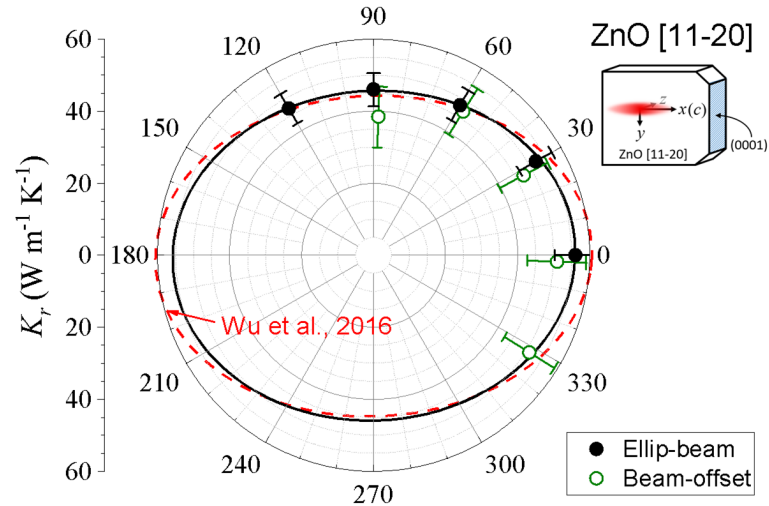

Figure 21. In-plane thermal conductivity tensor of ZnO [11-20] determined by the elliptical-beam method (solid symbols) and the beam-offset method (open symbols), compared with the results from the first-principles calculations (dashed line) from Ref. [195]. (Reproduced with permission from Jiang *et al*. Copyright 2018 by AIP Publishing.)

Figure 21 shows the in-plane thermal conductivity tensor of ZnO [11-20] measured using the elliptical-beam method, compared with those measured using the beam-offset method. A modulation frequency of 0.35 MHz was chosen for both methods. The beam-offset method used a circular laser spot with radius $w_0 = 4.7$ μm, while the elliptical-beam method used an elliptical laser spot with a long radius of $w_x = 17.3$ μm and a short radius of $w_y = 4.5$ μm. Both measurements were conducted under their optimal experimental configurations. Overall, these two methods compare relatively well with each other, with the measured in-plane thermal conductivities within the error bars. However, the data by the beam-offset method scatter more significantly whereas the data by the elliptical-beam method show better consistency. From the elliptical-beam experiments, the thermal conductivities along the in-plane directions perpendicular to and parallel to the *c*-axis of ZnO are determined to be $K_a$ = 46 W m$^{-1}$ K$^{-1}$ and $K_c$ = 56 W m$^{-1}$ K$^{-1}$, respectively. The $K_c$ of ZnO determined from the elliptical-beam method (56 W m$^{-1}$ K$^{-1}$) is slightly lower than a first-principles calculation in literature (62 W m$^{-1}$ K$^{-1}$)[195] but is consistent with another TDTR measurement of $K_z$ of ZnO [0001] (55 W m$^{-1}$ K$^{-1}$)[37], which has its *c*-axis along the through-plane direction.

### D. Probing phonon mean free paths

The knowledge of phonon mean free paths (MFPs) is important in developing microscopic pictures of the heat conduction processes, especially in micro/nano-structures and in ultrafast processes where Fourier's law of heat conduction fails. Recent advances in atomistic simulations have enabled the calculations of phonon properties including phonon dispersions and phonon lifetimes with good accuracy, even without input parameters.[196-200] However, their direct measurements remain very challenging. Based on the simple idea that the measured apparent thermal properties (thermal conductivity and interface thermal conductance) would deviate from the predictions of Fourier's law when the characteristic length is comparable to or even smaller than the phonon MFPs, some experimental work based on TDTR and its variations has been proposed to directly probe the phonon MFPs. Such characteristic thermal length scales can be either the heater size or the thermal penetration depth in TDTR experiments. This section focuses on previous attempts in the construction of phonon MFPs using TDTR and its variations.

The very first experimental observation of quasi-ballistic phonon transport induced by nano-sized heaters was realized by Siemens *et al.*[88] using pump-probe measurements based on an ultrafast soft X-ray detection scheme. In their experiment, periodic nickel nano-lines fabricated on sample surfaces acted as nanoscale heat sources when pumped by an infrared laser beam. When the size of the nano-lines decreased from 1000 to 50 nm, the measured apparent boundary thermal resistance between the nickel lines and the sapphire substrate increased by as much as three times. The authors attributed the increased thermal resistance to the ballistic effect induced by the small heater sizes being comparable to the phonon MFPs. Following this pioneering work, Minnich *et al.*[91] realized that such a heater-size dependence of thermal conductivity could be utilized to reconstruct the spectrum of phonon MFPs. They found that the apparent thermal conductivity of crystalline Si, which was determined from the best fit of TDTR experimental data by a thermal model prediction based on the heat diffusion equation, deviated from the bulk thermal conductivity values and decreased with a smaller laser spot size, as shown in Figure 22(a).[91] With the assumption that phonons with MFPs longer than the $1/e^2$ diameter ($D$) of the pump laser spot failed to establish thermal equilibrium in the heated region and do not contribute to the measured thermal conductivity, they correlated the reduced thermal conductivity of crystalline Si they obtain with the phonon MFP distribution.[91,195] Mathematically, the above statement can be expressed as:

$$\frac{K_A(D)}{K_{bulk}} = \alpha(\Lambda = D) = \int_0^D \phi(\Lambda)d\Lambda \qquad (4.13)$$

where $K_{bulk}$ is the bulk thermal conductivity, $\Lambda$ is the phonon MFP, $\alpha(\Lambda)$ is the so called accumulation function,[201] and $\phi(\Lambda)$ is the differential contribution to thermal conductivity of phonons with MFPs between $\Lambda$ and $\Lambda + d\Lambda$. Using Eq.(4.13), the laser spot size-dependent thermal conductivity were thus converted to the "MFP spectrum" of Si, as shown by the symbols in Figure 22(b), which qualitatively agrees with the first-principles predictions of thermal conductivity accumulation function (shown as the curves in Figure 22(b)).[91]

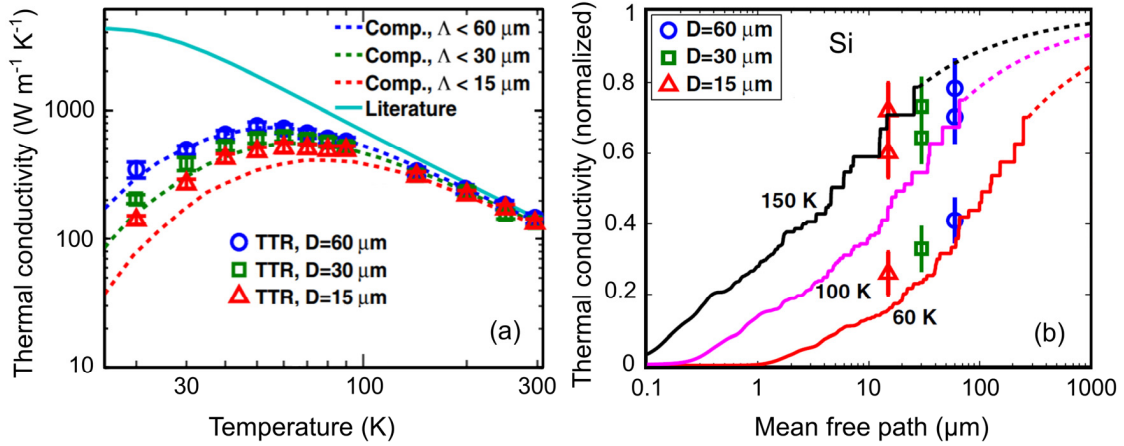

Figure 22. (a) Temperature-dependent thermal conductivity of Si measured using TDTR with different laser spot sizes (symbols) were found to deviate from the literature values (solid line) but in agreement with first-principles predictions (dashed lines) of the thermal conductivity contributed by phonons with $\Lambda < D$. (b) Apparent thermal conductivity of Si as a function of laser spot size $D$ agree with the accumulation functions from first-principles predictions. (Reproduced with permission from Minnich et al.[91] Copyright 2011 by American Physical Society.)

Despite the initial apparent success of Minnich et al.[91], their interpretations, however, are subject to many assumptions. For example, Wilson and Cahill[112] pointed out that the RMS average of the pump and probe spot sizes rather than the pump spot size alone should be the characteristic length scale due to the exchange symmetry between the heater and the thermometer, as shown in Eq. (2.18). Wilson and Cahill[112] also pointed out that the connection between the spot size-dependent thermal conductivity and the "MFP spectrum" should not be that simple, as the interface between the metal transducer and the substrate also played an important role in TDTR experiments. Besides, the assumption of a single cutoff length scale is too abrupt. Yang and Dames[202] later introduced a suppression function with a more realistic model for the apparent thermal conductivity $K_A$ as:

$$\frac{K_A(L_c)}{K_{bulk}} = \int_0^{L_c} \phi(\Lambda) S\left(\frac{\Lambda}{L_c}\right) d\Lambda = \int_0^{L_c} \alpha(\Lambda) \frac{dS\left(\frac{\Lambda}{L_c}\right)}{d\Lambda} d\Lambda \qquad (4.14)$$

where $S\left(\frac{\Lambda}{L_c}\right)$ is called the suppression function. The suppression function $S\left(\frac{\Lambda}{L_c}\right)$ can be understood as the fraction of phonons with MFPs being effectively scattered within the characteristic length $L_c$ of a heat source. For purely ballistic transport with a large $\frac{\Lambda}{L_c}$ the suppression function approaches zero, whereas for fully diffusive phonon transport with a small $\frac{\Lambda}{L_c}$ the suppression function approaches unity. The suppression function $S\left(\frac{\Lambda}{L_c}\right)$ depends on the experimental configuration and can be derived by solving Boltzmann transport equations. Reconstructing the phonon MFP spectrum [$i.e.$ $\phi(\Lambda)$ or $\alpha(\Lambda)$] from the measured $K_A(L_c)$ and the derived $S\left(\frac{\Lambda}{L_c}\right)$ is thus an inverse problem of Eq.(4.14) that involves complex optimization.[203-205]

Understanding that the spot size-dependent TDTR approach has a limit on the smallest laser spot size achievable due to the diffraction limit, Hu *et al.*[206] returned to nanostructured samples similar to Siemens *et al.*[88] to reconstruct the MFP spectrum by performing TDTR measurements of samples using nano-patterned heaters with size $D$ varying from 30 nm to 60 μm. The schematic of the sample structure is shown in Figure 23(a), and an example of the nano-patterned heater is shown in Figure 23(b). Figure 23(c) shows that the apparent thermal conductivity of SiGe determined from TDTR experiments with the hybrid nano-patterned transducers depends on the nano-heater size $D$. The MFP spectrum $\alpha(\Lambda)$ of SiGe was reconstructed based on Eq.(4.14) from the measured apparent thermal conductivity values, shown as the symbols in Figure 23(d), which are in good agreement with the first-principles prediction of thermal conductivity accumulation function (the dashed line in Figure 23(d)). This method has also been applied to other samples including sapphire, GaN and GaAs. The similar approach of reconstructing phonon MFPs has also been applied in other pump-probe techniques like transient thermal grating and ultrafast X-ray based pump-probe measurements. Interested readers can refer to Refs.[89, 207, 208] for more details.

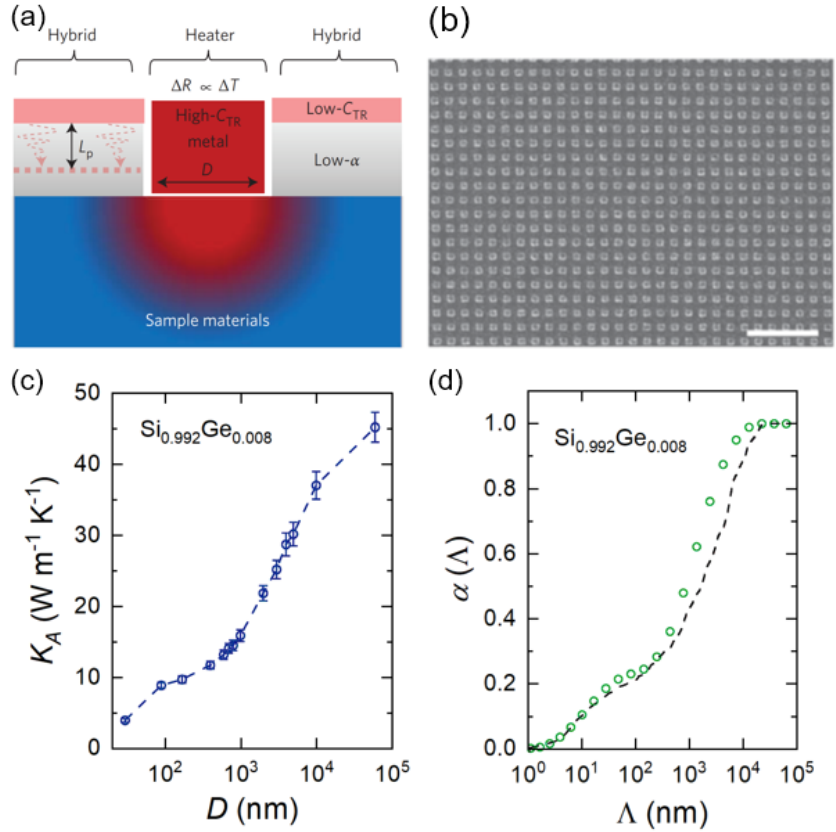

Figure 23. (a) Schematic of hybrid nanostructures for local heating and temperature detection for TDTR experiments. The sample material is heated up locally by the heater and its surface temperature change ($\Delta T$) is detected as the thermoreflectance signal ($\Delta R$). (b) SEM image of a nano-patterned sample. (c) The apparent thermal conductivity of $Si_{0.992}Ge_{0.008}$ measured as a function of the heater size $D$. (c) Reconstructed accumulation function of $Si_{0.992}Ge_{0.008}$ (circles) compared with first principles calculations (dashed line). (Reproduced with permission from Hu et al.[206] Copyright 2015 by Nature Group.)

In addition to the heater size, another important length scale that affects the phonon transport regime is the thermal penetration depth $d_p$, which describes the characteristic length of the temperature field when the surface of a sample is periodically heated at the frequency $f$. Phonon transport becomes quasi-ballistic when $d_p$ is comparable to the phonon MFPs. Koh and Cahill[90] were the first to observe a modulation-frequency-dependent thermal conductivity in some semiconductor alloys measured using TDTR, and they attributed it to the non-diffusive phonon transport, *i.e.*, phonons with $\Lambda > d_p$ (and $\Lambda > 2d_p$ in their later publication[209]) do not contribute to the apparent thermal conductivity measured by TDTR. This simple interpretation, however, works as a good approximation only for semiconductor alloys due to their extremely broad MFP distributions[209] but not for pure crystals, as frequency dependence was not observed in TDTR measurements of crystalline Si despite its very long MFPs of phonons[185, 187]. Assuming the same cutoff length of $L_c = d_p$, Regner *et al.*[92] measured the accumulation function of silicon using a variation of TDTR, the broadband frequency-domain thermoreflectance (BB-FDTR), as described in Section III(A). However, their results failed to compare well with first-principles predictions at cryogenic temperatures. Wilson and Cahill[112] later pointed out that the discrepancy could be due to an artifact created by the transducer thermalization. In the BB-FDTR experiments by Regner *et al.*[92], a 55 nm Au/5 nm Cr bilayer was used as the metal transducer, and both the heat deposition and temperature sensing were assumed to happen at the surface of the Au layer. However, Wilson and Cahill[112] pointed out that a significant amount of the laser energy was deposited in the Cr layer instead of the surface of the Au layer due to the weak electron-phonon coupling in Au, neglect of which would cause a considerable error. A later publication by Regner *et al.*[210] confirmed the non-negligible effect of the transducer type on TDTR/FDTR experiments. To better understand the frequency-dependent thermal conductivity in TDTR and FDTR experiments, Yang and Dames[211] proposed a refined model based on Boltzmann transport equation and suggested that the relaxation time $\tau$ as compared to the heating period $1/f$, represented by the dimensionless number $f\tau$, rather than the Knudsen number $\Lambda/L_p$, is the key quantity to determine the phonon transport regime.[211] Yang and Dames[211] then reconstructed the accumulation function using the relaxation times of Si, and achieved much better agreement with first-principles calculations.

While both the spot size dependence and modulation frequency dependence in TDTR experiments have been utilized to reconstruct phonon MFPs, it was puzzling to understand why the limiting length scale sometimes depends on the spot size but not the modulation frequency, and other times on the frequency but not the spot size. Wilson and Cahill[112] provided a unifying picture on the failure of Fourier's law in TDTR experiments. By comparing TDTR measurements of crystalline Si, $Si_{0.99}Ge_{0.01}$ alloy, and boron-doped Si, they pointed out that Fourier's law can fail to predict the heat flux correctly in either in-plane or through-plane directions or both, depending on different circumstances.[112] In measurements using small spot sizes, the primary effect of the small spot size is to cause a reduced in-plane heat flux than Fourier's law prediction. The thermal diffusivity of high-wavevector phonons of the substrate determines whether Fourier's law fails in in-plane or through-plane direction. Fourier's law fails more readily in through-plane direction (manifested as a frequency dependence in the measured thermal conductivity) for materials with a low thermal diffusivity for the high-wavevector phonons, and Fourier's law fails in in-plane direction (manifested as a spot size dependence) for materials with a high thermal diffusivity for the high-wavevector phonons. Besides the work by Wilson and Cahill,[112] there are also a bunch of other theoretical and numerical studies attempting to elucidate the nature of frequency and spot size dependences in TDTR experiments.[209-215] Interested readers can refer to these references.

In addition to the frequency-dependent thermal conductivity observed in TDTR measurements, the transducer/sample interface conductance $G$ was also reported to significantly depend on the modulation frequency in SiGe and some layered 2D materials like $MoS_2$[35] coated with Al transducers. The simultaneous frequency-dependence of $K_z$ and $G$ was explained by a different physical picture based on non-equilibrium phonon transport. Interested readers can also refer to Ref.[35, 192, 216] for more details. Meanwhile, Vermeersch *et al.*[109] also observed frequency dependence in both the apparent $K_z$ and $G$ for an InGaAs sample coated with a 50-nm-thick Al transducer, and they suggested that such deviations from the Fourier's law of heat conduction were due to the Lévy-type super-diffusive processes. Interested readers can refer to Ref. 110, 111 for more details.

To close, the knowledge of phonon mean free paths (MFPs) is important in developing microscopic pictures of heat conduction process. Variations of TDTR enabled many attempts in the construction of phonon MFPs with the assistance of recent developments in atomistic simulations. However, a convincing approach has not appeared.

## V. SUMMARY AND OUTLOOKS

During the past two decades, TDTR has been developed into a powerful and versatile tool for the measurements of thermal transport properties of bulk and thin film materials. In this tutorial, we have discussed the basic principles and implementation of the TDTR technique as well as its various configurations for measuring novel materials. By utilizing different laser spot sizes and modulation frequencies for the measurements, the heat transfer regime in TDTR experiments and consequently the sensitivity of TDTR signals to different parameters can be controlled, enabling TDTR measurements of multiple thermal transport properties such as the through-plane thermal conductivity, in-plane thermal conductivity, heat capacity, and the thermal boundary conductance of interfaces. TDTR has been demonstrated to measure a wide range of through-plane thermal conductivity from a high value of 2000 W m$^{-1}$ K$^{-1}$ of diamond to the ultralow thermal conductivity of ~0.03 W m$^{-1}$ K$^{-1}$ for disordered WSe$_2$ thin films. Featured by a sub-picosecond time resolution due to the ultrafast pulsed laser used, TDTR can also be applied for the study of heat carrier dynamics, such as phonon mean free paths and electron-phonon coupling. However, while being powerful and reliable, TDTR is still subject to several limitations, which also offers great opportunities for future developments of the TDTR technique.

1) As an optical-based pump-probe technique, TDTR requires the sample surface to be optically smooth to have the probe beam specularly reflected into the detector. A hypothesis is that the diffusely scattered probe light is most likely modulated by thermoelastic effects and would have an erroneous contribution to the signal if received by the detector.[12] However, there is still no systematic study on the effect of diffusely scattered probe light yet. A general rule-of-thumb is that the RMS surface roughness of the sample should be < 15 nm.[12] An alternative way to measure rough samples using TDTR is to deposit the metal transducer on a transparent substrate first, and then bond the rough sample onto the transducer and measure from the transparent substrate side. This, however, has a stringent requirement on the bonding between the metal transducer and the sample so that the additional thermal resistance introduced by the bonding would not dominate in the measurements. Meanwhile, TR-MOKE thermometry is expected to be more tolerant to the sample surface roughness because most of the spurious signals induced by the diffusely scattered pump beam, which should be independent of the magnetization states of the transducer, can be cancelled out by subtracting the two signals recorded for oppositely aligned magnetization states of the transducer. This speculation, however, has not been rigorously proven so far.

2) The range of thermal conductivity that can be measured by TDTR is limited by the workable modulation frequencies and laser spot sizes. The modulation frequency $f$ in TDTR is usually in the range of 0.2-20 MHz. TDTR measurements at $f$ < 0.2 MHz are usually problematic due to the poor signal-to-noise ratio caused by the $1/f$ noise and the large uncertainty in determining the phase caused by the strong pulse accumulation. TDTR measurements at $f$ > 20 MHz are also challenging due to the weak out-of-phase signals and the high level of the radio-frequency noise picked up by the detector and the signal cables. Besides, the Nyquist criterion[172, 217, 218] also limits the modulation frequency to be less than half of the laser repetition rate, i.e., < 40 MHz. Likewise, the laser spot size $w_0$ in TDTR is usually limited in the range 1-30 μm. The lower value in $w_0$ is due to the diffraction limit (in fact, the diffraction limit cannot even be reached because the back focal plane of the objective lens cannot be filled due to the thermal lensing effect[219]) and the higher limit in $w_0$ is due to the requirement of a minimum laser power intensity. Given the limits in the lowest modulation frequency and the smallest spot size, it is difficult for TDTR to measure in-plane thermal conductivity $K_r$ < 5 W m$^{-1}$ K$^{-1}$.[37] Meanwhile, the limit in the highest modulation frequency also constrains TDTR measurements of thermally thin films (with thickness $h < d_{p,z}$). For example, it is very challenging for TDTR to measure the thermal conductivity of Si films with a thickness <1 μm at room temperature due to the insufficient sensitivity at modulation frequencies <10 MHz. Expanding the workable modulation frequency and laser spot size can greatly extend the measurable thermal conductivity range. For example, measurements of small in-plane thermal conductivity $K_r$ < 5 W m$^{-1}$ K$^{-1}$ usually requires a lower modulation frequency of < 0.2 MHz to induce sufficient in-plane heat diffusion. The challenge of a poor signal-to-noise ratio at low modulation frequencies can be overcome by averaging over many repeated measurements. The problem of strong pulse accumulation can be resolved by reducing the laser repetition rate. The advancement in laser technology with an adjustable repetition rate of laser sources can make it possible to extend the lowest possible modulation frequency so that TDTR can measure even lower in-plane thermal conductivities. On the other hand, the modulation frequency of TDTR can also be extended up to a few hundred MHz using a GHz repetition rate laser. The advantage of a high-frequency TDTR is that it has a much finer depth resolution and thus it can measure thermal conductivities of extremely thin films, interface thermal conductance of even higher values, and probe phonon dynamics over a broader range. Direct measurements at such high modulation frequencies are certainly challenging due to the strong coherent noise and the low signal level, which, however, can be overcome by using the heterodyne detection, which has been implemented successfully in BB-FDTR.[147] It is also possible to reduce the laser spot sizes even further below the diffraction limit through the use of near-field optics,[220] by applying the pump and probe light pulses to localized regions of the sample.

3) Current TDTR cannot be used to measure the thermal conductivity of monolayer or few-layer two-dimensional (2D) materials. So far, the thermal conductivities of monolayer or few-layer 2D materials have been dominantly measured by micro-Raman spectroscopy and thermal bridge method, both with their intrinsic limitations that impair the accuracy of their measurements.[221, 222] While many attempts have been made on direct TDTR measurements of semiconductors without metal transducers,[223-227] significant progress is needed in thermal conductivity measurements of monolayer and few-layer 2D materials.

4) Most analyses of TDTR data to-date are based on solutions of the diffusion equation with the assumption that all the thermal excitations are in equilibrium with each other and a temperature field is well defined. These assumptions are not rigorously justified unless the couplings between different heat carriers are sufficiently strong, and the mean-free-paths of all of the significant heat carriers are small compared to the characteristic lengths (e.g., thermal penetration depth, laser spot size, etc.). The thermal conductivity of some semiconductor materials extracted from TDTR experiments was found to depend on the modulation frequency or the laser spot size, suggesting that the heat diffusion equation cannot accurately describe the experiment. A more rigorous practice is to analyze the TDTR data based on phonon Boltzmann transport equation, which, despite some recent progress,[110, 111, 215, 228] inevitably complicates the data reduction process.

5) TDTR measurements at low temperatures < 30 K are usually challenging due to the small heat capacities of both the metal transducer and the experimental sample that do not allow a high enough laser power for the measurements without inducing too much temperature excursion. On the other hand, TDTR measurements at high temperatures are mainly limited by the chemical and physical stabilities of the metal transducer layer. TDTR measurements using Al transducer is usually limited to be at < 600 K due to the low melting point of Al. Researchers are looking for alternative metal transducers for TDTR measurements at higher temperatures (with Pt and HfN[229] as two examples of the potential candidates) or attempting to perform TDTR measurements without metal transducers at high temperatures.[142] Meanwhile, the advent of ultrafast laser sources with tunable wavelength over a wide range (such as a system based on optical parametric oscillator[230, 231]) would also be a great advantage for TDTR, as the probe wavelength could then be chosen to match a peak in the thermoreflectance of the metal transducer, thus increasing the signal-to-noise ratio significantly. TDTR with a tunable wavelength laser also has some other advantages. For example, Lozan *et al*.[232] used a red pump beam ($\lambda$=800 nm) in a TDTR system to generate a surface plasmon on a gold film and used a green probe beam ($\lambda$=532 nm) to detect its propagation because gold has a large thermoreflectance coefficient at the probe wavelength.

While TDTR has been extensively employed to measure thermal properties of bulk and thin film solid materials, its functionality has yet to be fully explored. For example, TDTR can also be applied to characterize thermal properties of liquids and fluids, such as the thermal conductivity/heat capacity of liquids,[103] thermal conductance of solid/liquid interfaces,[80, 233, 234] the heat transfer coefficient of fluids during evaporation,[235, 236] condensation,[237] microchannel cooling,[238] and flow boiling.[239] TDTR can also be easily adapted to transient absorption and be exploited to study thermophysical phenomena in nanoparticle solutions, such as the heat diffusion from particles to the surroundings and thermal transport across surfactants,[240-243] which can have important applications such as medical therapies using intensely heated nanoparticles. TDTR combined with x-ray diffraction[244] also provides insight into the nature of heat transfer in the quasi-ballistic regime. We anticipate that the transient thermoreflectance technique will continue to improve and be applied to study more open questions of heat transfer.

## ACKNOWLEDGMENTS

The authors sincerely thank Prof. David Cahill at UIUC for helpful comments and suggestions after the first draft of this tutorial was posted on arXiv. The authors are also very grateful to the three anonymous reviewers for their constructive comments. R.Y. acknowledges the financial support by National Science Foundation (Grant No. 0635579, 0846561, 1512776), DOD (Grant No. FA9550-08-1-0078, FA9550-11-1-0109, FA9550-11-C-0034, FA8650-15-1-7524) and ARPA-E (DE-AR DE-AR0000743) over the past decade on modeling and characterization of thermal transport in nanostructured materials. R.Y. also thanks the previous students and post-doctors in his research group who have made significant contributions to some of the works reviewed here, including Jie Zhu, Jun Liu, Wei Wang, Xiaokun Gu, and many collaborators worldwide who provided materials or other assistance to the experimental details.

## REFERENCES

1. E. Machlin, *Materials Science in Microelectronics I: The Relationships Between Thin Film Processing and Structure*. (Elsevier Science, 2010).
2. P. Peumans, A. Yakimov and S. R. Forrest, J. Appl. Phys. **93** (7), 3693-3723 (2003).
3. J. Q. Xi, M. F. Schubert, J. K. Kim, E. F. Schubert, M. Chen, S.-Y. Lin, W. Liu and J. A. Smart, Nature Photonics **1**, 176 (2007).
4. M. S. Dresselhaus, G. Chen, M. Y. Tang, R. G. Yang, H. Lee, D. Z. Wang, Z. F. Ren, J. P. Fleurial and P. Gogna, Adv. Mater. **19** (8), 1043-1053 (2007).
5. G. Chen, *Nanoscale Energy Transport and Conversion : A Parallel Treatment of Electrons, Molecules, Phonons, and Photons*. (Oxford University Press, USA, 2005).
6. Z. Zhang, *Nano/Microscale Heat Transfer*. (McGraw-Hill Education, 2007).
7. H. D. Wang, *Theoretical and Experimental Studies on Non-Fourier Heat Conduction Based on Thermomass Theory*. (Springer Science & Business, 2014).
8. C. Dames, in *Heat conduction*, edited by L. M. Jiji (2009), pp. 347-401.

9. E. S. Toberer, L. L. Baranowski and C. Dames, Annual Review of Materials Research **42** (1), 179-209 (2012).
10. A. Ziabari, M. Zebarjadi, D. Vashaee and A. Shakouri, Rep. Prog. Phys. **79** (9), 095901 (2016).
11. D. G. Cahill, W. K. Ford, K. E. Goodson, G. D. Mahan, A. Majumdar, H. J. Maris, R. Merlin and S. R. Phillpot, J. Appl. Phys. **93** (2), 793 (2003).
12. D. G. Cahill, P. V. Braun, G. Chen, D. R. Clarke, S. Fan, K. E. Goodson, P. Keblinski, W. P. King, G. D. Mahan, A. Majumdar, H. J. Maris, S. R. Phillpot, E. Pop and L. Shi, Appl. Phys. Rev. **1** (1), 011305 (2014).
13. T. Luo and G. Chen, Phys. Chem. Chem. Phys. **15** (10), 3389-3412 (2013).
14. A. J. Minnich, J Phys Condens Matter **27** (5), 053202 (2015).
15. L. Shi, C. Dames, J. R. Lukes, P. Reddy, J. Duda, D. G. Cahill, J. Lee, A. Marconnet, K. E. Goodson, J.-H. Bahk, A. Shakouri, R. S. Prasher, J. Felts, W. P. King, B. Han and J. C. Bischof, Nanoscale and Microscale Thermophysical Engineering **19** (2), 127-165 (2015).
16. T. M. Tritt, *Thermal Conductivity: Theory, Properties, and Applications*. (Springer, 2004).
17. R. L. Hamilton and O. K. Crosser, Industrial & Engineering Chemistry Fundamentals **1** (3), 187-191 (1962).
18. R. C. Zeller and R. O. Pohl, Phys. Rev. B **4** (6), 2029-2041 (1971).
19. M. G. Cooper, B. B. Mikic and M. M. Yovanovich, Int. J. Heat Mass Transfer **12** (3), 279-300 (1969).
20. C. V. Madhusudana and L. S. Fletcher, AIAA J. **24** (3), 510-523 (1986).
21. R. Prasher, Proc. IEEE **94** (8), 1571-1586 (2006).
22. D. G. Cahill, Rev. Sci. Instrum. **75** (12), 5119 (2004).
23. K. Kang, Y. K. Koh, C. Chiritescu, X. Zheng and D. G. Cahill, Rev. Sci. Instrum. **79** (11), 114901 (2008).
24. A. J. Schmidt, X. Chen and G. Chen, Rev. Sci. Instrum. **79** (11), 114902 (2008).
25. J. Zhu, D. Tang, W. Wang, J. Liu, K. W. Holub and R. Yang, J. Appl. Phys. **108** (9), 094315 (2010).
26. C. Chiritescu, D. G. Cahill, N. Nguyen, D. Johnson, A. Bodapati, P. Keblinski and P. Zschack, Science **315** (5810), 351-353 (2007).
27. M. N. Luckyanova, J. Garg, K. Esfarjani, A. Jandl, M. T. Bulsara, A. J. Schmidt, A. J. Minnich, S. Chen, M. S. Dresselhaus, Z. Ren, E. A. Fitzgerald and G. Chen, Science **338** (6109), 936-939 (2012).
28. D.-W. Oh, C. Ko, S. Ramanathan and D. G. Cahill, Appl. Phys. Lett. **96** (15), 151906 (2010).
29. R. Cheaito, J. C. Duda, T. E. Beechem, K. Hattar, J. F. Ihlefeld, D. L. Medlin, M. A. Rodriguez, M. J. Campion, E. S. Piekos and P. E. Hopkins, Phys. Rev. Lett. **109** (19), 195901 (2012).
30. Q. Zheng, P. V. Braun and D. G. Cahill, Advanced Materials Interfaces **3** (16), 1600234 (2016).
31. H. Zhang, X. Chen, Y. D. Jho and A. J. Minnich, Nano Lett. **16** (3), 1643-1649 (2016).
32. Z. Guo, A. Verma, X. Wu, F. Sun, A. Hickman, T. Masui, A. Kuramata, M. Higashiwaki, D. Jena and T. Luo, Appl. Phys. Lett. **106** (11), 111909 (2015).
33. X. Qian, P. Jiang and R. Yang, Materials Today Physics **3**, 70-75 (2017).
34. J. Liu, G.-M. Choi and D. G. Cahill, J. Appl. Phys. **116** (23), 233107 (2014).
35. P. Jiang, X. Qian, X. Gu and R. Yang, Adv. Mater. **29** (36), 1701068 (2017).
36. Z. Cheng, T. Bougher, T. Bai, S. Y. Wang, C. Li, L. Yates, B. M. Foley, M. Goorsky, B. A. Cola, F. Faili and S. Graham, ACS Appl Mater Interfaces **10** (5), 4808-4815 (2018).
37. P. Jiang, X. Qian and R. Yang, Rev. Sci. Instrum. **88** (7), 074901 (2017).
38. J. Liu, J. Zhu, M. Tian, X. Gu, A. Schmidt and R. Yang, Rev. Sci. Instrum. **84** (3), 034902 (2013).
39. C. Wei, X. Zheng, D. G. Cahill and J. C. Zhao, Rev. Sci. Instrum. **84** (7), 071301 (2013).
40. B. Gundrum, D. Cahill and R. Averback, Phys. Rev. B **72** (24), 245426 (2005).
41. H.-K. Lyeo and D. Cahill, Phys. Rev. B **73** (14), 144301 (2006).
42. B. F. Donovan, C. J. Szwejkowski, J. C. Duda, R. Cheaito, J. T. Gaskins, C. Y. Peter Yang, C. Constantin, R. E. Jones and P. E. Hopkins, Appl. Phys. Lett. **105** (20), 203502 (2014).
43. G. T. Hohensee, R. B. Wilson and D. G. Cahill, Nat. Commun. **6**, 6578 (2015).
44. R. B. Wilson, B. A. Apgar, W.-P. Hsieh, L. W. Martin and D. G. Cahill, Phys. Rev. B **91** (11), 115414 (2015).

45. L. S. Larkin, M. R. Redding, N. Q. Le and P. M. Norris, J. Heat Transfer **139** (3), 031301 (2016).
46. R. M. Costescu, M. A. Wall and D. G. Cahill, Phys. Rev. B **67** (5), 054302 (2003).
47. M. D. Losego, M. E. Grady, N. R. Sottos, D. G. Cahill and P. V. Braun, Nat Mater **11** (6), 502-506 (2012).
48. A. J. Schmidt, Annu. Rev. Heat Transfer **16** (1), 159 (2014).
49. A. Rosencwaig, Science **218** (4569), 223-228 (1982).
50. A. Rosencwaig and A. Gersho, J. Appl. Phys **47** (1), 64 (1976).
51. H. E. Elsayed-Ali, T. B. Norris, M. A. Pessot and G. A. Mourou, Phys. Rev. Lett. **58** (12), 1212-1215 (1987).
52. S. D. Brorson, A. Kazeroonian, J. S. Moodera, D. W. Face, T. K. Cheng, E. P. Ippen, M. S. Dresselhaus and G. Dresselhaus, Phys. Rev. Lett. **64** (18), 2172-2175 (1990).
53. R. H. M. Groeneveld, R. Sprik and A. Lagendijk, Phys. Rev. B **45** (9), 5079-5082 (1992).
54. R. H. M. Groeneveld, R. Sprik and A. Lagendijk, Phys. Rev. B **51** (17), 11433-11445 (1995).
55. A. Giri, J. T. Gaskins, B. F. Donovan, C. Szwejkowski, R. J. Warzoha, M. A. Rodriguez, J. Ihlefeld and P. E. Hopkins, J. Appl. Phys. **117** (10), 105105 (2015).
56. A. Giri, J. T. Gaskins, B. M. Foley, R. Cheaito and P. E. Hopkins, J. Appl. Phys. **117** (4), 044305 (2015).
57. O. B. Wright and K. Kawashima, Phys. Rev. Lett. **69** (11), 1668-1671 (1992).
58. M. Hase, K. Ishioka, J. Demsar, K. Ushida and M. Kitajima, Phys. Rev. B **71** (18), 184301 (2005).
59. J. Ravichandran, A. K. Yadav, R. Cheaito, P. B. Rossen, A. Soukiassian, S. J. Suresha, J. C. Duda, B. M. Foley, C. H. Lee, Y. Zhu, A. W. Lichtenberger, J. E. Moore, D. A. Muller, D. G. Schlom, P. E. Hopkins, A. Majumdar, R. Ramesh and M. A. Zurbuchen, Nat Mater **13** (2), 168-172 (2014).
60. W. Ma, T. Miao, X. Zhang, M. Kohno and Y. Takata, The Journal of Physical Chemistry C **119** (9), 5152-5159 (2015).
61. C. Thomsen, J. Strait, Z. Vardeny, H. J. Maris, J. Tauc and J. J. Hauser, Phys. Rev. Lett. **53** (10), 989-992 (1984).
62. C. Thomsen, H. J. Maris and J. Tauc, Thin Solid Films **154** (1–2), 217-223 (1987).
63. H. T. Grahn, H. J. Maris and J. Tauc, Quantum Electronics, IEEE Journal of **25** (12), 2562-2569 (1989).
64. H. N. Lin, R. J. Stoner, H. J. Maris and J. Tauc, J. Appl. Phys. **69** (7), 3816 (1991).
65. G. Tas and H. Maris, Phys. Rev. B **49** (21), 15046-15054 (1994).
66. C. J. Morath and H. J. Maris, Phys. Rev. B **54** (1), 203-213 (1996).
67. H. Y. Hao, H. J. Maris and D. K. Sadana, Electrochem. Solid-State Lett. **1** (1), 54-55 (1998).
68. H. Y. Hao and H. J. Maris, Phys. Rev. Lett. **84** (24), 5556-5559 (2000).
69. G. A. Antonelli, H. J. Maris, S. G. Malhotra and J. M. E. Harper, J. Appl. Phys. **91** (5), 3261 (2002).
70. Y. Li, Q. Miao, A. V. Nurmikko and H. J. Maris, J. Appl. Phys. **105** (8), 083516 (2009).
71. F. Yang, T. J. Grimsley, S. Che, G. A. Antonelli, H. J. Maris and A. V. Nurmikko, J. Appl. Phys. **107** (10), 103537 (2010).
72. G. T. Hohensee, W. P. Hsieh, M. D. Losego and D. G. Cahill, Rev. Sci. Instrum. **83** (11), 114902 (2012).
73. R. H. Magruder, L. Yang, R. F. Haglund, C. W. White, L. Yang, R. Dorsinville and R. R. Alfano, Appl. Phys. Lett. **62** (15), 1730-1732 (1993).
74. Y. C. Chen, N. R. Raravikar, L. S. Schadler, P. M. Ajayan, Y. P. Zhao, T. M. Lu, G. C. Wang and X. C. Zhang, Appl. Phys. Lett. **81** (6), 975-977 (2002).
75. P. A. George, J. Strait, J. Dawlaty, S. Shivaraman, M. Chandrashekhar, F. Rana and M. G. Spencer, Nano Lett. **8** (12), 4248-4251 (2008).
76. X. Zheng, D. G. Cahill, R. Weaver and J.-C. Zhao, J. Appl. Phys. **104** (7), 073509 (2008).
77. C. A., Paddock and G. L. Eesley, J. Appl. Phys. **60** (1), 285 (1986).
78. W. S. Capinski, H. J. Maris, T. Ruf, M. Cardona, K. Ploog and D. S. Katzer, Phys. Rev. B **59** (12), 8105 (1999).
79. R. M. Costescu, D. G. Cahill, F. H. Fabreguette, Z. A. Sechrist and S. M. George, Science **303** (5660), 989-990 (2004).

80. Z. Ge, D. Cahill and P. Braun, Phys. Rev. Lett. **96** (18), 186101 (2006).
81. Y. K. Koh, M. H. Bae, D. G. Cahill and E. Pop, Nano Lett. **10** (11), 4363-4368 (2010).
82. E. Ziade, J. Yang, G. Brummer, D. Nothern, T. Moustakas and A. J. Schmidt, Appl. Phys. Lett. **107** (9), 091605 (2015).
83. Y. K. Koh, A. S. Lyons, M. H. Bae, B. Huang, V. E. Dorgan, D. G. Cahill and E. Pop, Nano Lett. **16** (10), 6014-6020 (2016).
84. R. Stoner and H. Maris, Phys. Rev. B **48** (22), 16373-16387 (1993).
85. F. Krahl, A. Giri, J. A. Tomko, T. Tynell, P. E. Hopkins and M. Karppinen, Advanced Materials Interfaces **5** (11), 1701692 (2018).
86. J. P. Feser and D. G. Cahill, Rev. Sci. Instrum. **83** (10), 104901 (2012).
87. J. P. Feser, J. Liu and D. G. Cahill, Rev. Sci. Instrum. **85** (10), 104903 (2014).
88. M. E. Siemens, Q. Li, R. Yang, K. A. Nelson, E. H. Anderson, M. M. Murnane and H. C. Kapteyn, Nat Mater **9** (1), 26-30 (2010).
89. K. M. Hoogeboom-Pot, J. N. Hernandez-Charpak, X. Gu, T. D. Frazer, E. H. Anderson, W. Chao, R. W. Falcone, R. Yang, M. M. Murnane, H. C. Kapteyn and D. Nardi, Proc. Natl. Acad. Sci. U.S.A. **112** (16), 4846-4851 (2015).
90. Y. K. Koh and D. G. Cahill, Phys. Rev. B **76** (7), 075207 (2007).
91. A. J. Minnich, J. A. Johnson, A. J. Schmidt, K. Esfarjani, M. S. Dresselhaus, K. A. Nelson and G. Chen, Phys. Rev. Lett. **107** (9), 095901 (2011).
92. K. T. Regner, D. P. Sellan, Z. Su, C. H. Amon, A. J. McGaughey and J. A. Malen, Nat. Commun. **4**, 1640 (2013).
93. A. J. Schmidt, R. Cheaito and M. Chiesa, Rev. Sci. Instrum. **80** (9), 094901 (2009).
94. J. A. Malen, K. Baheti, T. Tong, Y. Zhao, J. A. Hudgings and A. Majumdar, J. Heat Transfer **133** (8), 081601 (2011).
95. B. Sun and Y. K. Koh, Rev. Sci. Instrum. **87** (6), 064901 (2016).
96. L. S. Larkin, J. L. Smoyer and P. M. Norris, Int. J. Heat Mass Transfer **109**, 786-790 (2017).
97. W. S. Capinski and H. J. Maris, Rev. Sci. Instrum. **67** (8), 2720-2726 (1996).
98. R. B. Wilson, B. A. Apgar, L. W. Martin and D. G. Cahill, Opt. Express **20** (27), 28829-28838 (2012).
99. A. N. Smith, J. L. Hostetler and P. M. Norris, Microscale Thermophys. Eng. **4** (1), 51-60 (2000).
100. T. Naoyuki, B. Tetsuya and O. Akira, Meas. Sci. Technol. **12** (12), 2064 (2001).
101. A. J. Schmidt, PhD thesis, MASSACHUSETTS INSTITUTE OF TECHNOLOGY, 2008.
102. G. Pernot, H. Michel, B. Vermeersch, P. Burke, H. Lu, J.-M. Rampnoux, S. Dilhaire, Y. Ezzahri, A. Gossard and A. Shakouri, MRS Proceedings **1347**, mrss11-1347-bb1307-1307 (2011).
103. A. Schmidt, M. Chiesa, X. Chen and G. Chen, Rev. Sci. Instrum. **79** (6), 064902 (2008).
104. E. D. Palik, *Handbook of Optical Constants of Solids*. (Academic Press, 1991).
105. Y. Wang, J. Y. Park, Y. K. Koh and D. G. Cahill, J. Appl. Phys. **108** (4), 043507 (2010).
106. T. Favaloro, J. H. Bahk and A. Shakouri, Rev. Sci. Instrum. **86** (2), 024903 (2015).
107. J. Yang, C. Maragliano and A. J. Schmidt, Rev. Sci. Instrum. **84** (10), 104904 (2013).
108. J. L. Braun, C. J. Szwejkowski, A. Giri and P. E. Hopkins, J. Heat Transfer **140** (5), 052801 (2018).
109. B. Vermeersch, A. M. S. Mohammed, G. Pernot, Y. R. Koh and A. Shakouri, Phys. Rev. B **90** (1), 014306 (2014).
110. B. Vermeersch, J. Carrete, N. Mingo and A. Shakouri, Phys. Rev. B **91** (8), 085202 (2015).
111. B. Vermeersch, A. M. S. Mohammed, G. Pernot, Y. R. Koh and A. Shakouri, Phys. Rev. B **91** (8), 085203 (2015).
112. R. B. Wilson and D. G. Cahill, Nat. Commun. **5**, 5075 (2014).
113. H. S. Carslaw and J. C. Jaeger, *Conduction of heat in solids*. (Clarendon Press, 1959).
114. A. Feldman, High Temp. - High Press. **31**, 293 (1999).
115. J. H. Kim, A. Feldman and D. Novotny, J. Appl. Phys. **86** (7), 3959-3963 (1999).
116. D. Maillet, *Thermal quadrupoles: solving the heat equation through integral transforms*. (Wiley, 2000).
117. R. N. Bracewell, *The Fourier Transform and its applications*, 3, illustrated ed. (McGraw Hill, 2000).

118. L. A. Barragán, J. I. Artigas, R. Alonso and F. Villuendas, Rev. Sci. Instrum. **72** (1), 247-251 (2001).
119. P. Jiang, X. Qian and R. Yang, Rev. Sci. Instrum. **89** (9), 094902 (2018).
120. D. G. Cahill, K. Goodson and A. Majumdar, J. Heat Transfer **124** (2), 223 (2002).
121. D. Cahill, F. Watanabe, A. Rockett and C. Vining, Phys. Rev. B **71** (23), 235202 (2005).
122. X. Zheng, D. G. Cahill, P. Krasnochtchekov, R. S. Averback and J. C. Zhao, Acta Mater. **55** (15), 5177-5185 (2007).
123. V. Rawat, Y. K. Koh, D. G. Cahill and T. D. Sands, J. Appl. Phys. **105** (2), 024909 (2009).
124. D. W. Oh, S. Kim, J. A. Rogers, D. G. Cahill and S. Sinha, Adv. Mater. **23** (43), 5028 (2011).
125. X. Wang, C. D. Liman, N. D. Treat, M. L. Chabinyc and D. G. Cahill, Phys. Rev. B **88** (7), 075310 (2013).
126. D. Hamby, Environ. Monit. Assess. **32** (2), 135-154 (1994).
127. J. Yang, E. Ziade and A. J. Schmidt, Rev. Sci. Instrum. **87** (1), 014901 (2016).
128. R. B. Wilson and D. G. Cahill, Appl. Phys. Lett. **107** (20), 203112 (2015).
129. X. Xie, D. Li, T.-H. Tsai, J. Liu, P. V. Braun and D. G. Cahill, Macromolecules **49** (3), 972-978 (2016).
130. M. J. Assael, S. Botsios, K. Gialou and I. N. Metaxa, Int. J. Thermophys. **26** (5), 1595-1605 (2005).
131. W. R. Thurber and A. J. H. Mante, Phys. Rev. **139** (5A), A1655-A1665 (1965).
132. M. N. Touzelbaev, P. Zhou, R. Venkatasubramanian and K. E. Goodson, J. Appl. Phys. **90** (2), 763 (2001).
133. Y. Ezzahri, S. Dilhaire, S. Grauby, J. M. Rampnoux, W. Claeys, Y. Zhang, G. Zeng and A. Shakouri, Appl. Phys. Lett. **87** (10), 103506 (2005).
134. Y. K. Koh, Y. Cao, D. G. Cahill and D. Jena, Adv. Funct. Mater. **19** (4), 610-615 (2009).
135. M. N. Luckyanova, J. A. Johnson, A. A. Maznev, J. Garg, A. Jandl, M. T. Bulsara, E. A. Fitzgerald, K. A. Nelson and G. Chen, Nano Lett. **13** (9), 3973-3977 (2013).
136. A. Sood, J. A. Rowlette, C. G. Caneau, E. Bozorg-Grayeli, M. Asheghi and K. E. Goodson, Appl. Phys. Lett. **105** (5), 051909 (2014).
137. R. Cheaito, C. A. Polanco, S. Addamane, J. Zhang, A. W. Ghosh, G. Balakrishnan and P. E. Hopkins, Phys. Rev. B **97** (8), 085306 (2018).
138. F. Krahl, A. Giri, J. A. Tomko, T. Tynell, P. E. Hopkins and M. Karppinen, Advanced Materials Interfaces **5** (11), 1701692 (2018).
139. H.-S. Yang, D. G. Cahill, X. Liu, J. L. Feldman, R. S. Crandall, B. A. Sperling and J. R. Abelson, Phys. Rev. B **81** (10), 104203 (2010).
140. D. G. Cahill, A. Melville, D. G. Schlom and M. A. Zurbuchen, Appl. Phys. Lett. **96** (12), 121903 (2010).
141. L. Li, X.-J. Yan, S.-T. Dong, Y.-Y. Lv, X. Li, S.-H. Yao, Y.-B. Chen, S.-T. Zhang, J. Zhou, H. Lu, M.-H. Lu and Y.-F. Chen, Appl. Phys. Lett. **111** (3), 033902 (2017).
142. Q. Zheng, A. B. Mei, M. Tuteja, D. G. Sangiovanni, L. Hultman, I. Petrov, J. E. Greene and D. G. Cahill, Physical Review Materials **1** (6), 065002 (2017).
143. W. Wang and D. G. Cahill, Phys. Rev. Lett. **109** (17), 175503 (2012).
144. A. J. Schmidt, K. C. Collins, A. J. Minnich and G. Chen, J. Appl. Phys. **107** (10), 104907 (2010).
145. M. Panzer, G. Zhang, D. Mann, X. Hu, E. Pop, H. Dai and K. Goodson, J. Heat Transfer **130** (5), 052401 (2008).
146. R. Garrelts, A. Marconnet and X. Xu, Nanoscale and Microscale Thermophysical Engineering **19** (4), 245-257 (2015).
147. K. T. Regner, S. Majumdar and J. A. Malen, Rev. Sci. Instrum. **84** (6), 064901 (2013).
148. J. P. Freedman, J. H. Leach, E. A. Preble, Z. Sitar, R. F. Davis and J. A. Malen, Sci Rep **3**, 2963 (2013).
149. J. Y. Chen, J. Zhu, D. Zhang, D. M. Lattery, M. Li, J. P. Wang and X. Wang, J Phys Chem Lett **7** (13), 2328-2332 (2016).
150. J. Kimling, A. Philippi-Kobs, J. Jacobsohn, H. P. Oepen and D. G. Cahill, Phys. Rev. B **95** (18), 184305 (2017).
151. J. Zhu, H. Park, J.-Y. Chen, X. Gu, H. Zhang, S. Karthikeyan, N. Wendel, S. A. Campbell, M. Dawber, X. Du, M. Li, J.-P. Wang, R. Yang and X. Wang, Advanced Electronic Materials **2** (5), 1600040 (2016).

152. K. E. O'Hara, X. Hu and D. G. Cahill, J. Appl. Phys. **90** (9), 4852 (2001).
153. O. Florez, P. F. Jarschel, Y. A. Espinel, C. M. Cordeiro, T. P. Mayer Alegre, G. S. Wiederhecker and P. Dainese, Nat. Commun. **7**, 11759 (2016).
154. M. Cinchetti, M. Sanchez Albaneda, D. Hoffmann, T. Roth, J. P. Wustenberg, M. Krauss, O. Andreyev, H. C. Schneider, M. Bauer and M. Aeschlimann, Phys. Rev. Lett. **97** (17), 177201 (2006).
155. E. Carpene, E. Mancini, C. Dallera, M. Brenna, E. Puppin and S. De Silvestri, Phys. Rev. B **78** (17), 174422 (2008).
156. M. Krauß, T. Roth, S. Alebrand, D. Steil, M. Cinchetti, M. Aeschlimann and H. C. Schneider, Phys. Rev. B **80** (18), 180407 (2009).
157. J. Qi, Y. Xu, A. Steigerwald, X. Liu, J. K. Furdyna, I. E. Perakis and N. H. Tolk, Phys. Rev. B **79** (8), 085304 (2009).
158. A. Weber, F. Pressacco, S. Günther, E. Mancini, P. M. Oppeneer and C. H. Back, Phys. Rev. B **84** (13), 132412 (2011).
159. G.-M. Choi, B.-C. Min, K.-J. Lee and D. G. Cahill, Nat. Commun. **5**, 4334 (2014).
160. J. Kimling, J. Kimling, R. B. Wilson, B. Hebler, M. Albrecht and D. G. Cahill, Phys. Rev. B **90** (22), 224408 (2014).
161. G.-M. Choi, C.-H. Moon, B.-C. Min, K.-J. Lee and D. G. Cahill, Nature Physics **11** (7), 576-581 (2015).
162. R. Gebs, G. Klatt, C. Janke, T. Dekorsy and A. Bartels, Opt. Express **18** (6), 5974-5983 (2010).
163. A. E. Paul, E. L. Fred, J. Yanan, B. K. Galen and M. L. Normand, Appl. Spectmsc. **41** (1), 2-4 (1987).
164. G. J. Fiechtner, G. B. King, N. M. Laurendeau and F. E. Lytle, Appl. Opt. **31** (15), 2849-2864 (1992).
165. R. J. Kneisler, F. E. Lytle, G. J. Fiechtner, Y. Jiang, G. B. King and N. M. Laurendeau, Opt. Lett. **14** (5), 260-262 (1989).
166. J. D. Kafka, J. W. Pieterse and M. L. Watts, Opt. Lett. **17** (18), 1286-1288 (1992).
167. T. Yasui, E. Saneyoshi and T. Araki, Appl. Phys. Lett. **87** (6), 061101 (2005).
168. A. Bartels, R. Cerna, C. Kistner, A. Thoma, F. Hudert, C. Janke and T. Dekorsy, Rev. Sci. Instrum. **78** (3), 035107 (2007).
169. G. Klatt, R. Gebs, C. Janke, T. Dekorsy and A. Bartels, Opt. Express **17** (25), 22847-22854 (2009).
170. T. Yasui, K. Kawamoto, Y.-D. Hsieh, Y. Sakaguchi, M. Jewariya, H. Inaba, K. Minoshima, F. Hindle and T. Araki, Opt. Express **20** (14), 15071-15078 (2012).
171. J. Cuffe, O. Ristow, E. Chávez, A. Shchepetov, P. O. Chapuis, F. Alzina, M. Hettich, M. Prunnila, J. Ahopelto, T. Dekorsy and C. Sotomayor Torres, Phys. Rev. Lett. **110** (9), 095503 (2013).
172. S. Dilhaire, G. Pernot, G. Calbris, J. M. Rampnoux and S. Grauby, J. Appl. Phys. **110** (11), 114314 (2011).
173. Q. d'Acremont, G. Pernot, J.-M. Rampnoux, A. Furlan, D. Lacroix, A. Ludwig and S. Dilhaire, Rev. Sci. Instrum. **88** (7), 074902 (2017).
174. G. Pernot, M. Stoffel, I. Savic, F. Pezzoli, P. Chen, G. Savelli, A. Jacquot, J. Schumann, U. Denker, I. Monch, C. Deneke, O. G. Schmidt, J. M. Rampnoux, S. Wang, M. Plissonnier, A. Rastelli, S. Dilhaire and N. Mingo, Nat Mater **9** (6), 491-495 (2010).
175. C. Pradere, L. Clerjaud, J. C. Batsale and S. Dilhaire, Rev. Sci. Instrum. **82** (5), 054901 (2011).
176. S. L. Lai, G. Ramanath, L. H. Allen and P. Infante, Appl. Phys. Lett. **70** (1), 43-45 (1997).
177. D. R. Queen and F. Hellman, Rev. Sci. Instrum. **80** (6), 063901 (2009).
178. Y. Zhang, J. X. Cao, Y. Xiao and X. H. Yan, J. Appl. Phys. **102** (10), 104303 (2007).
179. J. S. Kurtz, R. R. Johnson, M. Tian, N. Kumar, Z. Ma, S. Xu and M. H. Chan, Phys. Rev. Lett. **98** (24), 247001 (2007).
180. N. R. Pradhan, H. Duan, J. Liang and G. S. Iannacchione, Nanotechnology **19** (48), 485712 (2008).
181. J. Liu, B. Yoon, E. Kuhlmann, M. Tian, J. Zhu, S. M. George, Y. C. Lee and R. Yang, Nano Lett. **13** (11), 5594-5599 (2013).
182. X. Xie, K. Yang, D. Li, T.-H. Tsai, J. Shin, P. V. Braun and D. G. Cahill, Phys. Rev. B **95** (3), 035406 (2017).
183. J. Liu, S. Ju, Y. Ding and R. Yang, Appl. Phys. Lett. **104** (15), 153110 (2014).
184. P. Jiang, B. Huang and Y. K. Koh, Rev. Sci. Instrum. **87** (7), 075101 (2016).

185. P. Jiang, L. Lindsay and Y. K. Koh, J. Appl. Phys. **119** (24), 245705 (2016).
186. P. Jiang, L. Lindsay, X. Huang and Y. K. Koh, Phys. Rev. B **97** (19), 195308 (2018).
187. K. Esfarjani, G. Chen and H. T. Stokes, Phys. Rev. B **84** (8), 085204 (2011).
188. A. T. Ramu and J. E. Bowers, Rev. Sci. Instrum. **83** (12), 124903 (2012).
189. V. Mishra, C. L. Hardin, J. E. Garay and C. Dames, Rev. Sci. Instrum. **86** (5), 054902 (2015).
190. O. W. Käding, H. Skurk, A. A. Maznev and E. Matthias, Appl. Phys. A **61** (3), 253-261 (1995).
191. M. N. Luckyanova, J. A. Johnson, A. A. Maznev, J. Garg, A. Jandl, M. T. Bulsara, E. A. Fitzgerald, K. A. Nelson and G. Chen, Nano Lett. **13** (9), 3973-3977 (2013).
192. X. Qian, P. Jiang, P. Yu, X. Gu, Z. Liu and R. Yang, Appl. Phys. Lett. **112** (24), 241901 (2018).
193. Y. Wang, L. Xu, Z. Yang, H. Xie, P. Jiang, J. Dai, W. Luo, Y. Yao, E. Hitz, R. Yang, B. Yang and L. Hu, Nanoscale **10** (1), 167-173 (2018).
194. Although not stated in the original publication in Ref. 87, through a private communication Prof. David Cahill noted that it is also helpful to cross-check the measurements by analyzing FWHM of the in-phase signal acquired at the same negative delay time, as it contains information of the spot size, which is critically important for the accuracy of the beam-offset TDTR experiments.
195. X. Wu, J. Lee, V. Varshney, J. L. Wohlwend, A. K. Roy and T. Luo, Sci Rep **6**, 22504 (2016).
196. D. A. Broido, M. Malorny, G. Birner, N. Mingo and D. A. Stewart, Appl. Phys. Lett. **91** (23), 231922 (2007).
197. W. Li, J. Carrete, N. A. Katcho and N. Mingo, Comput. Phys. Commun. **185** (6), 1747-1758 (2014).
198. N. Mingo, D. A. Stewart, D. A. Broido, L. Lindsay and W. Li, in *Length-Scale Dependent Phonon Interactions* (Springer, 2014), pp. 137-173.
199. L. Lindsay, Nanoscale and Microscale Thermophysical Engineering **20** (2), 67-84 (2016).
200. T. Feng, L. Lindsay and X. Ruan, Phys. Rev. B **96** (16), 161201 (2017).
201. A. S. Henry and G. Chen, J. Comput. Theor. Nanos. **5** (2), 141-152 (2008).
202. F. Yang and C. Dames, Phys. Rev. B **87** (3), 035437 (2013).
203. A. J. Minnich, Phys. Rev. Lett. **109** (20), 205901 (2012).
204. C. Hua and A. J. Minnich, Phys. Rev. B **90** (21), 214306 (2014).
205. V. Chiloyan, L. Zeng, S. Huberman, A. A. Maznev, K. A. Nelson and G. Chen, Phys. Rev. B **93** (15), 155201 (2016).
206. Y. Hu, L. Zeng, A. J. Minnich, M. S. Dresselhaus and G. Chen, Nat Nano **10** (8), 701-706 (2015).
207. J. A. Johnson, A. A. Maznev, J. Cuffe, J. K. Eliason, A. J. Minnich, T. Kehoe, C. M. S. Torres, G. Chen and K. A. Nelson, Phys. Rev. Lett. **110** (2), 025901 (2013).
208. A. A. Maznev, J. A. Johnson and K. A. Nelson, Phys. Rev. B **84** (19), 195206 (2011).
209. Y. K. Koh, D. G. Cahill and B. Sun, Phys. Rev. B **90** (20), 205412 (2014).
210. K. T. Regner, A. J. H. McGaughey and J. A. Malen, Phys. Rev. B **90** (6), 064302 (2014).
211. F. Yang and C. Dames, Phys. Rev. B **91** (16), 165311 (2015).
212. A. J. Minnich, G. Chen, S. Mansoor and B. S. Yilbas, Phys. Rev. B **84** (23), 235207 (2011).
213. C. A. da Cruz, W. Li, N. A. Katcho and N. Mingo, Appl. Phys. Lett. **101** (8), 083108 (2012).
214. D. Ding, X. Chen and A. J. Minnich, Appl. Phys. Lett. **104** (14), 143104 (2014).
215. S. A. Ali and S. Mazumder, Int. J. Heat Mass Transfer **107**, 607-621 (2017).
216. R. B. Wilson, J. P. Feser, G. T. Hohensee and D. G. Cahill, Phys. Rev. B **88** (14), 144305 (2013).
217. S. W. Smith, *The Scientist and engineer's guide to digital signal processing[*(California Technical Publishing, 2002).
218. A. J. Jerri, Proc. IEEE **65** (11), 1565-1596 (1977).
219. M. Stubenvoll, B. Schafer and K. Mann, Opt. Express **22** (21), 25385-25396 (2014).
220. E. G. Kenneth and A. Mehdi, Microscale Thermophys. Eng. **1** (3), 225-235 (1997).
221. X. Xu, J. Chen and B. Li, J Phys Condens Matter **28** (48), 483001 (2016).
222. A. K. Vallabhaneni, D. Singh, H. Bao, J. Murthy and X. Ruan, Phys. Rev. B **93** (12), 125432 (2016).
223. L. Wang, R. Cheaito, J. L. Braun, A. Giri and P. E. Hopkins, Rev. Sci. Instrum. **87** (9), 094902 (2016).
224. D. H. Hurley, O. B. Wright, O. Matsuda and S. L. Shinde, J. Appl. Phys. **107** (2), 023521 (2010).

225. M. Khafizov, C. Yablinsky, T. R. Allen and D. H. Hurley, Nuclear Instruments and Methods in Physics Research Section B: Beam Interactions with Materials and Atoms **325**, 11-14 (2014).
226. D. Fournier, M. Marangolo, M. Eddrief, N. N. Kolesnikov and C. Fretigny, J. Phys.: Condens. Matter **30** (11), 115701 (2018).
227. J. Yang, E. Ziade and A. J. Schmidt, J. Appl. Phys. **119** (9), 095107 (2016).
228. C. Hua, X. Chen, N. K. Ravichandran and A. J. Minnich, Phys. Rev. B **95** (20), 205423 (2017).
229. C. M. Rost, J. Braun, K. Ferri, L. Backman, A. Giri, E. J. Opila, J.-P. Maria and P. E. Hopkins, Appl. Phys. Lett. **111** (15), 151902 (2017).
230. V. Sinha, J. J. Gengler, C. Muratore and J. E. Spowart, J. Appl. Phys. **117** (7), 074305 (2015).
231. N. Savage, Nature Photonics **4** (2), 124-125 (2010).
232. O. Lozan, M. Perrin, B. Ea-Kim, J. M. Rampnoux, S. Dilhaire and P. Lalanne, Phys. Rev. Lett. **112** (19), 193903 (2014).
233. Z. Tian, A. Marconnet and G. Chen, Appl. Phys. Lett. **106** (21), 211602 (2015).
234. H. Harikrishna, W. A. Ducker and S. T. Huxtable, Appl. Phys. Lett. **102** (25), 251606 (2013).
235. S. A. Putnam, A. M. Briones, J. S. Ervin, M. S. Hanchak, L. W. Byrd and J. G. Jones, Int. J. Heat Mass Transfer **55** (23-24), 6307-6320 (2012).
236. J. Yong Park, A. Gardner, W. P. King and D. G. Cahill, J. Heat Transfer **136** (9), 092902 (2014).
237. J. Park, X. Xie, D. Li and D. G. Cahill, Nanoscale and Microscale Thermophysical Engineering **21** (2), 70-80 (2016).
238. M. Mehrvand and S. A. Putnam, J. Heat Transfer **139** (11), 112403 (2017).
239. M. Mehrvand and S. A. Putnam, Communications Physics **1** (1), 21 (2018).
240. A. J. Schmidt, J. D. Alper, M. Chiesa, G. Chen, S. K. Das and K. Hamad-Schifferli, The Journal of Physical Chemistry C **112** (35), 13320-13323 (2008).
241. Z. Ge, Y. Kang, T. A. Taton, P. V. Braun and D. G. Cahill, Nano Lett. **5** (3), 531-535 (2005).
242. J. Huang, J. Park, W. Wang, C. J. Murphy and D. G. Cahill, ACS Nano **7** (1), 589-597 (2013).
243. J. Park, J. Huang, W. Wang, C. J. Murphy and D. G. Cahill, The Journal of Physical Chemistry C **116** (50), 26335-26341 (2012).
244. M. Highland, B. Gundrum, Y. Koh, R. Averback, D. Cahill, V. Elarde, J. Coleman, D. Walko and E. Landahl, Phys. Rev. B **76** (7), 075337 (2007).